\documentclass{article}

\usepackage{microtype}
\usepackage{graphicx}
\usepackage{subfigure}
\usepackage{booktabs} 
\usepackage{hyperref}
\usepackage{arydshln}
\usepackage{longtable}
\usepackage{xspace}
\usepackage{nicematrix,adjustbox}
\newcommand{\train}{\includegraphics[height=1.1em]{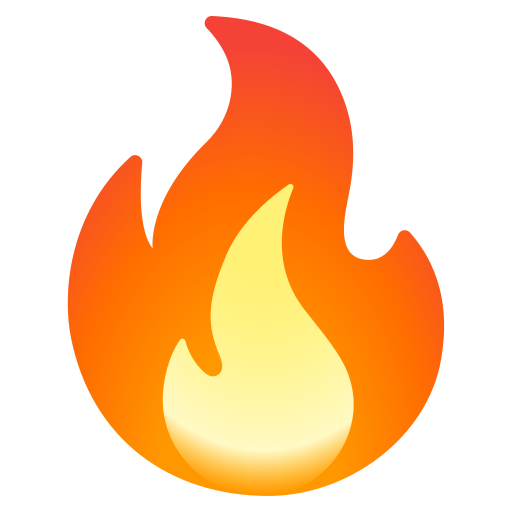}}
\newcommand{\freeze}{\includegraphics[height=1.1em]{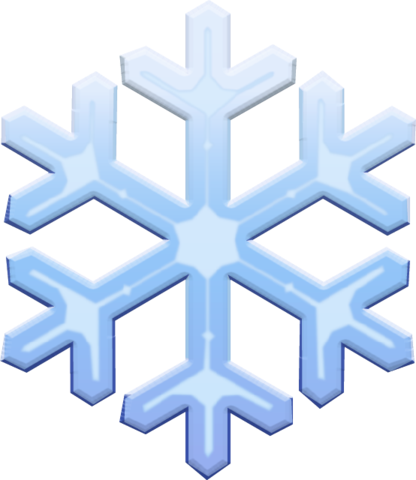}}
\newcommand{\lmmicon}{\includegraphics[height=1.1em]{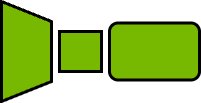}}
\definecolor{metabg}{HTML}{F1F4F7}

\usepackage[skins]{tcolorbox}

\newtcolorbox{mybox}[1]{enhanced,sharp corners=all,colback=white,colframe=gray,toprule=0pt,bottomrule=0pt,leftrule=1pt,rightrule=1pt,overlay={
            \draw[gray,line width=1pt] (frame.north west) -- ++(2cm,0pt);
            \draw[gray,line width=1pt] (frame.south east) -- ++(-2cm,0pt);
    },
}

\usepackage[accepted]{icml2025}

\usepackage{amsmath}
\usepackage{amssymb}
\usepackage{mathtools}
\usepackage{amsthm}
\usepackage{multicol, multirow}
\usepackage[framemethod=TikZ]{mdframed}
\usepackage[skip=2pt]{caption} 
\usepackage{makecell}

\usepackage[capitalize,noabbrev]{cleveref}
\usepackage{colortbl}
\PassOptionsToPackage{table}{xcolor}
\usepackage{xcolor}
\definecolor{revised}{RGB}{0, 0, 255}
\newcommand{\myccone}{\cellcolor[HTML]{f2f2f2}}

\definecolor{nvgreen}{HTML}{76B900}

\theoremstyle{plain}

\theoremstyle{definition}

\theoremstyle{remark}

\usepackage[textsize=tiny]{todonotes}
\usepackage{algorithm}
\usepackage{algpseudocode}
\icmltitlerunning{Audio Flamingo 2: An Audio-Language Model with Long-Audio Understanding and Expert Reasoning Abilities}

\begin{document}

\twocolumn[
\icmltitle{~\includegraphics[height=5pt,trim=0cm 4cm 0 10cm]{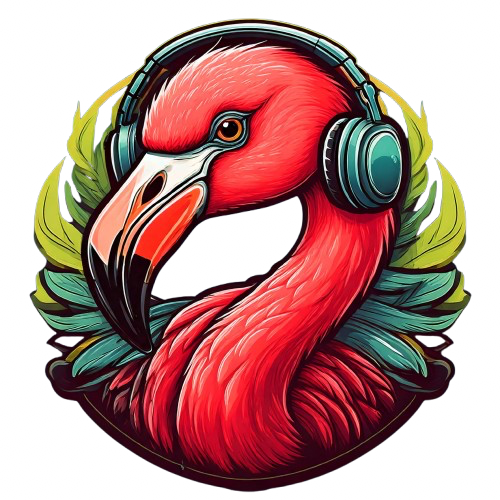} Audio Flamingo 2: An Audio-Language Model with Long-Audio Understanding and Expert Reasoning Abilities}

\icmlsetsymbol{equal}{*}

\begin{icmlauthorlist}
\icmlauthor{Sreyan Ghosh}{equal,nv,maryland}
\icmlauthor{Zhifeng Kong}{nv}
\icmlauthor{Sonal Kumar}{maryland}
\icmlauthor{S Sakshi}{maryland}
\icmlauthor{Jaehyeon Kim}{nv}
\icmlauthor{Wei Ping}{nv}
\icmlauthor{Rafael Valle}{nv}
\icmlauthor{Dinesh Manocha}{maryland}
\icmlauthor{Bryan Catanzaro}{nv}
\end{icmlauthorlist}

\icmlaffiliation{nv}{NVIDIA, Santa Clara, CA, USA}
\icmlaffiliation{maryland}{University of Maryland, College Park, MD, USA}

\icmlcorrespondingauthor{Sreyan Ghosh}{sreyang@umd.edu}
\icmlcorrespondingauthor{Zhifeng Kong}{zkong@nvidia.com}

\vskip 0.3in
]

\printAffiliationsAndNotice{\icmlEqualContribution}

\begin{abstract}
Understanding and reasoning over non-speech sounds and music are crucial for both humans and AI agents to interact effectively with their environments. In this paper, we introduce \textbf{Audio Flamingo 2} (AF2), an Audio-Language Model (ALM) with advanced audio understanding and reasoning capabilities. AF2 leverages (i) a custom CLAP model, (ii) synthetic Audio QA data for fine-grained audio reasoning, and (iii) a multi-stage curriculum learning strategy. AF2 achieves state-of-the-art performance with only a 3B parameter small language model, surpassing large open-source and proprietary models across over 20 benchmarks. Next, for the first time, we extend audio understanding to long audio segments (30 secs to 5 mins) and propose \textbf{LongAudio}, a large and novel dataset for training ALMs on long audio captioning and question-answering tasks. Fine-tuning AF2 on LongAudio leads to exceptional performance on our proposed \textbf{LongAudioBench}, an expert annotated benchmark for evaluating ALMs on long audio understanding capabilities. We conduct extensive ablation studies to confirm the efficacy of our approach. Project Website: \url{https://research.nvidia.com/labs/adlr/AF2/}
\end{abstract}

\vspace{-6mm}
\section{Introduction}
\label{sec:intro}

Understanding non-speech sounds, non-verbal speech, and music (collectively referred to as ``audio'' in this paper) is essential for real-world applications such as detecting anomalies in industrial environments, recognizing emotional cues, and improving assistive technologies for the impaired. While Large Language Models (LLMs) have demonstrated remarkable reasoning capabilities through language, extending these systems to comprehend audio is key to building intelligent systems capable of reasoning with contextual auditory cues~\cite{kong2024audio}. Verbal speech, inherently tied to language, benefits significantly from (L)LM advancements~\cite{watanabe2018espnet,chen2024hyporadise}; however, the potential to enhance perception and reasoning over non-verbal audio remains largely under-explored~\cite{ghosh-etal-2024-gama}.\begin{figure}
    \centering
    \includegraphics[width=0.9\linewidth,trim=1cm 0.25cm 0cm 0cm]{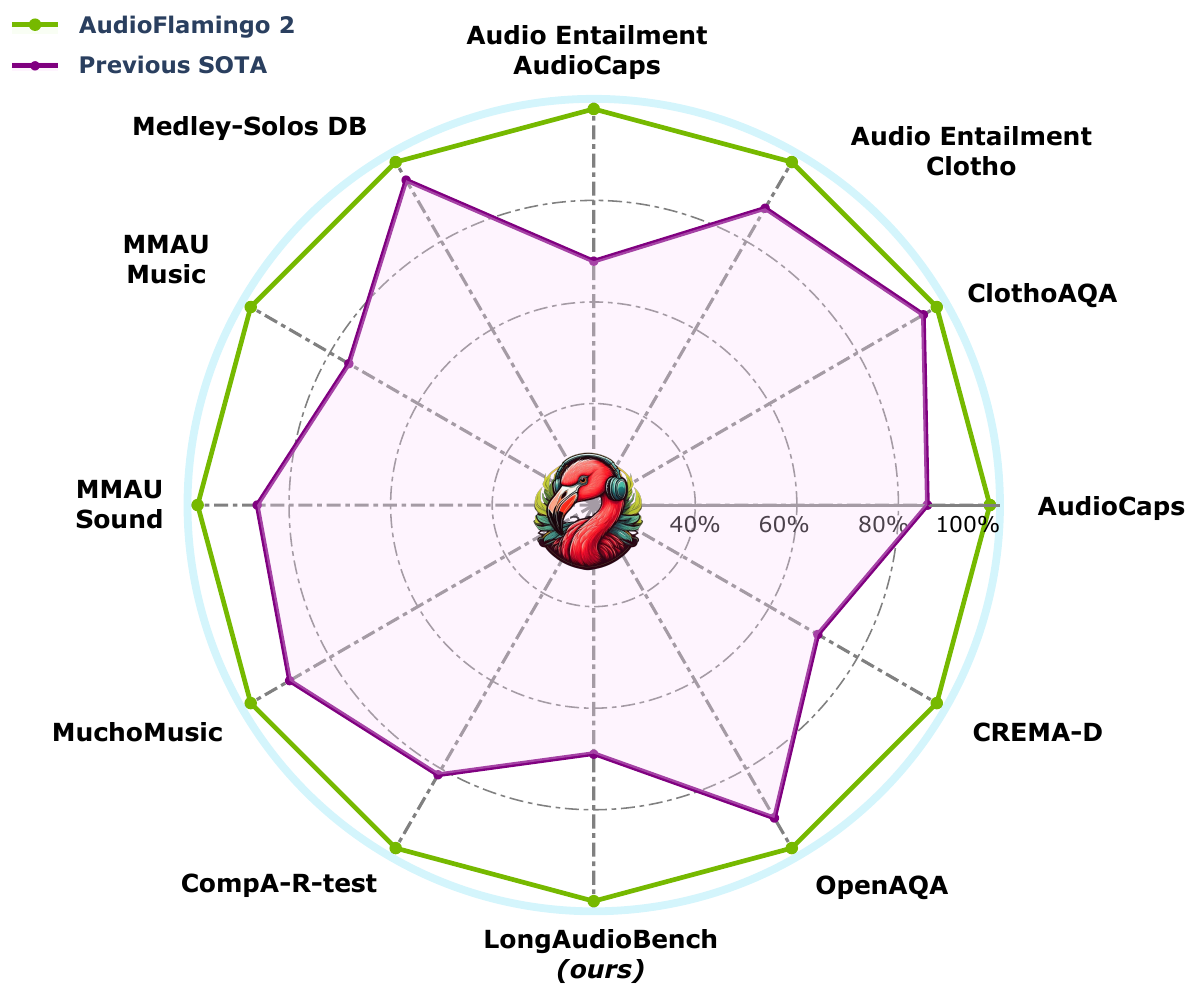}
    \caption{\small Audio Flamingo 2 versus previous SOTA ALMs on audio understanding and reasoning benchmarks (values normalized). AF2 outperforms all baselines and has smaller model footprints.}
    \label{fig:radar_af2}
    \vspace{-5mm}
\end{figure} Audio-Language Models (ALMs) extend language models with audio understanding capabilities. Contrastive Language-Audio Pre-training (CLAP)~\cite{elizalde2023clap} was among the first encoder-only ALMs to bridge audio and language with contrastive learning. Building on this, subsequent efforts introduced Large ALMs (LALMs), which integrate audio encoders with pre-trained decoder-based LLMs, enabling open-ended Audio Question Answering (AQA) and free-form response generation~\cite{gong2021ast,tang2024salmonn,ghosh-etal-2024-gama}. However, despite these developments, even the most advanced LALMs continue to underperform on expert-level reasoning tasks compared to foundational tasks like event classification. For example, Gemini-1.5-Pro~\cite{team2024gemini}, one of the most advanced models, achieves only 54.4\% and 48.5\% on the MMAU sound and music subsets~\cite{sakshi2024mmau}, a benchmark for evaluating expert-level audio reasoning. This underscores the challenges of improving an LLM's ability to understand and reason over audio, which we attribute to the lack of high-quality training data and robust audio representations.

{\noindent \textbf{Main Contributions:}} In this paper, we propose \textbf{Audio Flamingo 2} (AF2), a parameter-efficient ALM that combines a 3B-parameter small decoder LM with a 203M-parameter audio encoder, achieving state-of-the-art audio understanding and reasoning capabilities. There are three major innovations in our method: {$(1)$} \textbf{Data: } Recent studies highlight that improving data quality can rival or even surpass the performance gains achieved by scaling compute and model size~\cite{abdin2024phi}. To this end, we propose \textbf{AudioSkills} (Section~\ref{sec:af2_audioskills}), a large-scale, skill-specific AQA training dataset featuring complex, reasoning-intensive questions paired with each audio. We design questions that target \textit{seven} distinct skills, with the primary aim of improving fine-grained reasoning capabilities in ALMs. {$(2)$} \textbf{Audio Encoding: } We propose AF-CLAP (Section~\ref{sec:af2_clap}), where we scale CLAP training to over 8M audio-caption pairs, incorporating synthetic data, and propose an improved contrastive loss for better representational quality and robustness. {$(3)$} \textbf{Training Strategy: } We propose a novel 3-stage curriculum training strategy (Section~\ref{sec:af2_strat}) for improved performance.

Additionally, for the first time, we extend audio understanding to \textit{long audios}, moving beyond 30-second clips to audios lasting up to 5 minutes. To enable the model to comprehend and reason over long audio, we introduce \textbf{LongAudio}, a novel dataset comprising over 260k carefully curated AQA instances with audios ranging from 30 seconds to 5 minutes. LongAudio spans 10+ audio categories and supports 6 tasks, including captioning and 5 reasoning-based QA tasks. Next, we introduce \textbf{LongAudioBench}, an expert-annotated benchmark for evaluating ALMs on long audio understanding. AF2, trained on LongAudio, significantly outperforms our baseline. In summary, our main contributions are:

\begin{enumerate}
\vspace{-3mm}
\setlength\parskip{0em}
    \item We present Audio Flamingo 2, a SOTA ALM with advanced audio understanding and reasoning capabilities. 
    
    \item We propose innovations in data generation, architecture design, representation learning, and training strategies.

    \item We introduce the long audio understanding task and create dedicated training and evaluation datasets to drive progress in this area.

    \item Audio Flamingo 2 outperforms larger and proprietary LALMs across over 20 benchmarks, despite being smaller and trained exclusively on public datasets.

    \item We conduct systematic ablation studies to demonstrate the impact of each design choice.
\end{enumerate}

\vspace{-5mm}
\section{Related Work}
\label{sec:related}
\vspace{-1mm}

{\noindent \textbf{Audio-Language Models:}} ALMs can be classified into two broad categories: \textit{\textbf{1) Encoder-only ALMs:}} Encoder-only ALMs are a class of Multi-Modal Language Models (MLLM) that learn a shared space between the audio and language modalities with an encoder-only LM and an audio encoder. CLAP, a pioneering encoder-based ALM inspired by CLIP~\cite{radford2021learning}, showed state-of-the-art performance on audio-language tasks like retrieval, zero-shot classification, etc. Following this, several attempts have been made to improve CLAP by scaling data~\cite{Wu2022LargeScaleCL}, incorporating additional training objectives~\cite{ghosh2024compa}, or with synthetic data~\cite{ghosh2024reclap}. Other notable works include Wav2CLIP~\cite{wav2clip}, AudioClip~\cite{guzhov2022audioclip} and CoLLAT~\cite{silva2023collat}. \textit{\textbf{2) Decoder-based ALMs:}} With the advent of LLMs, Pengi~\cite{deshmukh2023pengi}, a pioneering decoder-based ALM, achieved SOTA results on a variety of audio classification tasks. Following Pengi, a large number of Large ALMs were introduced, including fully open-source models like LTU~\cite{gong2024listen}, LTU-AS~\cite{10389742}, SALMONN~\cite{tang2024salmonn}, AudioGPT~\cite{huang2024audiogpt}, GAMA~\cite{ghosh-etal-2024-gama}, Audio Flamingo~\cite{kong2024audio}, and open-access models like Qwen-Audio~\cite{chu2023qwen} and Qwen-2-Audio~\cite{chu2024qwen2}. A majority of the advances have focused on scaling model size and datasets, with very little advancements in data quality or audio encoder representations. This has eventually translated to performance advancing on foundational tasks like classification and captioning but under-performing on expert-level reasoning, the skill required for advancing towards Artificial General Intelligence (AGI)~\cite{morrisposition}.
\begin{figure*}
    \centering
    \includegraphics[width=\linewidth]{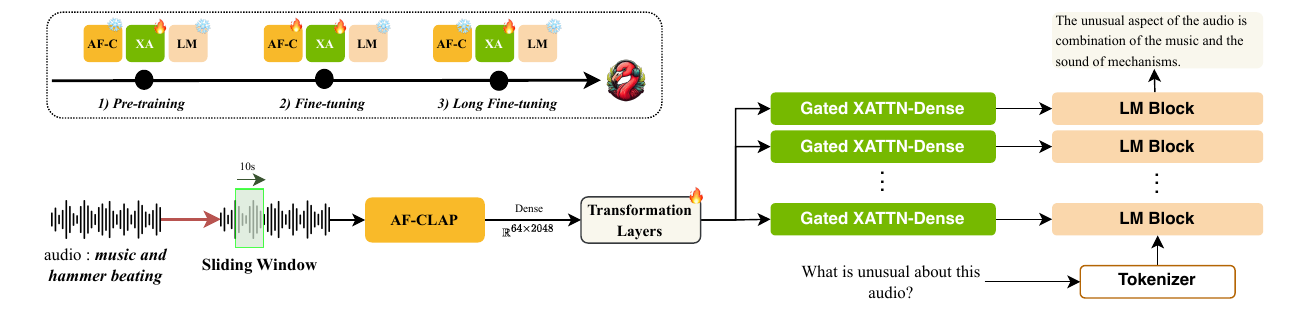}
    \caption{\small Overview of \textbf{Audio Flamingo 2}'s cross-attention architecture and three-stage curriculum training.}
    \label{fig:train_pipe}
    \vspace{-4mm}
\end{figure*}

{\noindent \textbf{Long Audio Understanding:}} Current ALMs are limited to perceiving at most 30 seconds of audio~\cite{chu2024qwen2}, with a majority confined to at most 10 seconds~\cite{ghosh-etal-2024-gama,gong2024listen}. This can be attributed to the fact that these models employ audio encoders that support only short audio encoding and datasets where a majority of the data is only at most 10 seconds. While long speech has received some attention~\cite{chiu2019comparison,arumugam2023improved}, and the field on long video understanding has seen advancements recently~\cite{weng2025longvlm,xue2024longvila}, to the best of our knowledge, no current work or datasets exist for long audio understanding and reasoning.

\vspace{-3mm}
\section{Audio Flamingo 2 Architecture}
\label{sec:af2}
\vspace{-1mm}

Fig.~\ref{fig:train_pipe} summarizes the AF2 architecture, consisting of four primary components: \textit{i)} AF-CLAP: a CLAP-based audio encoder with sliding window feature extraction, \textit{ii)} audio representation transformation layers for additional capacity, \textit{iii)} a decoder-only language model, and  \textit{iv)} gated cross-attention (XATTN-Dense) layers for audio conditioning.

\subsection{AF-CLAP Audio Encoder}
\label{sec:af2_clap}

CLAP, pre-trained with contrastive loss on audio-caption pairs, shows strong audio understanding and natural language alignment \citep{Wu2022LargeScaleCL,elizalde2023clap}. These make CLAP a suitable choice for building ALMs. However, CLAP is less favored in prior works \citep{tang2023salmonn,chu2023qwen} due to its under-performance compared to SSL pre-trained audio encoders~\cite{gong2023contrastive,li2024mert,ghosh-etal-2024-gama}. We hypothesize this is due to the limited availability of high-quality audio-caption pairs, which causes CLAP representations to struggle with compositional reasoning~\cite{ghosh2024compa} and linguistic variations in captions~\cite{selvakumar2024audio}.

In this section, we address these issues and introduce an improved version of CLAP called AF-CLAP. Specifically, we focus on: \textit{i)} constructing a large-scale, high-quality training dataset, and \textit{ii)} improving the training objective to for better representational robustness.

\subsubsection{AF-CLAP Training Dataset}
\label{sec:af2_clap_data}

We scale the training dataset for AF-CLAP to over 8M (10-second) audio-caption pairs. We collect these data from open-source audio and video datasets, with an emphasis on audio diversity and caption accuracy.

Existing models like Laion-CLAP~\citet{Wu2022LargeScaleCL} and MS-CLAP~\cite{CLAP2023} rely heavily on single-label audio classification datasets for captions. This limits their ability to generalize to complex, real-world audio with diverse compositions. Automated captioning efforts (e.g., ~\citet{yuan2024sound}) yield only about 1.4M pairs, lack diversity, and under-represent critical domains like home sounds.

Inspired by the recent success of training vision LMs on images from long videos~\cite{venkataramanan2024is}, we collect audio from open long-video datasets. Specifically, we select 100k diverse videos from MiraData~\cite{ju2024miradata} and Video Recap~\cite{10657851} (see Appendix~\ref{subsec:app_video_topic} for selection details). We segment these videos into 10-second clips and generate video captions using Qwen2-VL-2B-Instruct and audio captions using Qwen2-Audio. To ensure diversity and reduce redundancy, we filter out segments with audio-visual similarity above a threshold $p$. We then prompt GPT-4o(2024-05-13) (Prompts~\ref{fig:mira_short_prompt} and \ref{fig:audio_caption_recap_prompt}) to generate audio-centric captions that exclude visual attributes and emphasize sound events.  Using this approach, we collect approximately \textbf{5.5M} new audio-caption pairs, including 4M from unlabeled short audios and 1.5M from long videos. Detailed dataset statistics are provided in Table~\ref{tab:clap_datasets}.

\subsubsection{AF-CLAP Training Objective}
\label{sec:af2_clap_training}
Compared to the standard contrastive loss in audio-language
pre-training~\cite{Wu2022LargeScaleCL,elizalde2023clap}, we improve the training objective for better robustness to linguistic variations and compositional reasoning abilities.

\noindent \textbf{Improving Linguistic Invariance:}  CLAP-like models struggle to generalize to linguistic variations in captions that humans easily understand~\cite{selvakumar2024audio} (e.g., failing to equate \textit{helicopter} and \textit{chopper}). To address this, for every caption in the dataset, we generate $M-1$ linguistically varied captions with identical semantics and composition. These variations, along with the ground-truth caption, are treated as \textit{positives}.

\noindent \textbf{Improving Compositional Reasoning:} Captions with different word orders or structures often convey distinct relationships between acoustic events, such as temporal sequencing or attribute binding. However, CLAP-like models struggle to capture these nuances (e.g., differentiating whether one sound follows or precedes another)~\cite{ghosh2024compa}. To address this, we introduce composition-aware negatives. For every $M$ positive captions, we generate $N$ variations with modified temporal or attribute compositions and use them as \textit{negatives}. An example is below:

\begin{mybox}
\noindent\textbf{Original Caption:} A dog barking followed by the sound of a train approaching.

\noindent\textbf{Positive:} A dog barking followed by the sound of a railcar approaching.

\noindent\textbf{Negative:} A dog barking preceded by the sound of a railcar approaching.
\end{mybox}

\begin{figure}
    \includegraphics[width=\columnwidth,trim=0.5cm 0.5cm 0.75cm 0]{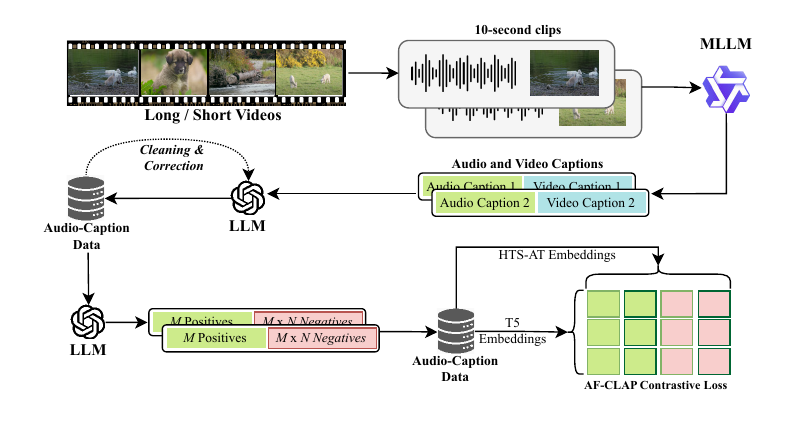}
    \caption{\small Illustration of AF-CLAP training process. We collect long and short videos, segment it into 10-second clips, caption it, and prompt an LLM to generate audio captions. The data is then used to train CLAP with a modified contrastive loss.}
    \label{fig:enter-label}
\vspace{-4mm}
\end{figure}

\noindent \textbf{Contrastive Loss.} Each sample $x$ in the training data $\mathcal{X}$ now has $M$ positives $\{\mathcal{P}(x)_m\}_{m=1}^M$ and $M \times N$ negatives $\{\mathcal{N}(x)_{m,n}\}_{m=1,n=1}^{M,N}$. Let $\mathcal{A}(x)$ be audio embedding from the HTSAT\textsubscript{large} audio encoder~\cite{htsatke2022} with MLP projection, and $\mathcal{T}(x)$ be the text embedding from Flan-T5 text encoder \citep{raffel2020exploring,chung2024scaling} with MLP projection. Let $\texttt{s}(u,v)=\exp(u^{\top}v/\tau)$ be the similarity function with temperature $\tau$. Our training objective is:
\vspace{-1.5mm}
\[
\mathcal{L} = -\frac{1}{B} \sum_{i=1}^{B} \log \frac{S(i,i)}{S_{\text{neg}}(i) + \sum_{j=1}^B S(j,i)},
\]
\vspace{-2mm}
where  
\vspace{-1mm}
\[
S(j,i) = \sum_{m} \texttt{s}(\mathcal{T}(\mathcal{P}(x_{j})_m), \mathcal{A}(x_i)),
\]
\vspace{-2mm}
\[
S_{\text{neg}}(i) = \sum_{m,n} \texttt{s}(\mathcal{T}(\mathcal{N}(x_{i})_{m,n}), \mathcal{A}(x_i)),
\]
\vspace{-2mm}

where $B$ is the batch size.
Our approach encourages CLAP to learn both linguistic invariance and compositional reasoning, aligning its capabilities more closely with human-like understanding and thus providing more human-aligned representations. AF-CLAP achieves SOTA performance on various retrieval and audio classification benchmarks. Since comparing CLAP performance is beyond our scope, we compare results in Tables~\ref{tab:exp-t2a-retrieval} and ~\ref{tab:zshot_res}.

\vspace{-1mm}
\subsection{Audio Conditioning Architecture \& LLM}
\label{sec:af2_arch}

\textbf{Audio Feature Extraction.}  
For each 10-second segment, we extract dense audio features $h \in \mathbb{R}^{64 \times 2048}$ from the penultimate layer of AF-CLAP. This approach yields higher-quality features compared to mean-pooled representations from the final layer (see Appendix~\ref{app:feature_extract}). For longer audio, we use non-overlapping sliding windows to compute and concatenate audio features. The maximum number of sliding windows varies across training stages, with up to 30 windows (5 minutes) when training on LongAudio. Once the sliding window features are obtained, we apply RoPE~\cite{su2023roformerenhancedtransformerrotary} with a base of 4096 to encode temporal information into the features.

\textbf{Representation Transformation Layers.} 
To expand model capacity, we apply three self-attention layers to the audio feature representations, each with 8 heads and an inner dimension of 2048 \citep{kong2024audio,vaswani2017attention}.

\textbf{Gated Cross-Attention.} Following Audio Flamingo, we use gated cross-attention dense (XATTN-Dense) layers from Flamingo~\cite{alayrac2022flamingo} to condition audio representations on the LLM. Each layer consists of two blocks: (1) a residual block with cross-attention and tanh gating and (2) a residual block with a dense layer and tanh gating. These layers are inserted before each LLM block.

The XATTN-Dense layers reduces the quadratic attention complexity in prefix tuning to linear complexity. For instance, let $l_1=80$ be the text token length and $l_2=30\times64$ be the number of audio embeddings. Prefix tuning requires a self-attention complexity of $(l_1+l_2)^2=4\times10^6$, whereas our cross-attention complexity is around $l_1\times l_2\approx 1.5\times10^5$. 

\textbf{Frozen Language Model.} Our architecture uses Qwen2.5-3B~\citep{yang2024qwen2}, a decoder-only causal LLM with 3B parameters, 36 hidden layers, and 16 attention heads. We find this model to have the best cost-performance ratio, as it has enough capacity for audio understanding and is light enough (see Section~\ref{subsec:llm_performance} for a comparison of LLM sizes).

\begin{figure*}[t]
    \centering
    \includegraphics[width=\linewidth]{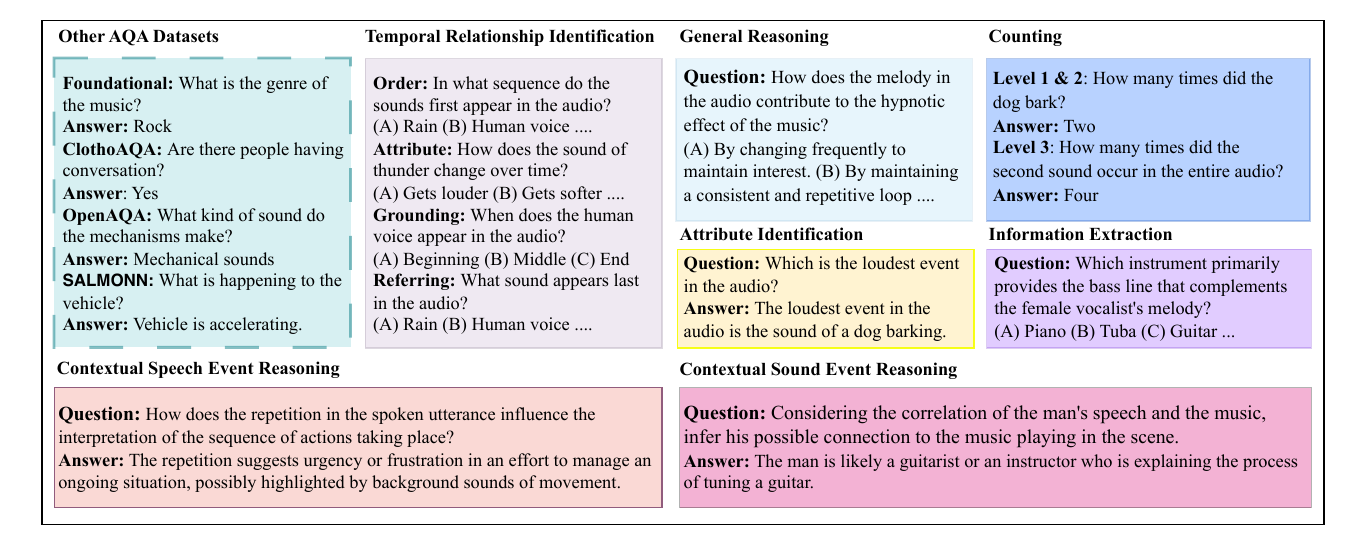}
    \caption{\small Examples from \textbf{AudioSkills}. Compared to other AQA datasets, questions in AudioSkills require deliberate reasoning.}
    \label{fig:audioskills_ex}
    \vspace{-4mm}
\end{figure*}

\vspace{-2mm}
\section{Audio Flamingo 2 Training Data}
\label{sec:af2_data}
\vspace{-1mm}

 Table~\ref{tab:dataset_by_type} summarizes the datasets used to train AF2. We first convert common benchmark datasets used in Audio Flamingo to AQA format (see Appendix~\ref{app:prompt_found}). In addition, we propose two datasets: AudioSkills for audio expert reasoning and LongAudio for long audio understanding.

\vspace{-2mm}
\subsection{AudioSkills: An Expert Audio Reasoning Dataset}
\label{sec:af2_audioskills}

Expert-level reasoning requires mastery of diverse and specialized skills~\cite{ericsson2003acquisition,huang2022towards}. However, most existing audio datasets focus on surface-level properties, such as acoustic events or category classification. This limitation extends to QA pairs derived from these datasets, which often fail to require expert-level reasoning.

To address this gap, we introduce \textbf{AudioSkills}, a high-quality, skill-specific synthetic dataset designed to prioritize the development of reasoning and problem-solving abilities. This dataset is carefully curated to ensure diversity and relevance, grounded in the hypothesis that expert reasoning emerges from the mastery of various relevant skills and world knowledge. Specifically, we use a combination of open-source sound and music datasets and synthetic audio and prompt GPT-4o, along with any available metadata, to create QA pairs. AudioSkills focuses on audios $\leq30$ seconds long to enable effective scaling on open-source datasets. It spans \textit{seven} distinct skills selected for their plausibility, relevance, and ablation insights as follows:

\begin{enumerate}
\vspace{-3mm}
\setlength\parskip{0em}
    \item \textbf{Temporal Reasoning:} Understanding temporal relationships between acoustic events in the audio. We generate 4 types of MCQ-type QA pairs: \textit{\textbf{i) Temporal Order:}} Understanding the sequential order of events or sounds; \textit{\textbf{ii) Temporal Attribute:}} Understanding how sound attributes change over time; \textit{\textbf{iii) Temporal Referring:}} Answering questions referring to sounds at specific temporal locations in the audio (e.g., start, middle, end); \textit{\textbf{iv) Temporal Grounding:}} Identifying the temporal location of specific acoustic events.

    \item \textbf{Attribute Identification:} Recognizing attributes of events in multi-event audio, such as sound characteristics (e.g., \textit{loud} bang) or gender (\textit{e.g., \textit{male} speech}). We generate QA pairs using attributes extracted via the method proposed by~\citet{kumar2024sila}.

    \item \textbf{Counting:} Counting occurrences of specific sounds in an audio. We use the Synthio TTA model~\cite{ghosh2024synthio} to create QA pairs at 3 difficulty levels: \textit{\textbf{i) Level 1:}} Single sounds concatenated; count the event occurrences. \textit{\textbf{ii) Level 2:}} Multiple interleaved sounds; count the main sound occurrences. \textit{\textbf{iii) Level 3:}} Same as Level 2 but referenced by their temporal position.
    
    \item \textbf{Contextual Sound Event Reasoning:} Identifying the purpose of a sound or acoustic event in the context of other sounds and events in the audio. This skill requires audio understanding, temporal reasoning, world knowledge, and various other skills. Inspired by CompA-R in GAMA~\cite{ghosh-etal-2024-gama}, we expand the dataset from $\approx$200k to $\approx$350k QA pairs.

    \item \textbf{Contextual Speech Event Reasoning:} Similar to Contextual Sound Event Reasoning but focused on identifying the purpose of spoken utterances in relation to other sounds or events.

    \item \textbf{Information Extraction:} Focuses on understanding characteristics of the audio beyond just surface-level properties, detailed content analysis, and the application of external world knowledge when necessary.

    \item \textbf{General Reasoning:} This category encompasses questions that do not fall into the specific skill types above but require a combination of multiple skills or unique abilities, such as identifying relationships between multiple events or interpreting complex scenarios.

\end{enumerate}
\vspace{-0.5em}
In total, we generate $\approx$4.2M QA pairs. Figure~\ref{fig:audioskills_ex} compares AudioSkills to other open-source AQA datasets, highlighting that these datasets lack the complexity required for deliberate reasoning. While training on such QA pairs is useful for alignment~\cite{zhou2024lima,wolf2023fundamental}, it does not equip models with the specialized skills needed to handle complex questions~\cite{sakshi2024mmau,ghosh2024closer}. \textit{Additional details, including statistics, metadata, and prompts, are provided in Appendix~\ref{subsec:app_skills_stats} and Table~\ref{tab:as_cat_tab}.}

\begin{figure*}[t]
    \centering
    \includegraphics[width=\linewidth, trim=0 0.25cm 0 0]{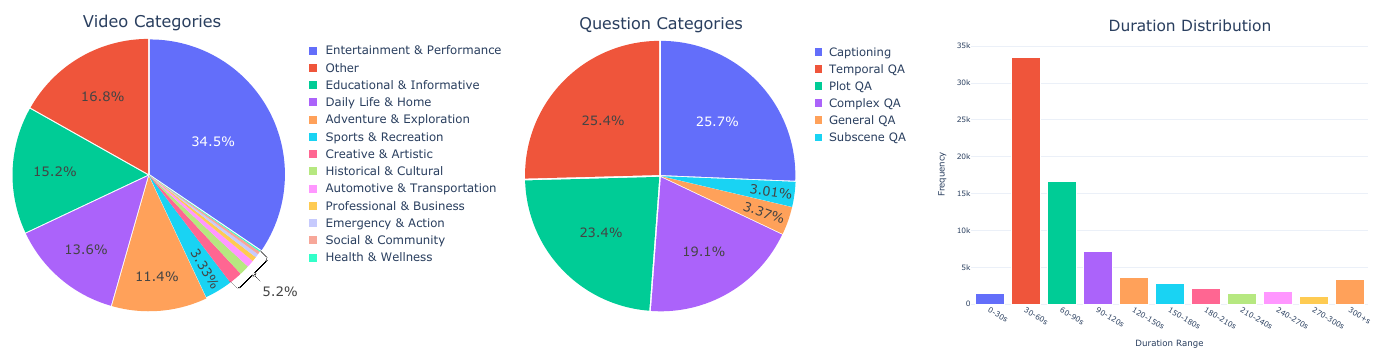}
    \caption{\small The proportion of video categories (for sourcing audios) (\textbf{left}), question categories (\textbf{middle}), and distribution of durations (\textbf{right}) for the LongAudio dataset with 262,928 unique AQA and 80k unique audios. We target captioning and reasoning tasks.}
    \label{fig:af2_charts}
\end{figure*}
\begin{figure*}[h]
    \centering
    \includegraphics[width=\linewidth,trim=0 0.25cm 0 0]{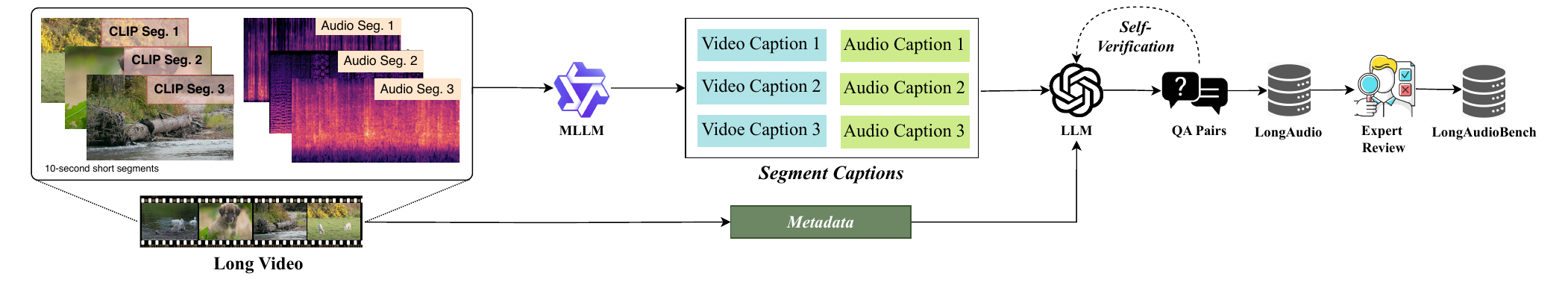}
    \caption{\small The pipeline for generating \textbf{LongAudio}. The process begins by segmenting the long video into short video and audio clips, each $\approx$10 seconds. These
clips are individually annotated with captions. Subsequently, an LLM
is employed to generate question-and-answer pairs based on the captions of these clips. A subset of the data goes through expert review to construct \textbf{LongAudioBench}.}
    \label{fig:long_qa_pipeline}
\end{figure*}

\subsection{LongAudio: A Long Audio Understanding Dataset}
\label{sec:af2_longaudio}

We construct \textbf{LongAudio}, the first large-scale long audio understanding dataset, which is comprised of over 80K unique audios and approximately 263K AQA pairs. Audios are sourced from open long-video datasets, including a subset of MiraData\citep{ju2024miradata} (featuring diverse content from natural scenes to gaming) and the entire Video ReCap\citep{10657851} (egocentric videos of daily activities). To ensure diversity, we filter MiraData by clustering videos and carefully selecting samples from each cluster (see Appendix~\ref{subsec:app_video_topic}). Fig.~\ref{fig:af2_charts} (left) categorizes these videos and topics across several domains. We generate captions for video and audio segments using Qwen2-VL-2B-Instruct and Qwen2-Audio, respectively. GPT-4o is then used to generate questions that ask for captions, challenge models to reason, or extract information from long audios. Fig.~\ref{fig:long_qa_pipeline} outlines the LongAudio creation process. Fig.\ref{fig:af2_charts} (middle) shows the distribution of these categories as outlined below:

\begin{enumerate}
\vspace{-3mm}
\setlength\parskip{0em}
    \item \textbf{Captioning:} The task is to generate detailed and accurate descriptions of long audio, focusing on capturing the essence of the entire audio.

    \item \textbf{Plot QA:} The task is to answer questions related to the overarching narrative or storyline of the audio, requiring an understanding of temporal and causal relationships between individual acoustic events.

    \item \textbf{Temporal QA:} The task is to identify the temporal positions of acoustic events and their relationships, such as order or overlap within the audio, including how certain attributes change with time.   
    
    \item \textbf{Needle QA:} The task is to locate and reason about a specific ``needle'' segment embedded within a longer audio ``haystack,'' ensuring the response is tied explicitly to the needle.

    \item \textbf{Subscene QA:} The task is to answer questions about a specific subscene within the audio, requiring identification and understanding of localized events. 

    \item \textbf{General QA:} The task is to address broad, open-ended questions about the audio that may span multiple events, themes, or contexts, demonstrating overall comprehension and reasoning.
\end{enumerate}
\vspace{-0.5em}
\begin{table*}[t]
    \centering
    \caption{\small Comparison of Audio Flamingo 2 with previous SOTA LALMs on foundational audio understanding benchmarks. CAP = Captioning, AQA = Audio Question Answering, CLS = Classification, ZS=Zero-Shot. Also, see note in Section~\ref{app_sec:clap_limitations}.}
    \label{tab:in-domain-compact}
    \resizebox{\linewidth}{!}{
    \begin{tabular}{l|c|l|l|l|c|l|l}
    \toprule
    \textbf{Dataset} & \multicolumn{1}{c|}{\textbf{Task}} & \textbf{Previous SOTA} & \textbf{Ours} & \textbf{Dataset} & \multicolumn{1}{c|}{\textbf{Task}} & \textbf{Previous SOTA} & \textbf{Ours} \\
    \midrule
    ClothoAQA\textsubscript{unan.} & AQA - ACC & 74.9\% - Qwen2-A & \textbf{86.9\%}$_{\textcolor{nvgreen}{+12.0\%}}$ & Clotho-v2 & {CAP - CIDEr} & 0.45 - Qwen2-A & \textbf{0.46}$_{\textcolor{nvgreen}{+0.01}}$  \\
    ClothoAQA\textsubscript{non-bin} & AQA - ACC & 49.5\% - AF & \textbf{52.6\%}$_{\textcolor{nvgreen}{+3.1\%}}$ & AudioCaps$^{\text{ZS}}$ & CAP - CIDEr & 0.46 - AF & $\textbf{0.58}$$_{\textcolor{nvgreen}{+0.12}}$ \\
    MusicAVQA\textsubscript{audio} & AQA - ACC & 72.1\% - Qwen-A & \textbf{72.3\%}$_{\textcolor{nvgreen}{+0.2\%}}$ & CREMA-D$^{\text{ZS}}$ & CLS - ACC & $26.5\%$ - AF & \textbf{36.6\%}$_{\textcolor{nvgreen}{+10.1\%}}$ \\ 
    NonSpeech7k & CLS - ACC & 83.9\% - AF & \textbf{84.3\%}$_{\textcolor{nvgreen}{+0.4\%}}$ & Ravdess$^{\text{ZS}}$ & CLS - ACC & 20.9\% - AF & \textbf{26.3\%}$_{\textcolor{nvgreen}{+5.4\%}}$ \\
    CochlScene & CLS - ACC & \textbf{91.6\%} - Pengi & 82.1\%$_{\textcolor{red}{-9.5\%}}$ & GTZAN$^{\text{ZS}}$ & CLS - ACC & 65.2\% - AF & \textbf{69.1\%}$_{\textcolor{nvgreen}{+3.9\%}}$ \\
    NS\textsubscript{source} & CLS - ACC & 60.1\% - Pengi & \textbf{62.0\%}$_{\textcolor{nvgreen}{+1.9\%}}$ & Medley-solos-DB$^{\text{ZS}}$ & CLS - ACC & $85.6\%$ - GAMA & $\textbf{85.8\%}$$_{\textcolor{nvgreen}{+0.2\%}}$ \\
    NS\textsubscript{instrument} & CLS - ACC & \textbf{78.8\%} - Qwen-A & 71.1\%$_{\textcolor{red}{-7.7\%}}$ & US8K$^{\text{ZS}}$ & CLS - ACC & \textbf{71.2\%} - AF & 68.0\%$_{\textcolor{red}{-3.2\%}}$ \\
    FSD50k & CLS - mAP & 47.9\% - GAMA & \textbf{49.2\%}$_{\textcolor{nvgreen}{+1.3\%}}$ & ESC50$^{\text{ZS}}$ & CLS - ACC & 80.6\% - GAMA & \textbf{83.9\%}$_{\textcolor{nvgreen}{+3.3\%}}$ \\
    \bottomrule
    \end{tabular}}
\end{table*}

Fig.\ref{fig:af2_charts} (right) shows the distribution of audio lengths, and 
Table~\ref{tab:longbench_ex} shows examples of each category.

{\noindent \textbf{LongAudioBench:}} We sample 10\% ($\approx$3k) from LongAudio, proportionally across QA types, for expert human verification. Annotators manually verify and correct these instances, discarding any incorrect ones. Subsequently, three authors conduct a quality check on the corrected samples. The full annotation process is detailed in Appendix~\ref{subsec:app_human_verification}. The final version of LongAudioBench has 2,429 instances. For evaluation, following a wealth of prior work~\cite{ghosh-etal-2024-gama,yang2024air}, we use an LLM-as-a-judge framework~\cite{zheng2023judging}: we prompt GPT-4o with the model response and ground truth answer using prompt~\ref{fig:evaluator_prompt} and score the response on a scale of 1 to 10.

\section{Audio Flamingo 2 Training Strategy}
\label{sec:af2_strat}

We train AF2 using a 3-stage curriculum learning strategy, progressively increasing the audio context length and improving data quality:  \textbf{\textit{Stage 1: Pre-training.}} This stage focuses on multi-modal alignment to align audio representations with the LLM. For training, we leverage large-scale and plausibly noisy classification and captioning datasets. Inspired by~\cite{cheng2024instruction}, we include a small amount of high-quality QA data to improve pre-training. During this stage, CLAP and LLM layers are frozen, and only the audio representation transformation and gated cross-attention layers are trainable. The audio context is restricted to $w=3$ windows (30 seconds). \textbf{\textit{Stage 2: Fine-tuning.}} This stage focuses on improving audio understanding and reasoning by learning skills. For training, we use high-quality short-audio classification, captioning, and QA datasets. This stage is often referred to as \textit{instruction-tuning}~\cite{chu2024qwen2,ghosh-etal-2024-gama}.  During this stage, only the LLM layers are frozen, and the CLAP model, audio representation transformation layers, and gated cross-attention layers are trainable. The audio context length is increased to $w=9$ windows (1.5 minutes). \textbf{\textit{Stage 3: Long Fine-tuning.}} This stage focuses on context-length extension and teaching skills specific to enabling reasoning on long audios. For training, we employ our proposed LongAudio dataset, keep the audio representation transformation and gated cross-attention layers trainable, and increase $w=30$ windows (5 minutes).

\section{Experiments}
\label{sec:experimental_setup}

\subsection{Experimental Setup}
\label{subsec:three_stage_training}

We train our model using 128 NVIDIA A100 80GB GPUs. During pre-training, we use an effective batch size of 1024, the AdamW optimizer (learning rate = \(10^{-4}\), weight decay = 0.1), and bf16 with automatic mixed precision for efficiency. For fine-tuning and long fine-tuning, we adopt dynamic batching based on audio length, ensuring batch sizes are multiples of 2, with effective batch sizes ranging from 128 to 1024 (see Appendix~\ref{subsec:data_loader}). We benchmark our model against recent SOTA LALMs, including GAMA, Audio Flamingo, Qwen-Audio, Qwen2-Audio, LTU, LTU-AS, SALMONN, AudioGPT, and Gemini (Flash~v2 and Pro~v1.5), GPT-4o-audio and report results for the best model. For zero-shot evaluations, we exclude corresponding datasets from all training stages, consistent with prior work~\cite{gong2024listen,ghosh-etal-2024-gama}. For LongAudioBench, as current LALMs (except Gemini) do not support audio inputs of length $\geq$30 seconds, we adopt a two-step cascaded approach. First, we generate captions for short segments of the input audio, and then we prompt the same LALM to answer the question using these captions. Our experiments with passing the original long audio consistently outperformed our cascaded approach. For Gemini, we prompt it with the entire original long audio.

\subsection{Smaller but Better}
\label{subsec:perform_found}

We employ standard benchmark datasets for evaluation (Table~\ref{tab:inference_data}). Foundational benchmarks include ClothoAQA, MusicAVQA (audio-only), NonSpeech7k, NSynth, CREMA-D, Ravdess, GTZAN, Medley-solos-DB and USD8K and Clotho-v2 and AudioCaps. Reasoning benchmarks include MMAU (sound and music subsets), Audio Entailment, OpenAQA-test, MuchoMusic, CompA-R-\textit{test}, MusicInstruct (Long subset), MusicQA, CMM (audio-language subset) and LongAudioBench.

{\noindent \textbf{Foundational Audio Understanding:}} Table~\ref{tab:in-domain-compact} presents the performance of AF2 on foundational audio understanding benchmarks. For evaluation, we follow the similarity-based retrieval approach using our CLAP model, proposed by ~\citet{deshmukh2023pengi} and widely adopted. While foundational audio understanding is not our primary focus, AF2 shows competitive results against SOTA LALMs while having half of their size (e.g., Qwen(2)-A, GAMA, and LTU are equipped with 7B LLMs and large audio encoders).

{\noindent \textbf{Expert Reasoning:}} Table~\ref{tab:reasoning_benchmarks} compares the performance of AF2 on audio reasoning benchmarks. We employ the original evaluation strategy and metrics. With a much smaller LLM, AF2 outperforms all LALMs by large margins. We provide fine-grained results and failure cases of AF2 on LongAudioBench in Table~\ref{tab:long_fine_grained_values} and Table~\ref{tab:reasoning_scores}.

\begin{table}[t]
    \centering
    \caption{\small Comparison of Audio Flamingo 2 with previous SOTA LALMs on audio reasoning benchmarks, all AQA-based.}
    \resizebox{\columnwidth}{!}{
    \begin{tabular}{l|l|l}
    \toprule
    \textbf{Dataset} & \multicolumn{1}{c|}{\textbf{Previous SOTA}} & \textbf{Ours} \\ \midrule
         MMAU Sound  & 61.7\% - Gemini F v2 & \textbf{65.1\%}$_{\textcolor{nvgreen}{+3.4\%}}$ \\
         MMAU Music   &  56.5\% - Gemini F v2 &  \textbf{72.9\%}$_{\textcolor{nvgreen}{+16.4\%}}$\\
         AE Clotho   & 83.3\% - Qwen-A & \textbf{92.5\%}$_{\textcolor{nvgreen}{+9.2\%}}$ \\
         AE AudioCaps   & 64.2\%  - Qwen-A & \textbf{93.3\%}$_{\textcolor{nvgreen}{+29.1\%}}$\\
         CompA-R-\textit{test}  & 80.0\% - GAMA-IT & \textbf{96.4\%}$_{\textcolor{nvgreen}{+16.4\%}}$\\
         MuchoMusic  & 51.4\% - Qwen-A & \textbf{56.5\%}$_{\textcolor{nvgreen}{+5.1\%}}$\\
         OpenAQA  & 80.0\% - GAMA-IT & \textbf{86.0\%}$_{\textcolor{nvgreen}{+6.0\%}}$ \\
         MusicInstruct (Long)  & 86.1\% - MusiLingo & \textbf{90.2\%}$_{\textcolor{nvgreen}{+4.1\%}}$ \\
          MusicQA  & 90.0\% - MusiLingo & \textbf{93.0\%}$_{\textcolor{nvgreen}{+3.0\%}}$ \\
          CMM Hallucination & 76.0\% - SALMONN & \textbf{82.0\%}$_{\textcolor{nvgreen}{+6.0\%}}$ \\
         LongAudioBench \textit{(ours)} & 45.3\% - Gemini F v2 & \textbf{64.2\%}$_{\textcolor{nvgreen}{+18.9\%}}$ \\ \bottomrule
    \end{tabular}}
\label{tab:reasoning_benchmarks}
\vspace{-4mm}
\end{table}

\subsection{Enhanced Audio Features Boost Performance}
\label{subsec:audio_rep_matter}

Table~\ref{tab:clap_comparison_transposed} compares the performance of AF2 using AF-CLAP against various SOTA CLAP models on benchmark datasets. The MMAU score is averaged across the sound and music subsets. ``AF-CLAP 630k'' refers to AF-CLAP trained using the same strategy but with Laion-CLAP's 630k audio-text pairs. ``AF-CLAP w/ con.'' represents our model trained with the same data but using the standard contrastive loss formulation. The results show that replacing AF-CLAP with other CLAP models results in a performance drop across benchmarks, highlighting the importance of robust audio representations for improving performance.

\begin{table}[h!]
\centering
\caption{\small Benchmark results on various CLAP as audio encoders.}
\resizebox{\columnwidth}{!}{
\begin{tabular}{lcccc}
\toprule
\textbf{Model}  & \textbf{AudioCaps} & \textbf{GTZAN} & \textbf{MuchoMusic} & \textbf{MMAU (avg)} \\
\midrule
{Laion-CLAP}    & 0.51 & 65.2 & 52.3 & 63.8\\
{MS-CLAP}        & \underline{0.55} & 65.8 & 53.1 & 64.3\\
\midrule
{AF-CLAP 630k}  & 0.51 & 66.0 & \underline{54.6} & 63.9 \\
{AF-CLAP w/ con.}  & {0.53} & \underline{66.2} & 54.4 & \underline{66.8}\\
\myccone {AF-CLAP}  & \myccone \textbf{0.58} & \myccone \textbf{69.1} & \myccone \textbf{56.5} & \myccone \textbf{69.0}\\
\bottomrule
\end{tabular}}
\vspace{-1mm}
\label{tab:clap_comparison_transposed}
\end{table}

\subsection{High Quality Data Boosts Reasoning Abilities}
\label{subsec:effect_of_data}

Table~\ref{tab:data_comparison} compares the impact of training data composition on performance. "w/ 1/2 data" and "w/ 1/3 data" refer to experiments using random subsets comprising half and one-third of the total instances from each dataset. Key findings include: (1) AudioSkills significantly enhances AF2's reasoning performance and shows notable improvements when combined with other datasets. (2) For challenging tasks like reasoning and zero-shot classification, data diversity is crucial, as using 1/2 of the data outperforms using only OpenAQA. (3) More data consistently improves performance.

\begin{table}[h!]
\vspace{-1.5mm}
\centering
\caption{\small Comparison of training on different data compositions.}
\resizebox{\columnwidth}{!}{
\begin{tabular}{lcccc}
\toprule
\textbf{Model}  & \textbf{AudioCaps} & \textbf{GTZAN} & \textbf{MuchoMusic} & \textbf{MMAU (avg)}\\
\midrule
w/ 1/2 data  & 0.48 & 51.3 & 46.7 & 49.3\\
w/ 1/3 data  & 0.41 &  47.7 & 43.9 & 45.1\\
w/o AudioSkills  & \textbf{0.58} & \underline{68.8} & 42.6 & 48.6\\
w/ OpenAQA  & \underline{}{0.55} & 19.3 & 39.8 & 38.2\\
\quad +AudioSkills  & \underline{0.55} & 22.6 & \underline{51.3} & \underline{57.5}\\ \midrule
\myccone Original  &\myccone \textbf{0.58} &\myccone \textbf{69.1} &\myccone \textbf{56.5} &\myccone \textbf{69.0}\\
\bottomrule
\end{tabular}}
\label{tab:data_comparison}
\vspace{-3mm}
\end{table}

\subsection{Cross-Attention Outperforms Prefix-Tuning}
\label{subsec:data_performance}

Table~\ref{tab:data_performance} compares GAMA, a SOTA 7B LALM using prefix tuning, trained on the same data and strategy (excluding stage 3), with AF2. AF2 significantly outperforms GAMA, attributed to superior audio representations and the shift cross-attention-based conditioning.

\begin{table}[h!]
\vspace{-1.5mm}
\centering
\caption{\small Comparison of AF2 with GAMA, a SOTA LALM, trained on our same data and same training recipe.}
\resizebox{\columnwidth}{!}{
\begin{tabular}{lcccc}
\toprule
\textbf{Model}  & \textbf{AudioCaps} & \textbf{GTZAN} & \textbf{MuchoMusic} & \textbf{MMAU (avg)} \\
\midrule
{GAMA}~\textit{(orig.)}    & \textcolor{black!50}{0.67} & 13.8 & 33.7 & 38.1\\
{GAMA}~\textit{(ours)}     & 0.41 & 54.7 & 40.3 & 52.8\\
\myccone {AF2}        &\myccone \textbf{0.58} &\myccone \textbf{69.1} &\myccone \textbf{56.5} &\myccone \textbf{69.0}\\
\bottomrule
\end{tabular}}
\label{tab:data_performance}
\vspace{-3mm}
\end{table}

\vspace{-2mm}
\subsection{Effect of Scaling LLM}
\label{subsec:llm_performance}

\begin{figure}[h]
    \centering \includegraphics[width=\columnwidth,trim=0 0.25cm 0 0]{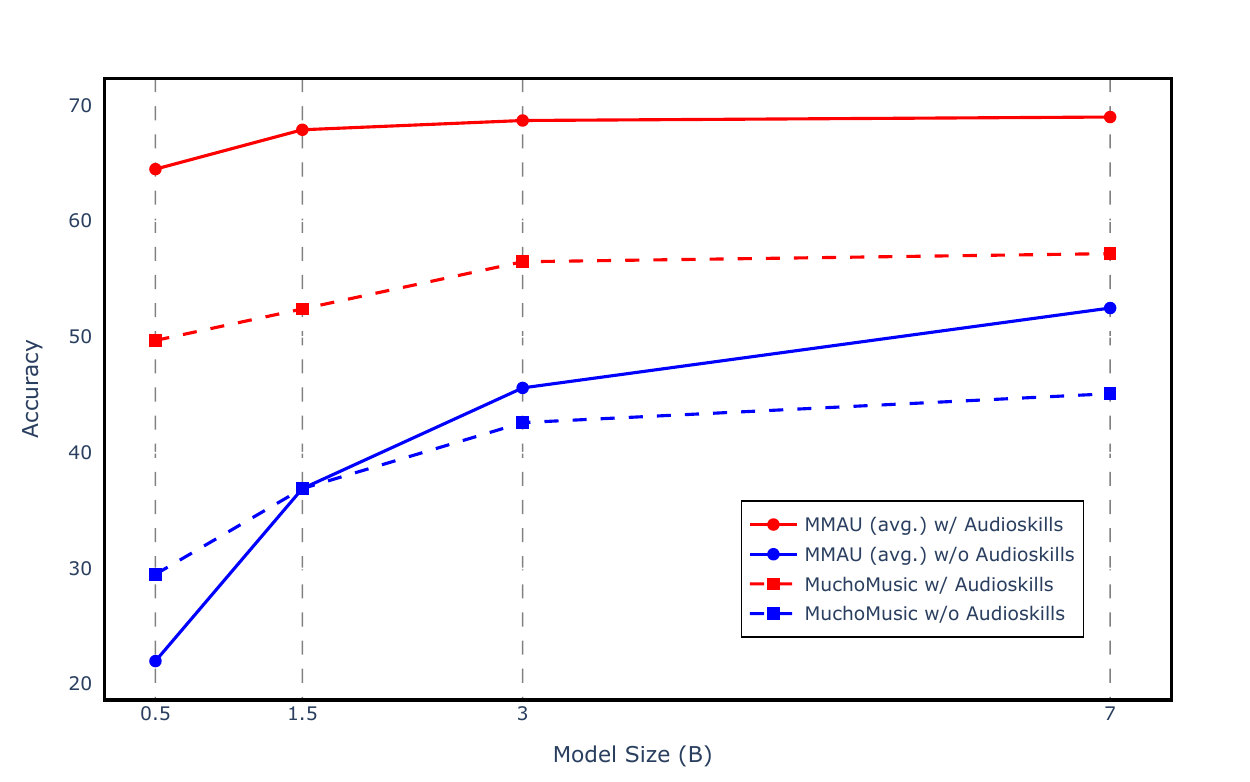}
    \caption{\small Performance comparison of AF2 on different LLM sizes, w/ and w/o AudioSkills. More results in Table~\ref{tab:8col_4rows}.}
\label{fig:scale_plot}
\vspace{-5mm}
\end{figure}

Fig.~\ref{fig:scale_plot} illustrates the performance of AF2 across various LLM sizes, ranging from 0.5B to 7B, trained with and without AudioSkills. The results demonstrate that data quality often surpasses the performance gains achieved by simply scaling compute. Training with AudioSkills not only delivers superior performance overall but also significantly boosts reasoning capabilities, even at smaller LLM sizes. In contrast, when AudioSkills is excluded, reasoning performance heavily depends on model size, with performance scaling more gradually as parameters increase. This highlights the critical role of high-quality, skill-specific data like AudioSkills in driving robust reasoning capabilities, regardless of model size.

\vspace{-2mm}
\subsection{Effect of Training Schedules}
\label{subsec:training_schedule}
Table~\ref{table:training_stages} compares 10 training schedules and their impact on AF2 performance. For 1- and 2-stage training, data from later stages is combined with earlier stages. Additionally, we evaluate a 4-stage curriculum where stage 2 is repeated with LLM fine-tuning and long fine-tuning is shifted to stage 4. Key findings include: (1) Fine-tuning the LLM improves classification and captioning tasks due to the style memorization effect~\cite{ghosh2024closer}, which benefits retrieval-based evaluation (see Appendix~\ref{app_sec:clap_limitations}) but reduces performance on reasoning tasks. (2) Pre-training alignment is essential before fine-tuning. (3) Gradual context-length extension is critical for effective long-audio understanding.

\begin{table}[h!]
\vspace{-1mm}
    \centering
    \caption{\small Overview of 10 training schedules detailing whether CLAP, XATTN, and LLM components (in order) are frozen or unfrozen during each stage and their impact on performance.}
    \setlength\tabcolsep{1.9mm}
    \adjustbox{width=\columnwidth}{
    \begin{NiceTabular}{lccccccccc}
    \CodeBefore
    \rectanglecolor{metabg}{2-10}{10-10}
    \Body
    \toprule
    & \multicolumn{4}{c}{Training Stages} & \multicolumn{6}{c}{Benchmarks} \\
    \cmidrule(lr){2-5} \cmidrule(lr){6-10}
    & \lmmicon &  \lmmicon &  \lmmicon & \lmmicon & AudioCaps & GTZAN & MMAU & LongAudioB \\
    \midrule
    \multirow{2}{*}{1 stage} 
         & \freeze\train\freeze & - & -  & -  & 0.50 & 56.7 & 58.8 & 5.1 \\
         & \train\train\train  & - & - & -  & 0.50 & 52.5 & 56.3 & 4.9  \\
        \midrule 
     \multirow{4}{*}{2 stage} 
         & \freeze\train\freeze & \train\train\freeze & -  & - & 0.56 & 68.9 & 68.0 & 5.1 \\ 
	& \freeze\train\freeze & \train\train\train & -  & - & 0.59 & 70.9 & 63.2 & 5.7 \\ 
         & \freeze\train\freeze & \freeze\train\train & -  & - & 0.59 & 72.3 & 60.7 & 5.7 \\
         & \train\train\freeze  & \train\train\freeze & -  & - & 0.52 & 65.7 & 63.4 & 5.3 \\ 

    \midrule 
    \multirow{2}{*}{3 stage} 
         & \freeze\train\freeze  & \train\train\freeze & \freeze\train\freeze & -   & \myccone 0.58 & \myccone 69.1 & \myccone \textbf{69.0} & \myccone \textbf{6.4}\\
         & \freeze\train\freeze  & \train\train\freeze & \train\train\train & -   & 0.45 & 62.6 & 59.1 & 6.2  \\

    \midrule 
    \multirow{2}{*}{\textcolor{black!50}{4 stage}} 
	& \freeze\train\freeze  & \train\train\freeze &  \freeze\train\train  &  \freeze\train\freeze & \textcolor{black!50}{\textbf{0.62}} & \textcolor{black!50}{\textbf{74.2}} &\textcolor{black!50}{64.5} & \textcolor{black!50}{\textbf{6.4}}\\
         & \freeze\train\freeze  & \train\train\freeze & \train\train\train  &  \freeze\train\freeze &\textcolor{black!50}{0.59} &\textcolor{black!50}{74.0} &\textcolor{black!50}{62.0} &\textcolor{black!50}{6.3}\\
    
    \bottomrule
    \end{NiceTabular}
    }
    \vspace{-2mm}
    \label{table:training_stages}
\end{table}

\vspace{-2mm}
\section{Conclusion, Limitations and Future Work}
\label{sec:conclusion}

In this paper, we introduce Audio Flamingo 2, an ALM designed for long audio understanding and expert reasoning. Our model leverages a custom CLAP, trained on a large-scale dataset with a novel objective function, and is trained on synthetic reasoning AQA data to develop unique skills essential for real-world reasoning. Audio Flamingo 2 achieves SOTA performance in audio understanding and reasoning despite being small in model footprints. Additionally, we propose LongAudio and LongAudioBench to advance the field of ALM reasoning over long audio contexts. 

For future work, we aim to address current limitations, including: (1) enhancing speech content understanding capabilities, (2) scaling AudioSkills to include more diverse datasets, and (3) developing audio encoders inherently capable of processing long audio.

\bibliography{example_paper}
\bibliographystyle{icml2025}

\clearpage
\onecolumn
\appendix
\twocolumn

\begin{table*}[ht!]
\centering
\caption{\small Performance comparison of AF-CLAP with baselines on Text-to-Audio and Audio-to-Text retrieval on AudioCaps and Clotho.}
\label{tab:exp-t2a-retrieval}
\resizebox{1.9\columnwidth}{!}{
\begin{tabular}{lccc|ccc|ccc|ccc}
\toprule \toprule
& \multicolumn{6}{c|}{{AudioCaps}} & \multicolumn{6}{c}{{Clotho}} \\ 
\multicolumn{1}{c}{{Model}} & \multicolumn{3}{c}{{Text-to-Audio}} & \multicolumn{3}{c|}{{Audio-to-Text}}
  & \multicolumn{3}{c}{{Text-to-Audio}} & \multicolumn{3}{c}{Audio-to-Text} \\ 
& R@1 & R@5 & R@10 & R@1 & R@5 & R@10 & R@1 & R@5 & R@10 & R@1 & R@5 & R@10 \\ 
\midrule
{MMT} & 
  36.1 & 72.0 & \textbf{84.5} & 39.6 & 76.8 & 86.7 & 6.7 & 21.6 & 33.2 & 7.0 & 22.7 & 34.6 \\

{ML-ACT} & 
  33.9 & 69.7 & 82.6 & 39.4 & 72.0 & 83.9 & 14.4 & 36.6 & 49.9 & 16.2 & 37.6 & 50.2 \\

{CLAP} & 
  34.6 & 70.2 & 82.0 & 41.9 & 41.9 & 84.6 & 16.7 & 41.1 & 54.1 & 20.0 & 44.9 & 58.7 \\

{CompA-CLAP} & 
  36.1 & \underline{72.6} & 81.6 & 45.2 & 80.1 & 86.7 & \underline{16.8} & \underline{43.5} & \underline{56.1} & 19.7 & 45.2 & 55.6 \\

{Laion-CLAP} & 
  36.1 & 71.8 & 83.9 & \underline{46.8} & 82.9 & 90.7 & 16.1 & 38.3 & 51.1 & 22.7 & 48.5 & 60.8 \\ 
\midrule
\myccone AF-CLAP-630k &\myccone  
  36.3 &\myccone  71.8 &\myccone  83.9 &\myccone  46.3 &\myccone  83.0 &\myccone  90.9 &\myccone  16.1 &\myccone  38.3 &\myccone  51.4 &\myccone  22.7 &\myccone  48.7 &\myccone  60.8 \\
\myccone AF-CLAP w/ van. con.  &\myccone  
  \underline{36.8} &\myccone  72.0 &\myccone  \underline{84.0} &\myccone  45.2 &\myccone  \underline{83.5} &\myccone  90.0 &\myccone  16.5 &\myccone  41.9 &\myccone  55.3 &\myccone  \underline{22.9} &\myccone  \underline{51.0} &\myccone  \underline{63.2} \\
\myccone AF-CLAP w/o noise red. &\myccone  
  35.4 &\myccone  70.7 &\myccone  81.6 &\myccone  45.0 &\myccone  82.6 &\myccone  \underline{91.0} &\myccone  16.3 &\myccone  40.5 &\myccone  52.7 &\myccone  22.8 &\myccone  \textbf{51.2} &\myccone  59.4 \\
\myccone AF-CLAP \textit{(ours)} &\myccone  
  \textbf{37.3} &\myccone  \textbf{72.9} &\myccone  \underline{84.0} &\myccone  \textbf{46.9} &\myccone  \textbf{84.1} &\myccone  \textbf{91.9} &\myccone  \textbf{17.3} &\myccone  \textbf{43.9} &\myccone  \textbf{56.8} &\myccone  \textbf{23.2} &\myccone  \textbf{51.2} &\myccone  \textbf{63.5} \\ 
\bottomrule
\end{tabular}}
\end{table*}

\section{Table of Contents}
\begin{enumerate}
\setlength\parskip{0em}
    \item~\ref{sec:af_clap} : AF-CLAP
    \item~\ref{app_sec:clap_limitations} : Limitations of CLAP-Retrieval-Based Evaluation
    \item~\ref{app_sec:audioskills} : AudioSkills
    \item~\ref{app_sec:long_audio} : LongAudio
    \item~\ref{app_sec:prompts} : Prompts
    \item~\ref{app_sec:design_analysis} : More Results and Design Analysis
\item~\ref{app_sec:design_rope} : Effect of RoPE
\item~\ref{app_sec:design_audio_transform} :  Effect of Audio Transformation Layers
\item~\ref{app_sec:cross_atten_freq} : Effect of Cross-Attention Frequency
\item~\ref{app_sec:more_details} : More Details
\end{enumerate}
\section{AF-CLAP}
\label{sec:af_clap}

\subsection{Training Hyper-parameters}

We train AF-CLAP on 8 A100 80GB GPUs. We train it with a learning rate of 5e-4, $M$=$N$=3, and an effective batch size of 256 for 12 epochs. This batch size is smaller than that in the literature, but we do so due to computational constraints.

\subsection{Feature Extraction Layer}
\label{app:feature_extract}

As described in Section~\ref{sec:af2_arch}, unlike Audio Flamingo which uses CLAP audio embeddings directly, AF2 discards the CLAP head and rather leverages dense features from the final layer of the audio encoder (HTS-AT in our case). Table~\ref{tab:clap_feature_comparison} compares the performance of AF2 trained with CLAP head features versus dense audio features from the encoder’s last layer. The results clearly demonstrate that using dense features significantly improves AF2's performance.

\begin{table}[h!]
\centering
\caption{\small Performance comparison of AF2 with different audio feature extraction methods from CLAP. Head refers to audio features extracted from the CLAP head, and Dense refers to dense features extracted from the last layer of the audio encoder.}
\resizebox{\columnwidth}{!}{
\begin{tabular}{lcccc}
\toprule
\textbf{Model}  & \textbf{AudioCaps} & \textbf{GTZAN} & \textbf{MuchoMusic} & \textbf{MMAU} \\
\midrule
{Head $\mathbb{R}^{1 \times 256}$}    & 0.50 & 54.7 & 42.6 & 47.0 \\
\myccone {Dense $\mathbb{R}^{64 \times 2048}$}        &\myccone \textbf{0.58} &\myccone \textbf{69.1} &\myccone \textbf{56.5} &\myccone \textbf{69.0}\\
\bottomrule
\end{tabular}}
\label{tab:clap_feature_comparison}
\end{table}

\subsection{Training Datasets}
Table~\ref{tab:clap_datasets} provides detailed statistics of datasets used for training AF-CLAP.
\begin{table}[!h]
    \centering
    \caption{\small Statistics of audio-caption datasets used for CLAP training. $*$ indicates the dataset was collected by us. Furthermore, all datasets with captions were made to go through our cleaning stage to reduce noise, as mentioned in Section~\ref{sec:af2_clap_data}.}
    \resizebox{\columnwidth}{!}{
    \begin{tabular}{cc}
    \toprule
        Dataset & \#Audio-Text Pairs \\ \midrule
        YouTube-8M$^*$~\cite{abu2016youtube} & 3,947,057  \\
        Sound-VECaps~\cite{yuan2024sound} & 1,657,029  \\
        MiraData$^*$~\cite{ju2024miradata} & 748,320 \\
        Action2sound$^*$~\cite{chen2024action2sound} & 306,602 \\
        NSynth~\cite{engel2017neural} & 289,205 \\
        Freesound~\cite{font2013freesound} & 256,695 \\
        AudioSet Strong$^*$~\cite{hershey2021benefit} & 216,622 \\
        VGGSound~\cite{chen2020vggsound} & 185,161 \\
        FMA~\cite{defferrard2016fma} & 106,412 \\
        Video Recap$^*$~\cite{10657851} & 64,627 \\
        CochlScene~\cite{jeong2022cochlscene} & 60,855 \\
        FSD50k~\cite{fonseca2022fsd50k} & 40,966 \\
        MACS~\cite{irene_martin_morato_2021_5114771} & 31,675 \\
        BBC {\footnote{\url{https://sound-effects.bbcrewind.co.uk/}}} & 31,201 \\
        MagnaTagATune~\cite{law2009evaluation} & 25,863 \\
        SoundDescs~\cite{koepke2022audio} & 23,085 \\
        Clotho~\cite{drossos2020clotho} & 19,195 \\
        TAU-Urban~\cite{heittolatau} & 14,400 \\
        MusicCaps~\cite{agostinelli2023musiclm} & 5,479 \\
        WavText5K~\cite{deshmukh2022audio} & 4,347 \\
        SONICS~\cite{rahman2024sonics} & 1,602 \\
        SoundBible{\footnote{\url{https://soundbible.com/}}}  & 935 \\
        MUSDB18~\cite{rafii2019musdb18} & 276 \\
        Medleybd-Pitch~\cite{bittner2014medleydb} & 103 \\
    \midrule
    \textbf{Total} & 8,037,712 \\
    \bottomrule
    \end{tabular}}
    \label{tab:clap_datasets}
\end{table}
\subsection{Comparison with prior-art}

Table~\ref{tab:exp-t2a-retrieval} presents the performance of our CLAP on audio-to-text and text-to-audio retrieval tasks using the Clotho and AudioCaps datasets. ``AF-CLAP 630k'' refers to AF-CLAP trained using our proposed strategy but with Laion CLAP's 630k audio-text pairs. ``AF-CLAP w/ van. con.'' represents AF-CLAP trained with the same data but using the standard contrastive loss formulation. ``AF-CLAP w/o noise red.'' represents our AF-CLAP trained using the same data and method but without our noise reduction step. Our final proposed AF-CLAP achieves state-of-the-art performance across all metrics. Overall, AF-CLAP 630k does not lead to any improvements over Laion-CLAP, as our data augmentation and cleaning are only applicable to captions obtained from complex real-world audios, which represents only a significantly small portion of Laion-CLAP.

Table~\ref{tab:zshot_res} highlights the performance of AF-CLAP on zero-shot audio classification benchmarks, where it consistently achieves SOTA results. We make the same conclusion for Laion-CLAP as previously stated.

We emphasize that benchmark datasets do not holistically evaluate a CLAP model's capabilities, as highlighted in several works~\cite{ghosh2024compa,wu2023audio,selvakumar2024audio,ghosh2024reclap}. While AF-CLAP outperforms baselines on benchmark datasets, its audio features are significantly more robust and enhance the audio perception capabilities of (L)ALMs (see also Table~\ref{tab:clap_comparison_transposed}).

\begin{table*}[t]
\centering
\caption{\small Performance comparison of our CLAP with baselines on Zero-shot Audio classification benchmarks.}
\label{tab:zshot_res}
\resizebox{1.5\columnwidth}{!}{
\begin{tabular}{lccccccc}
\toprule \toprule
{Model} & {ESC-50} & {US8K} & {VGGSound} & {FSD50K} & {TUT} & {AudioSet} & {NSynth} \\ 
\midrule
Wav2CLIP & 41.4 & 40.4 & 10.0 & 3.0 & 28.2 & 5.0 & 5.9 \\
AudioClip & 69.4 & 65.3 & 9.9 & 6.6 & 29.5 & 3.7 & 6.8 \\
CLAP & 82.6 & 73.2 & 16.4 & 14.0 & 29.6 & 5.1 & 9.9 \\
Laion-CLAP & {88.2} & 74.1 & 21.2 & {22.4} & {58.4} & 20.8 & {11.8} \\
CoLLAT & 84.0 & 77.0 & - & 19.0 & 29.0 & 9.0 & - \\
CompA-CLAP & 86.5 & {88.1} & {21.9} & 19.6 & 56.7 & {21.6} & {11.8} \\ \midrule
\myccone AF-CLAP-630k &\myccone {88.1} &\myccone 74.3 &\myccone 21.2 &\myccone 22.0 &\myccone 57.5 &\myccone 19.3 &\myccone 12.0 \\
\myccone AF-CLAP w/ van. con. &\myccone {91.0} &\myccone \underline{91.8} &\myccone 23.5 &\myccone \textbf{28.1} &\myccone \underline{63.0} &\myccone 21.8 &\myccone \textbf{17.4} \\
\myccone AF-CLAP w/o noise red. &\myccone \underline{91.1} &\myccone 91.3 &\myccone \underline{23.9} &\myccone {26.9} &\myccone \textbf{63.2} &\myccone \underline{22.8} &\myccone {15.5} \\
\myccone AF-CLAP \textit{(ours)} &\myccone \textbf{91.3} &\myccone \textbf{92.3} &\myccone \textbf{24.1} &\myccone \underline{27.2} &\myccone \textbf{63.2} &\myccone \textbf{23.7} &\myccone \underline{15.9} \\
\bottomrule
\end{tabular}}
\end{table*}
\section{Limitations of CLAP Retrieval Based Evaluation}
\label{app_sec:clap_limitations}
We attribute the suboptimal performance of AF2 on some datasets to the limitations of the CLAP-based evaluation method, which often fails to retrieve the correct label corresponding to the model's open-ended generation response. we follow the evaluation scheme introduced by~\citet{deshmukh2023pengi} and widely adopted in prior works~\cite{gong2024listen,ghosh-etal-2024-gama}. This approach uses a CLAP model to retrieve a label from the label set by comparing the model's open-ended generation and assigning the label with the highest similarity as the predicted output. We show some examples below where even a correct prediction by AF2 leads to incorrect retrieval and therefore lower accuracy:

\begin{mdframed}[linewidth=1pt, linecolor=black, leftmargin=1pt, rightmargin=1pt, innerleftmargin=10pt, innerrightmargin=10pt, innertopmargin=4pt, innerbottommargin=4pt, backgroundcolor=gray!20, roundcorner=5pt]

\noindent\textbf{Dataset:} GTZAN, \textbf{Correct Label:} Rock, \textbf{Predicted Label:} Punk Metal, \textbf{Retrieved Label:} Metal 

\noindent\textbf{Dataset:} ESC50, \textbf{Correct Label:} Pouring water, \textbf{Predicted Label:} liquid, , \textbf{Retrieved Label:} Water drops 

\end{mdframed}

However, fine-tuning the LLM leads to style memorization, which favors this evaluation method. We show some examples of prediction shift and how this leads to increase in accuracy:

\begin{mdframed}[linewidth=1pt, linecolor=black, leftmargin=1pt, rightmargin=1pt, innerleftmargin=10pt, innerrightmargin=10pt, innertopmargin=4pt, innerbottommargin=4pt, backgroundcolor=gray!20, roundcorner=5pt]
\noindent\textbf{Dataset:} GTZAN, \textbf{Correct Label:} rock, \textbf{Predicted Label:} rock, \textbf{Retrieved Label:} rock

\noindent\textbf{Dataset:} ESC50, \textbf{Correct Label:} clock alarm, \textbf{Predicted Label:} clock alarm, \textbf{Retrieved Label:} clock alarm
\end{mdframed}

\section{AudioSkills}
\label{app_sec:audioskills}

\subsection{Dataset Statistics}
\label{subsec:app_skills_stats}
Table~\ref{tab:as_cat_tab} presents detailed category-wise statistics on our proposed AudioSkills dataset. We also list the meta-data used for data generation. For meta-data, transcripts were obtained from Whisper~\textsubscript{Large-v3}~\cite{radford2023robust}. We generate all audio captions from Qwen2-Audio and visual captions from Qwen2-VL.

\section{LongAudio}
\label{app_sec:long_audio}

\subsection{Detailed Statistics}
\label{subsec:longaudio_stats}

Table~\ref{tab:category_counts} presents detailed statistics of LongAudio and LongAudioBench, categorized into the various types of QAs. 
\begin{table}[!h]
\centering
\vspace{-2mm}
\caption{\small Dataset statistics for LongAudio and LongAudioBench.}
\label{tab:category_counts}
\resizebox{0.75\columnwidth}{!}{
\begin{tabular}{lcc}
\toprule
\textbf{Category} & \textbf{LongAudio} & \textbf{LongAudioBench} \\
\midrule
Captioning     & 67,498 & 917 \\
Plot QA        & 61,511 & 237 \\
Complex QA     & 50,315 & 361 \\
Subscene QA    & 7,908  & 176 \\
Temporal QA    & 66,836 & 253 \\
General QA     & 8,860  & 485 \\ \midrule
\textbf{Total} & 262,928 & 2,429 \\
\bottomrule
\end{tabular}}
\vspace{-1mm}
\end{table}

\subsection{Examples}
\label{subsec:app_longaudio_examples}

Table~\ref{tab:longbench_ex} shows category-wise examples from LongAudio.
\subsection{Comparison with other datasets}
\label{subsec:long_compare}

Table~\ref{tab:aqa_duration} compares the duration statistics of LongAudio with other AQA datasets. LongAudio stands out with the longest average audio durations.
\begin{table}[h!]
\centering
\caption{\small Comparison of duration statistics of LongAudio with other AQA datasets. LongAudio stands out with the longest average audio durations.}
\label{tab:aqa_duration}
\resizebox{\columnwidth}{!}{
\begin{tabular}{lrrr}
\toprule
\textbf{Method} & \textbf{Min} & \textbf{Max} & \textbf{Avg} \\
\midrule
Clotho-AQA~\cite{lipping2022clotho} & 15.00 & 30.00 & 22.53 \\
AudioEntailment~\cite{audioentail} & 15.00 & 30.00 & 22.49 \\
CompA-R~\cite{ghosh-etal-2024-gama} & 0.47 & 10.01 & 9.87 \\
OpenAQA~\cite{gong2024listen} & 0.06 & 180.00 & 12.01 \\
MU-LLAMA~\cite{liu2024music} & \textbf{29.12} & 29.12 & 29.12 \\
Salmonn~\cite{tang2024salmonn} & 0.47 & 1069.04 & 10.62 \\
\myccone LongAudio \textit{(ours)} & \myccone 5.00 & \myccone \textbf{1797.71} & \myccone \textbf{117.08} \\
\bottomrule
\end{tabular}}
\end{table}

\subsection{Fine-grained results for AF2}
\label{subsec:long_fine_grained}
Table~\ref{tab:long_fine_grained_values} presents category-wise fine-grained results of AF2 on LongAudioBench. While AF2 demonstrates strong performance, fine-tuning on LongAudio further improves scores across all categories, particularly in tasks unique to LongAudioBench, such as NeedleQA and SubsceneQA, which were not encountered during AF2's two-stage training. These results emphasize the significance of our proposed LongAudio dataset in effectively extending (L)ALM context length and improving long-audio reasoning.

\begin{table}[!h]
    \centering
    \caption{\small Fine-grained scores of AF2 on LongAudioBench w/ and w/o fine-tuning on our dataset LongAudio.}
    \label{tab:long_fine_grained_values}
    \resizebox{0.8\columnwidth}{!}{
    \begin{tabular}{l|c|c}
        \toprule
        \textbf{Category} & \textbf{AF2} & \textbf{AF2 w/o LongAudio} \\
        \midrule
        Captioning & 63.75\% & 46.02\%\\
        Plot QA & 68.02\% & 44.15\% \\
        Temporal QA & 62.61\% & 51.00\% \\
        Needle QA & 63.13\% & 35.92\%\\
        Subscene QA & 64.03\% & 33.61\%\\
        General QA & 63.61\% & 42.39\% \\ \midrule
        \myccone Avg &\myccone 64.19\% &\myccone 50.22\%\\
        \bottomrule
    \end{tabular}
    }
\end{table}

\subsection{Success and Failure Cases on LongAudioBench}
\label{subsec:sucess_failure}

Table~\ref{tab:reasoning_scores} presents success and failure cases of AF2 on LongAudioBench.

\subsection{Metadata}
\label{subsec:metadata_long}
For MiraData, metadata includes visual captions (e.g., main object, background, and style) as detailed in the original paper and shown in prompt~\ref{fig:temporal_prompt}. For Video ReCap, metadata focuses on action captions describing activities, as shown in prompt~\ref{fig:subscene_prompt}.
\begin{figure}[h!]
    \centering
    \includegraphics[width=0.9\columnwidth]{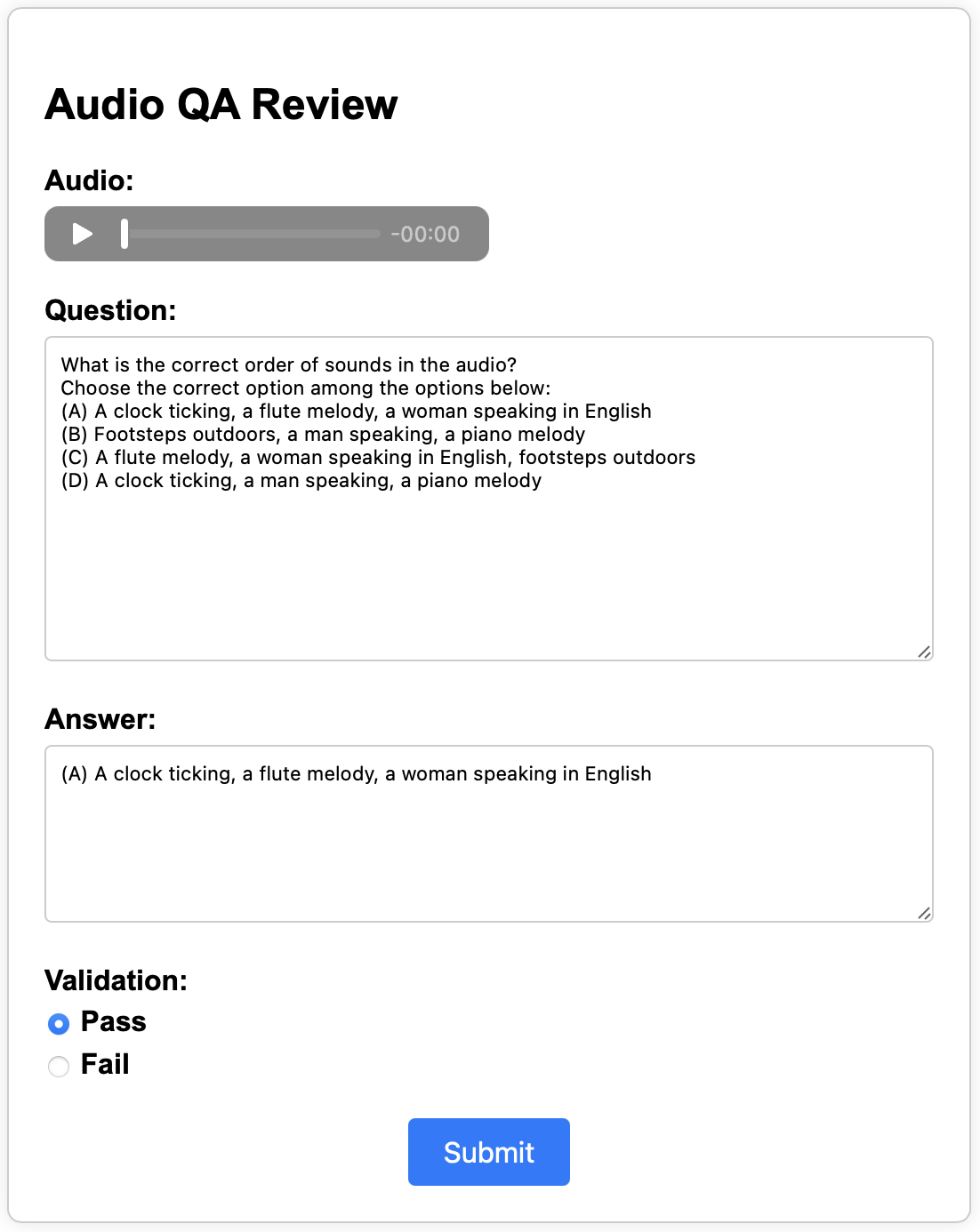}
    \caption{Snapshot of the annotation tool used for LongAudio annotation.}
    \label{fig:annot_tool}
\end{figure}
\subsection{Human Verification for LongAudioBench}
\label{subsec:app_human_verification}
The human verification process has been approved by our institution’s Institutional Review Board (IRB).
\begin{table*}[t]
\centering
\caption{\small Detailed statistics of AudioSkills, categorized into individual reasoning types, together with details on open-source datasets, additional meta-data, and prompt used for QA generation.}
\resizebox{\linewidth}{!}{
\begin{tabular}{lllll}
\toprule
Reasoning Type                    & Size    & Datasets Used           & Meta-Data Used               & Prompt Reference \\ \midrule
Temporal                          & 188,799  & AudioSet-SL, MusicBench & Time Stamped Events (GT), Caption &    Fig. \ref{fig:temporal_prompt}, \ref{fig:mmau_temporal_prompt}             \\
Attribute Identification          & 201,455   & AudioSet-SL             & Attribute values from ~\citet{kumar2024sila}, Caption    &     Fig. \ref{fig:attribute_prompt}             \\
Counting                           & 50,493   & Synthetic               & Transcript, Caption          &        pythonic        \\
Contextual Sound Event Reasoning  & 982,847  & AudioSet-SL, MusicBench, FMA & Time Stamped Events, Caption &       Fig. \ref{fig:aqa_prompt_gama},\ref{fig:sound_mmau_prompt}, \ref{fig:eco_acoustic_prompt}, \ref{fig:mmau_complex_3_prompt}             \\
Contextual Speech Event Reasoning & 1,272,782 & AudioSet-SL             & Transcript, Caption                 &        Fig. \ref{fig:mmau_speech_prompt}          \\
Information Extraction            & 858,639  & AudioSet-SL, MusicBench, MTG-Jamendo, MusicNet &  Time Stamped Events, Caption                         &     Fig.   
\ref{fig:harmony_mmau_prompt},
\ref{fig:texture_mmau_prompt},
\ref{fig:rhythm_prompt},
\ref{fig:sound_source_prompt},
\ref{fig:jamendo_reasoning_prompt},  \ref{fig:musicnet_captioning_prompt}         \\
General Reasoning                             & 704,040  & AudioSet-SL, MusicBench, MusicNet, FMA & Time Stamped Events, Caption           &    Fig. \ref{fig:sound_mmau_prompt}, 
\ref{fig:rhythm_prompt},
\ref{fig:sound_source_prompt},
\ref{fig:fma_mir_prompt}, \ref{fig:musicnet_captioning_prompt}             \\ \bottomrule
\end{tabular}}
\label{tab:as_cat_tab}
\end{table*}

\begin{table*}[!h]
\centering
\caption{\small Category-wise examples from \textbf{LongBench}.}
\begin{tabular}{|p{2cm}|p{7cm}|p{6.6cm}|}
\toprule
\multicolumn{1}{|c|}{Category}    & \multicolumn{1}{c|}{Question}  & \multicolumn{1}{c|}{Answer}  \\ \midrule
\multirow{6}{*}{Captioning}  & How would you describe the scene from the audio?          & Two women converse in a car with ambient bird chirping and occasional mechanical noises in the background.                  \\ \cmidrule{2-3}
            & What is taking place in the audio?                        & The audio features ambient wind, distant motor vehicle hums, and occasional background noises creating an eerie atmosphere. \\ \midrule
\multirow{6}{*}{Plot QA}     & What is the predominant mood conveyed by the background music and sound effects in the audio segments?                                      & Sad and suspenseful.  \\ \cmidrule{2-3}
            & What is the primary theme conveyed by the audio in the initial segments?                                           & The audio conveys a solemn and spiritual theme with classical music and religious references. \\ \midrule
\multirow{18}{*}{Temporal QA} & What is the correct order of sounds in the audio? Choose the correct option among the options below: & \\ 
& (A) A clock ticking, a flute melody, a woman speaking in English & \\ & (B) Footsteps outdoors, a man speaking, a piano melody & (A) A clock ticking, a flute melody, a woman speaking in English \\ & (C) A flute melody, a woman speaking in English, footsteps outdoors & \\ & (D) A clock ticking, a man speaking, a piano melody &   \\ \cmidrule{2-3}
            & When can the sound of a horse galloping be heard? Choose the correct option among the options below: & \\ & (A) At the beginning & (B) In the middle \\ & (B) In the middle & \\ & (C) Towards the end  &  \\ \midrule
\multirow{6}{*}{Needle QA}   & What audio event indicates an interruption during the conversation between the man and woman? & The background sound of a phone ringing from 0.53 seconds to 7.40 seconds. \\ \cmidrule{2-3}
            & What audio element creates a sense of suspense in the dark room scene?                                            & The cinematic strings playing a scary tune along with a church bell and distant horror scream. \\ \midrule
\multirow{5}{*}{Subscene QA} & What happened between the sound of ocean waves and the sound of mechanisms functioning? & The sound of a phone being dropped on a table followed by mechanical fan noises. \\ \cmidrule{2-3}
            & What sounds occur between brushing teeth and using a phone?                                                      & Music playing continuously with occasional running water and a shower sound. \\ \midrule
\multirow{4}{*}{General QA}  & What continuous background noise is heard while the person is using the laptop?                                                         & The sound of an engine idling consistently. \\ \cmidrule{2-3}
            & What was the continuous background noise throughout the audio?                                                       & Engine idling. \\ \bottomrule                  
\end{tabular}
\label{tab:longbench_ex}
\end{table*}
{\noindent \textbf{Annotation Method.}} The annotation process for LongAudioBench was carried out in two stages by a group of experts. In the first stage, experts corrected and verified each QA pair, including discarding completely erroneous pairs. In the second stage, experts reviewed QA pairs corrected by each other. Thus, each QA pair was at least annotated by 2 individuals. Their primary role in both stages was to annotate the data while ensuring accuracy, consistency, and adherence to the predefined guidelines (mentioned next). We build an annotation tool for this purpose, and Fig.~\ref{fig:annot_tool} provides a snapshot of the tool. This multi-stage process, involving annotation followed by a detailed review, was designed to enhance the reliability and depth of the annotated dataset, leveraging the combined expertise and experience of the annotators.

{\noindent \textbf{Annotation Guidelines.}} The authors of the paper put forward the following annotation guidelines for annotating and verifying LongAudioBench:

\begin{itemize}  
\vspace{-3mm}
\setlength\parskip{0em}
    \item \textbf{Accuracy and Consistency:} Ensure all annotations are accurate and consistent, strictly adhering to the predefined standards and guidelines.  

    \item \textbf{Two-Stage Process:} Ensure all QAs go through the 2-stage verification process mentioned above.

    \item \textbf{Listening Requirements:} Listen to the complete audio before annotating to ensure the QA pair is contextually accurate.  

    \item \textbf{QA Pair Validation:} \textit{i)} Ensure the audio in the QA pair is valid, not corrupt, and corresponds to the question. \textit{ii)} Discard or flag QA pairs containing irrelevant or ambiguous content.  

    \item \textbf{Question Format:} \textit{i)} All questions must be in English. \textit{ii)} Check if the tagged QA type (e.g., captioning, PlotQA) is correct and relevant.  \textit{iii)} Avoid mentioning identifiable information about the audio, such as names or metadata.  

    \item \textbf{Answer Format:}  \textit{i)} All answers must follow the MCQ or open-ended format, pre-defined for each question category. \textit{ii)} Additionally, annotators should ensure answers are comprehensive and free of ambiguities.  

    \item \textbf{Tool Usage:}  \textit{i)} Use the annotation tool (Fig.~\ref{fig:annot_tool}) for all corrections, reviews, and annotations. \textit{ii)} Document and report any technical issues encountered during the process.  

    \item \textbf{Quality Assurance:}  \textit{i)} Each QA pair must be reviewed by at least two individuals to ensure reliability and adherence to the guidelines. \textit{ii)} Maintain detailed logs of discarded QA pairs and reasons for exclusion.  \textit{iii)} Escalate ambiguous or complex QA pairs to the team for collective review.  

    \item \textbf{Collaboration:}  \textit{i)} Maintain open communication with other annotators to clarify task-specific doubts or resolve disagreements during review.  \textit{ii)} Share feedback on ambiguous QA pairs to iteratively improve the annotation process.  
\end{itemize}  

\section{Prompts}
\label{app_sec:prompts}
\subsection{Prompts for foundational datasets}
\label{app:prompt_found}

Tables~\ref{tab:task_prompts_1} and \ref{tab:task_prompts_2} list the prompts used to convert foundational audio classification and captioning datasets into the AQA format for training AF2. For the answers, only the label was retained, excluding any additional text. Initially, we experimented with adding a prefix to the labels to convert them into sentence format, but this approach did not yield favorable results.

\subsection{Prompts for other tasks}
\label{app:prompt_other_task}
Table.~\ref{tab:task_prompts_1} and \ref{tab:task_prompts_1} list prompts we used to convert foundational audio classification and captioning datasets into AQA format for training AF2.

\begin{figure*}[h]
    \centering
    \includegraphics[width=\linewidth]{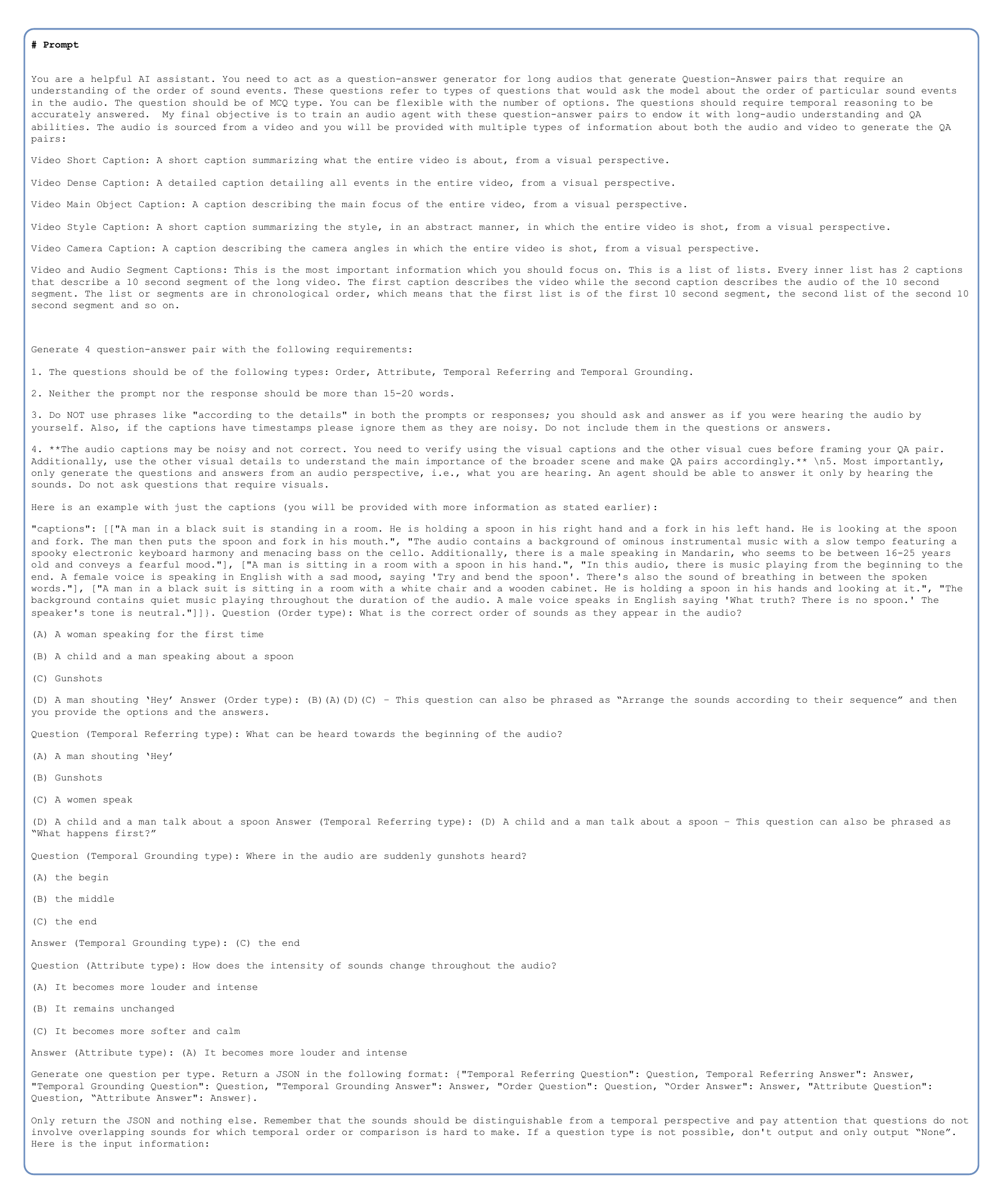}
    \caption{\small Prompt 1 used for generating \textbf{Temporal QA} for LongAudio from MiraData.}
    \label{fig:temporal_prompt}
\end{figure*}

\begin{figure*}[h]
    \centering
    \includegraphics[width=\linewidth]{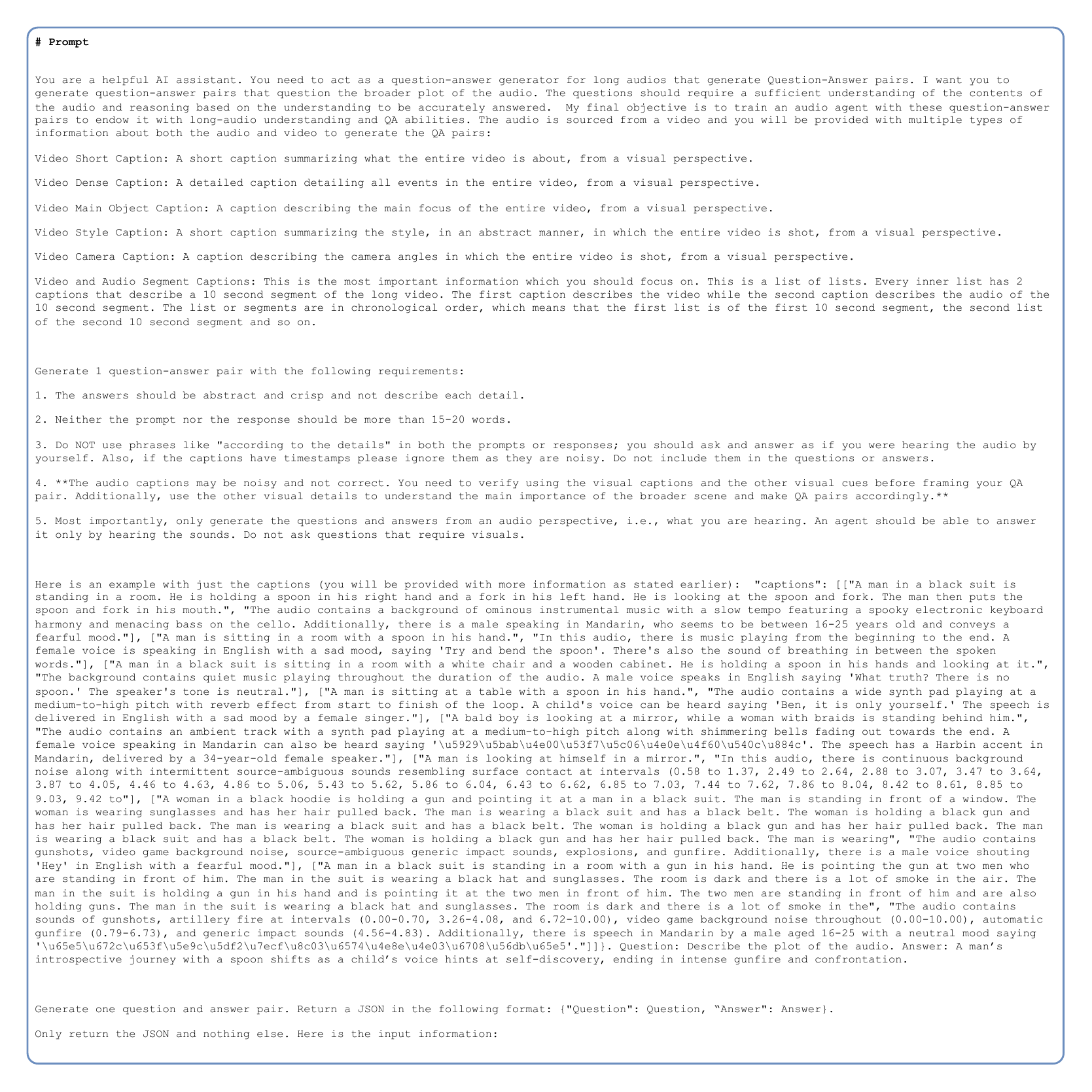}
    \caption{\small Prompt used for generating \textbf{Plot QA} for LongAudio from MiraData.}
    \label{fig:plot_prompt}
\end{figure*}

\begin{figure*}[h]
    \centering
    \includegraphics[width=\linewidth]{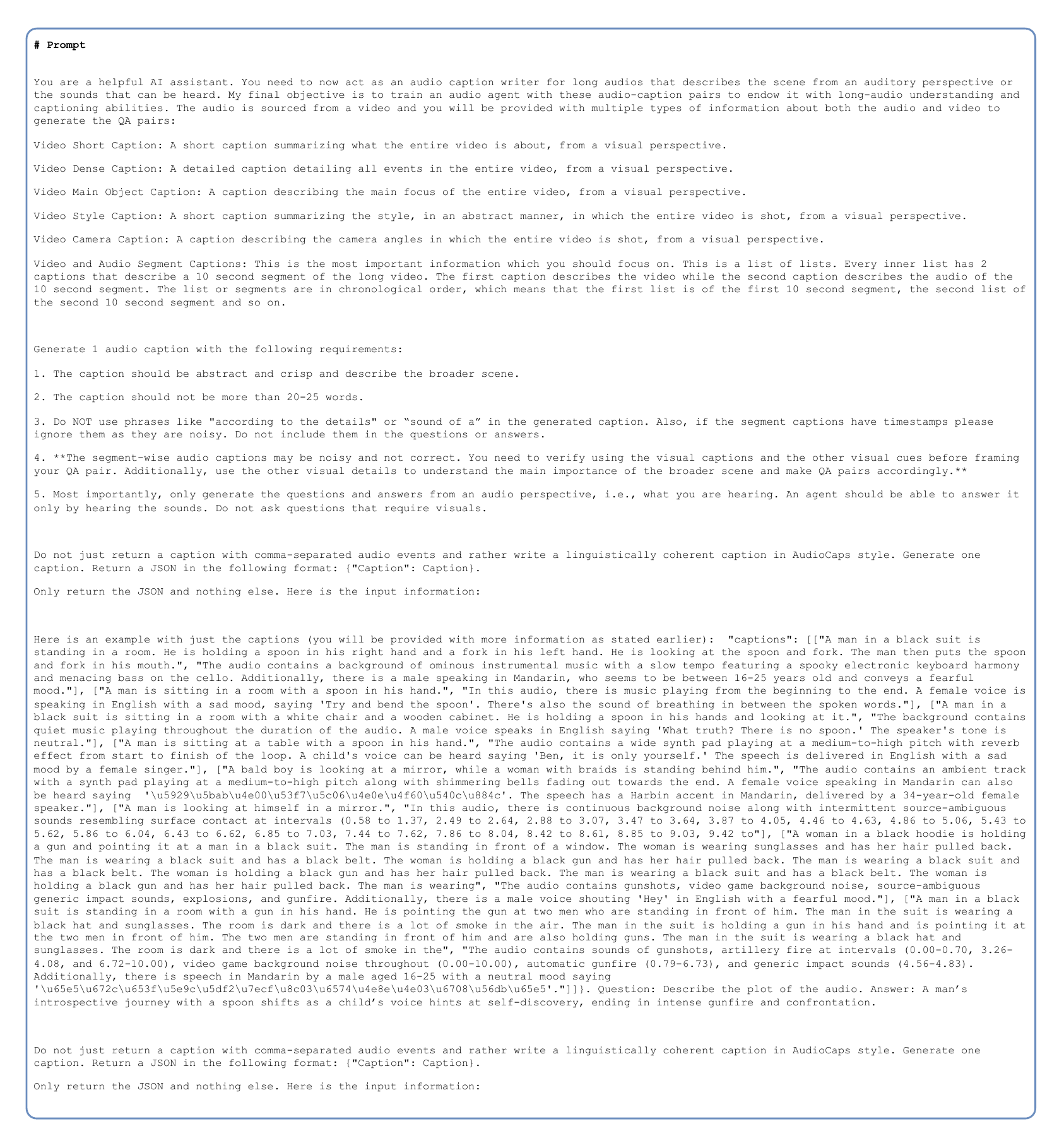}
    \caption{\small Prompt used for generating \textbf{Caption QA} for LongAudio from MiraData.}
    \label{fig:caption_prompt}
\end{figure*}

\begin{figure*}[h]
    \centering
    \includegraphics[width=\linewidth]{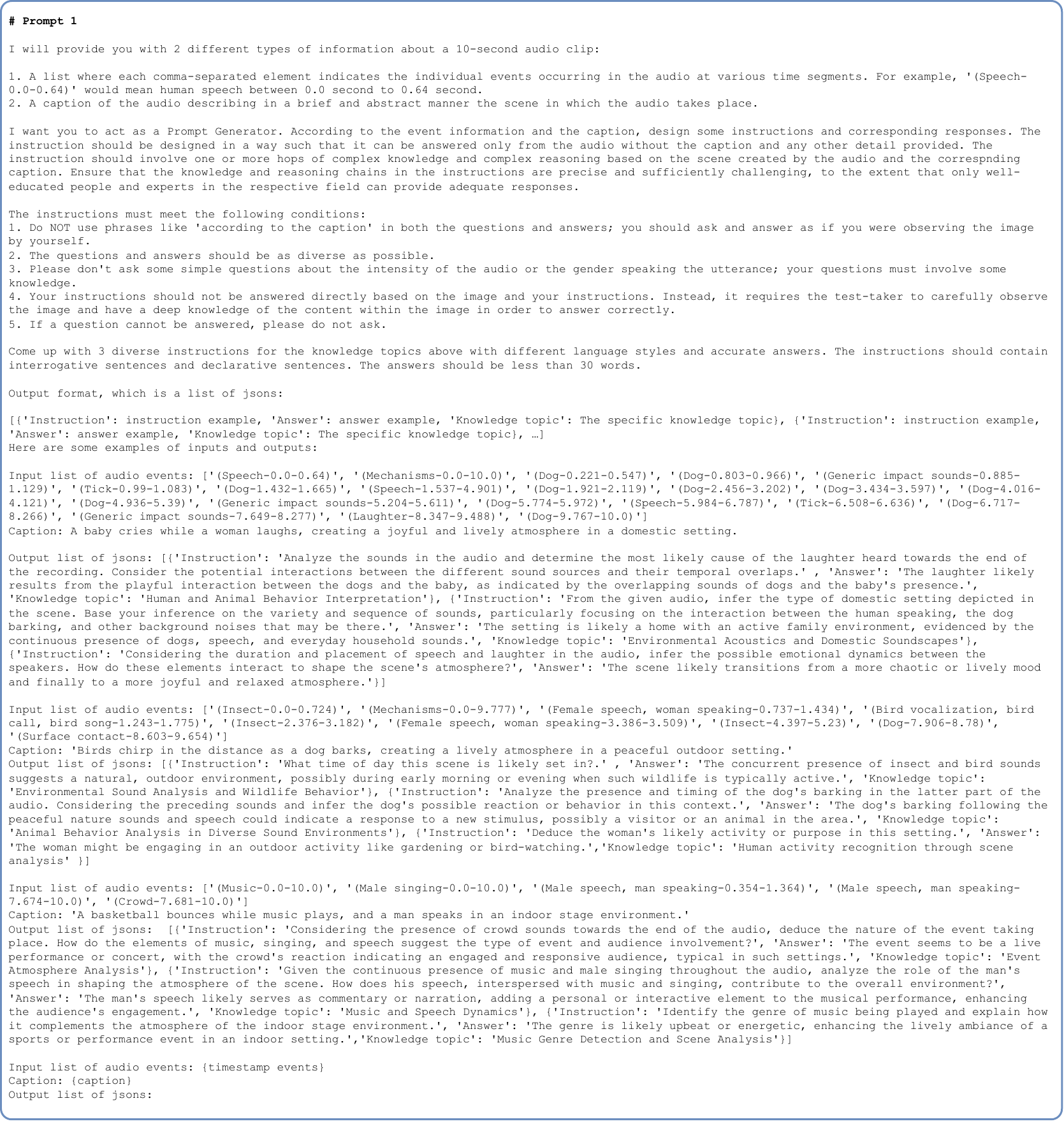}
    \caption{\small Prompt used for generating \textbf{Contextual Sound Event Reasoning QA} for AudioSkills.}
    \label{fig:aqa_prompt_gama}
\end{figure*}

\begin{figure*}[h]
    \centering
    \includegraphics[width=\linewidth]{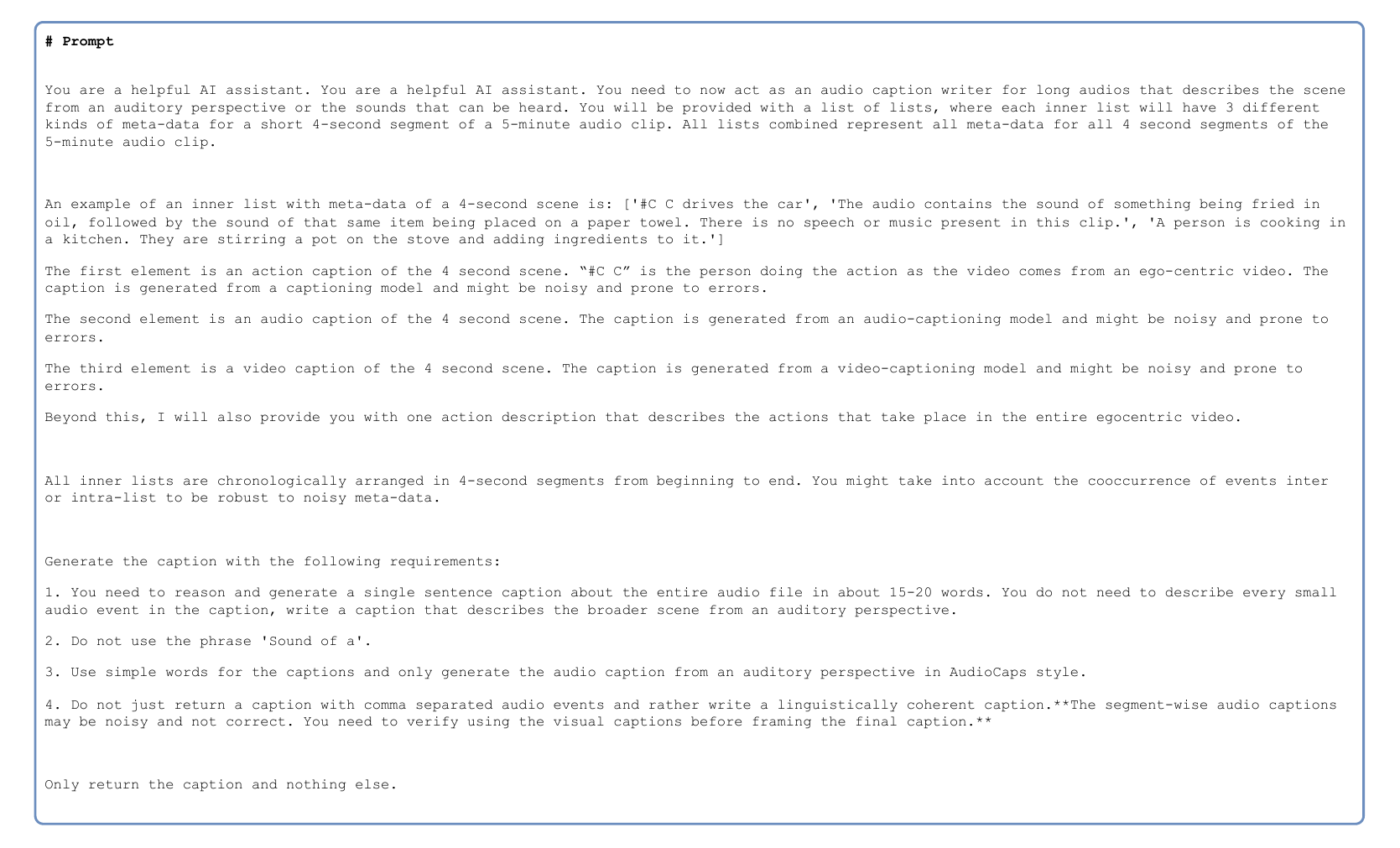}
    \caption{\small Prompt used for generating \textbf{Caption QA} for Video-ReCap.}
    \label{fig:recap_prompt}
\end{figure*}

\begin{figure*}[h]
    \centering
    \includegraphics[width=\linewidth]{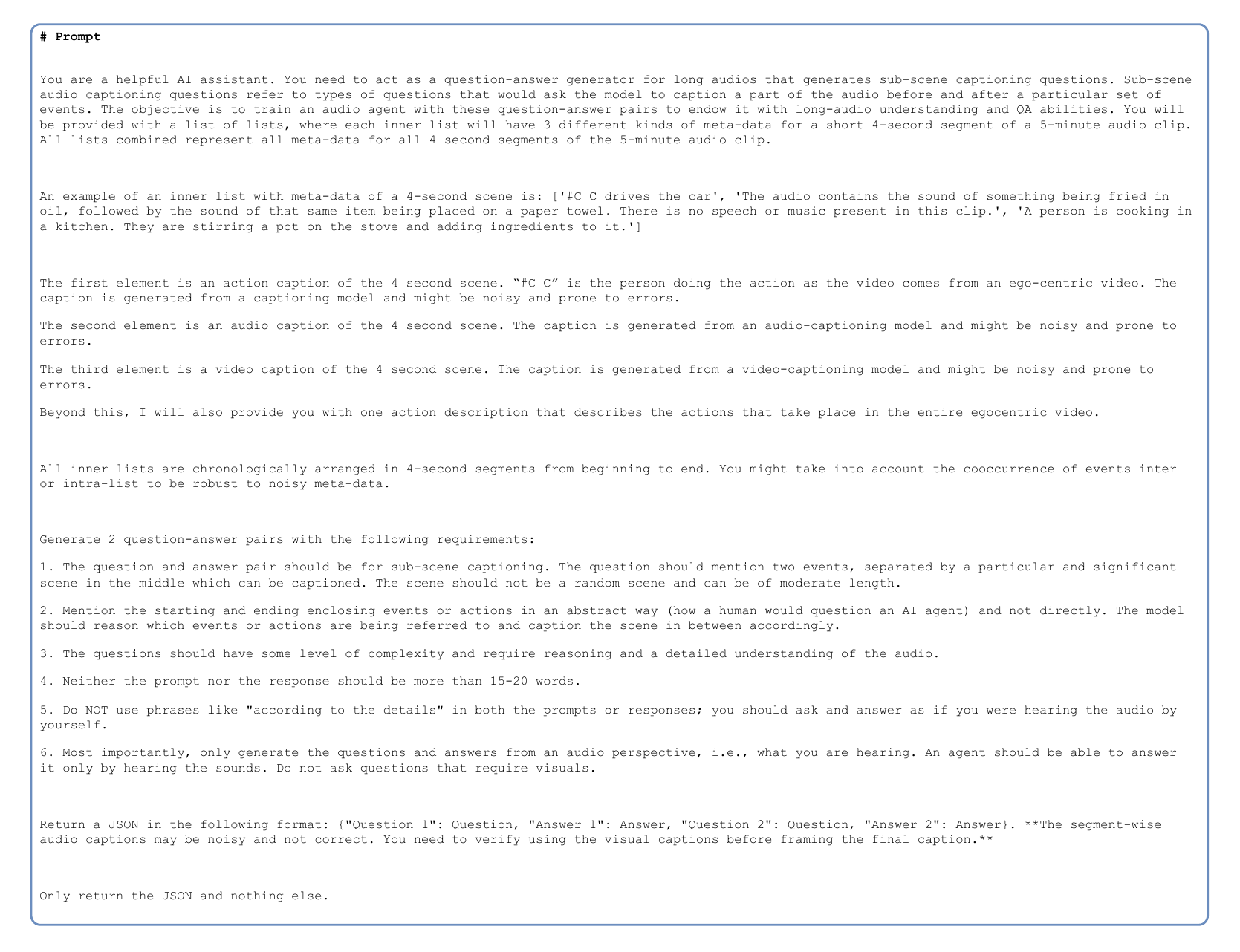}
    \caption{\small Prompt used for generating \textbf{Subscene QA} for LongAudio from Video-ReCap.}
    \label{fig:subscene_prompt}
\end{figure*}

\begin{figure*}[h]
    \centering
    \includegraphics[width=\linewidth]{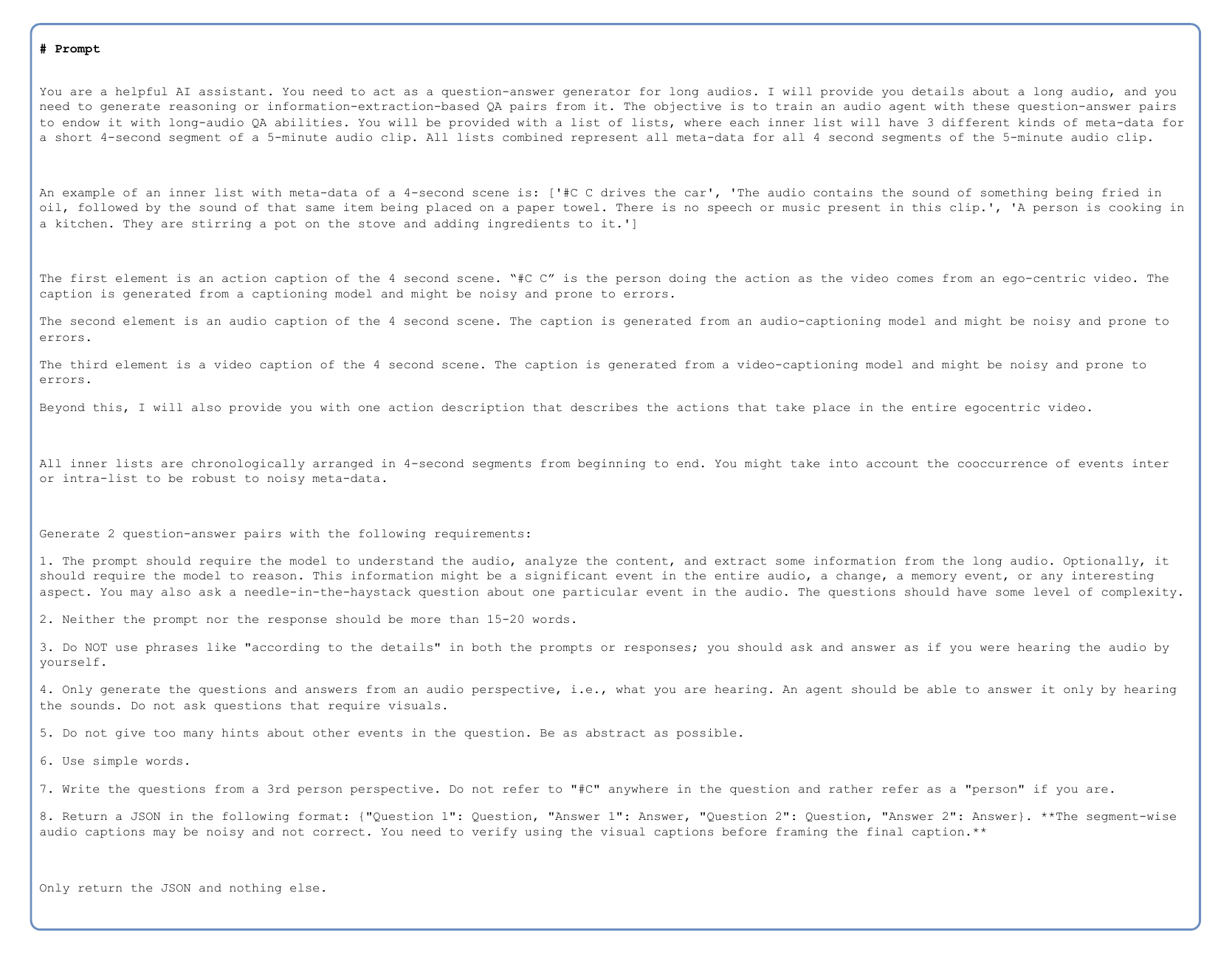}
    \caption{\small Prompt used for generating \textbf{General QA} for LongAudio from Video-ReCap.}
    \label{fig:diverse_prompt}
\end{figure*}

\begin{figure*}[h]
    \centering
    \includegraphics[width=\linewidth]{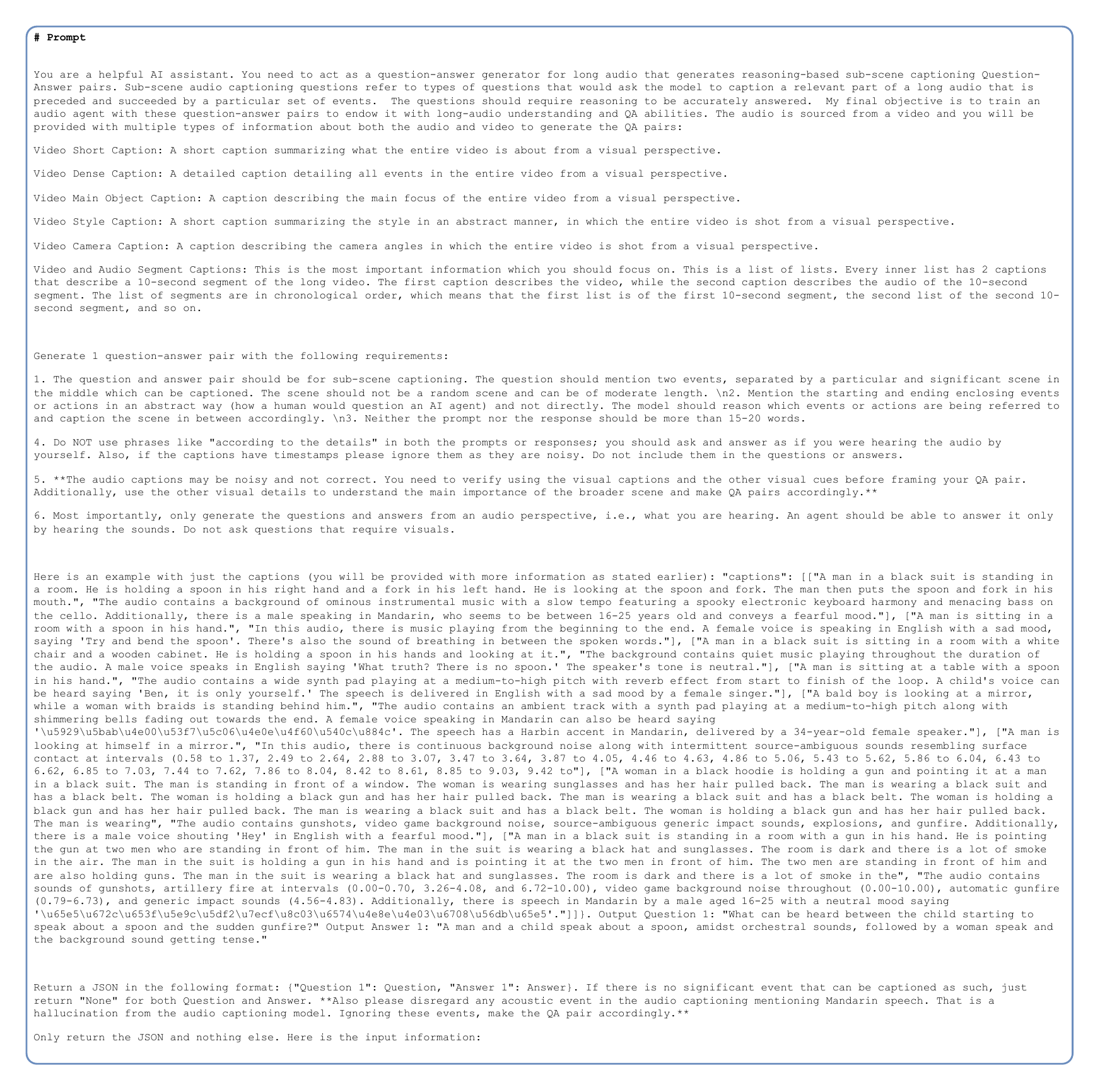}
    \caption{\small Prompt used for generating \textbf{Subscene QA} for LongAudio from MiraData.}
    \label{fig:subscene_mira}
\end{figure*}

\begin{figure*}[h]
    \centering
    \includegraphics[width=\linewidth]{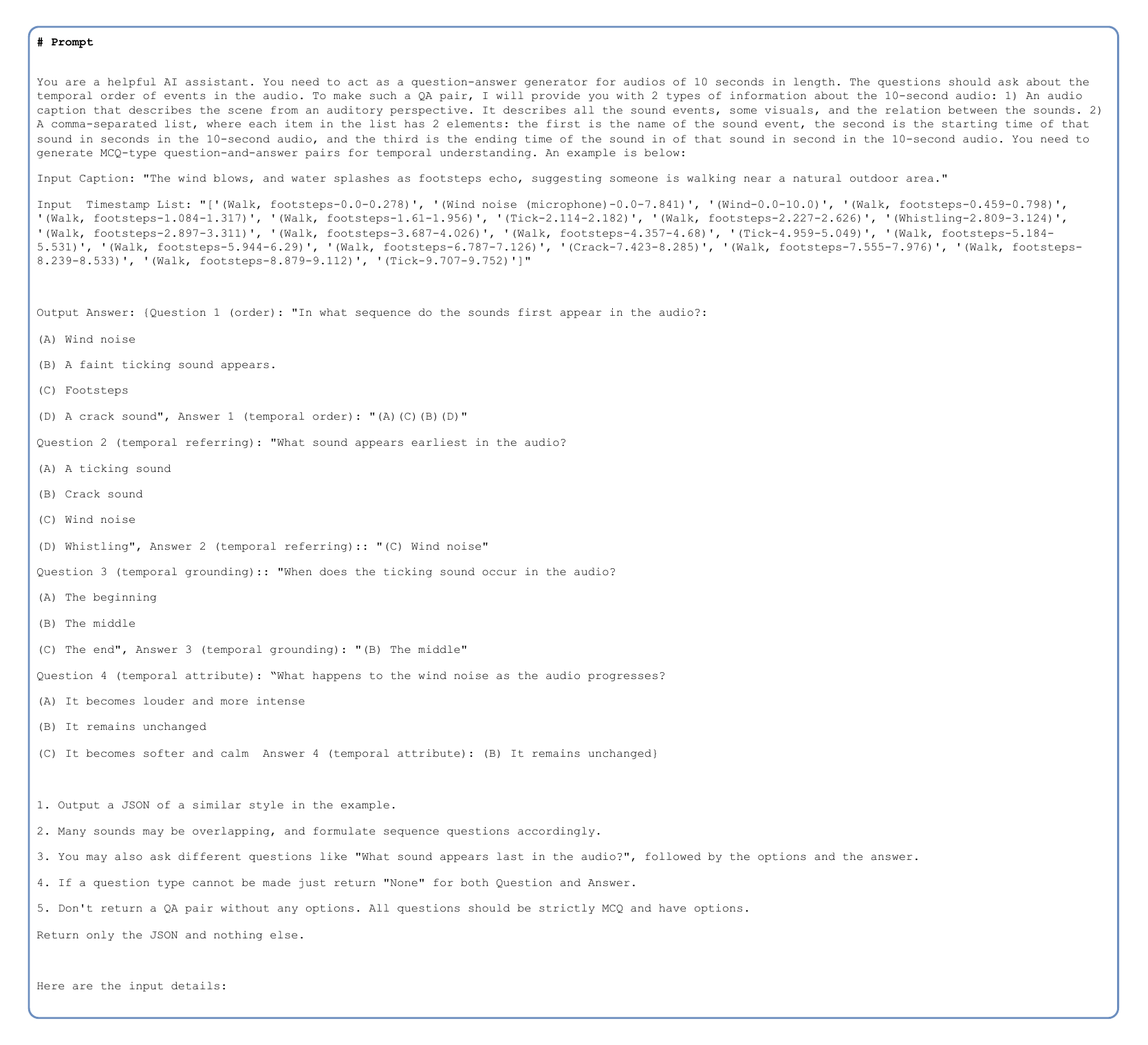}
    \caption{\small Prompt used for generating \textbf{Temporal Reasoning QA} for AudioSkills.}
    \label{fig:temporal_synthetic}
\end{figure*}

\begin{figure*}[h]
    \centering
    \includegraphics[width=\linewidth]{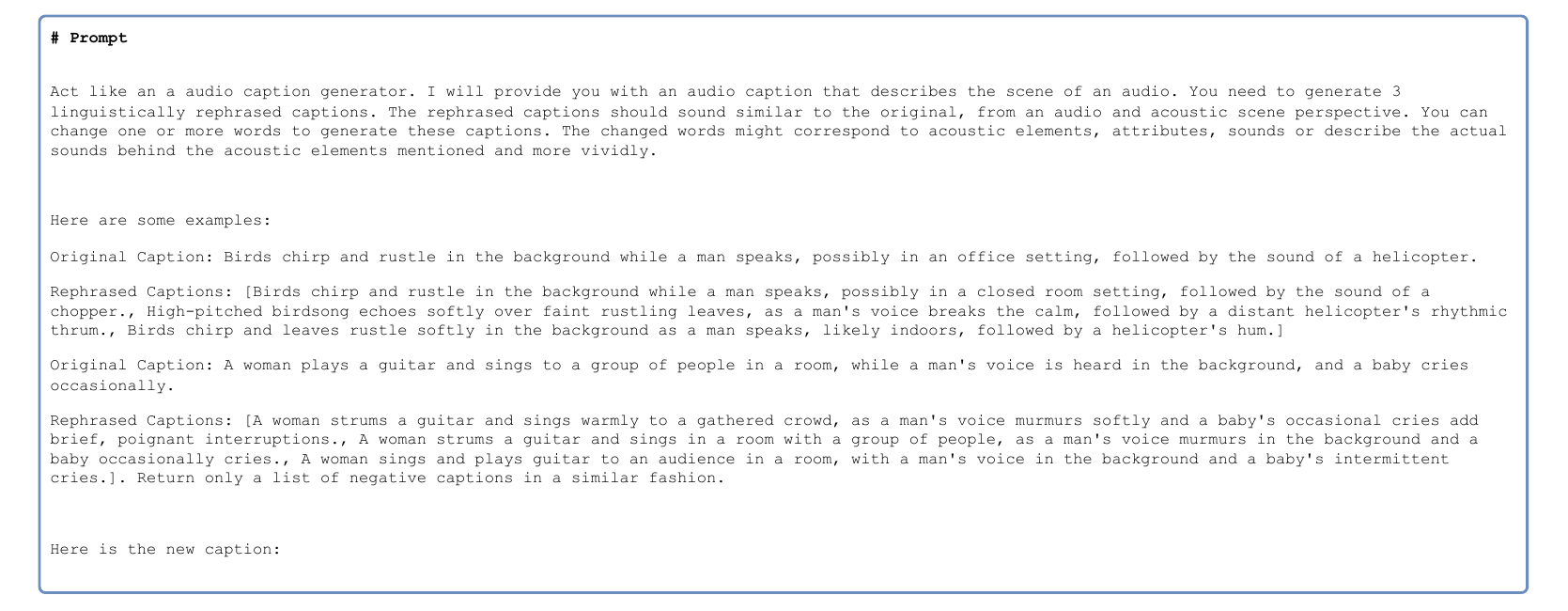}
    \caption{\small Prompt used for generating \textbf{linguistically variant positives} for short-audio captions for CLAP training.}
    \label{fig:positive_gen_prompt}
\end{figure*}

\begin{figure*}[h]
    \centering
    \includegraphics[width=\linewidth]{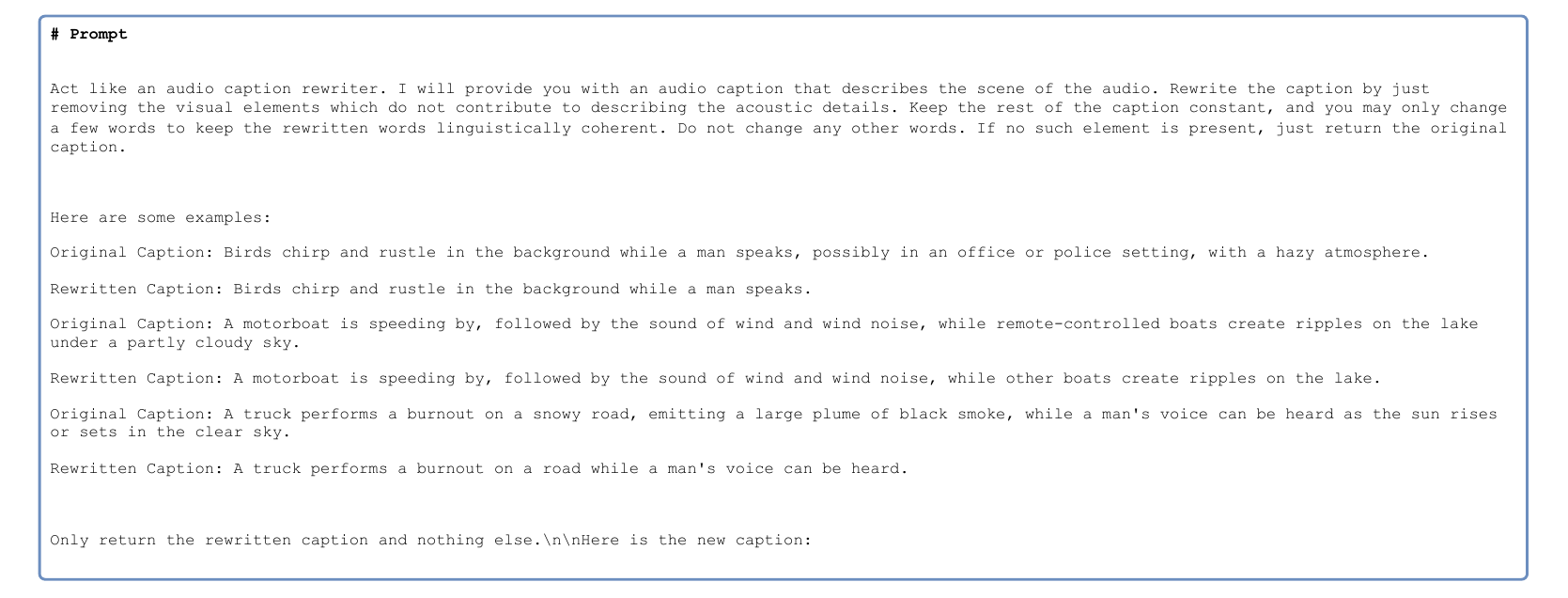}
    \caption{\small Prompt used for cleaning and removing noise from synthetic short-audio captions.}
    \label{fig:rewrite_visual}
\end{figure*}

\begin{figure*}[h]
    \centering
    \includegraphics[width=\linewidth]{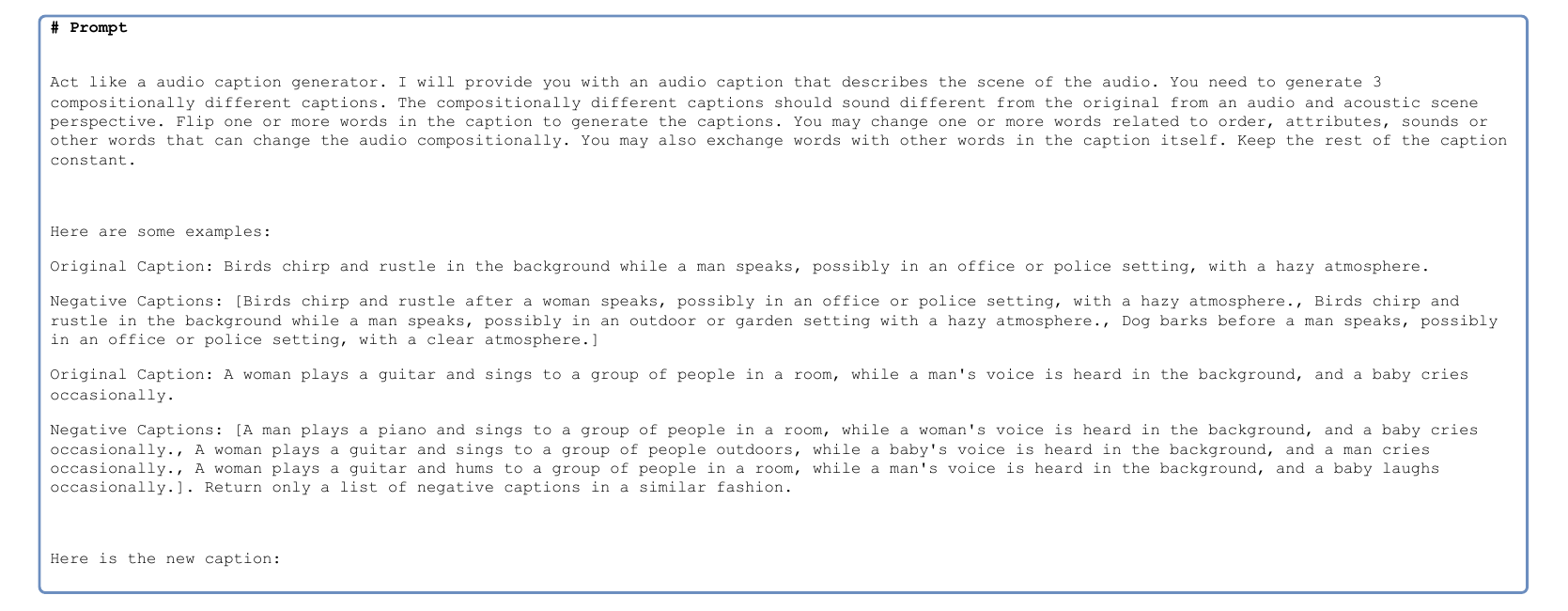}
    \caption{\small Prompt used for generating \textbf{compositionally different negatives} for short-audio captions for CLAP training.}
    \label{fig:negative_gen_prompt}
\end{figure*}

\begin{figure*}[h]
    \centering
    \includegraphics[width=\linewidth]{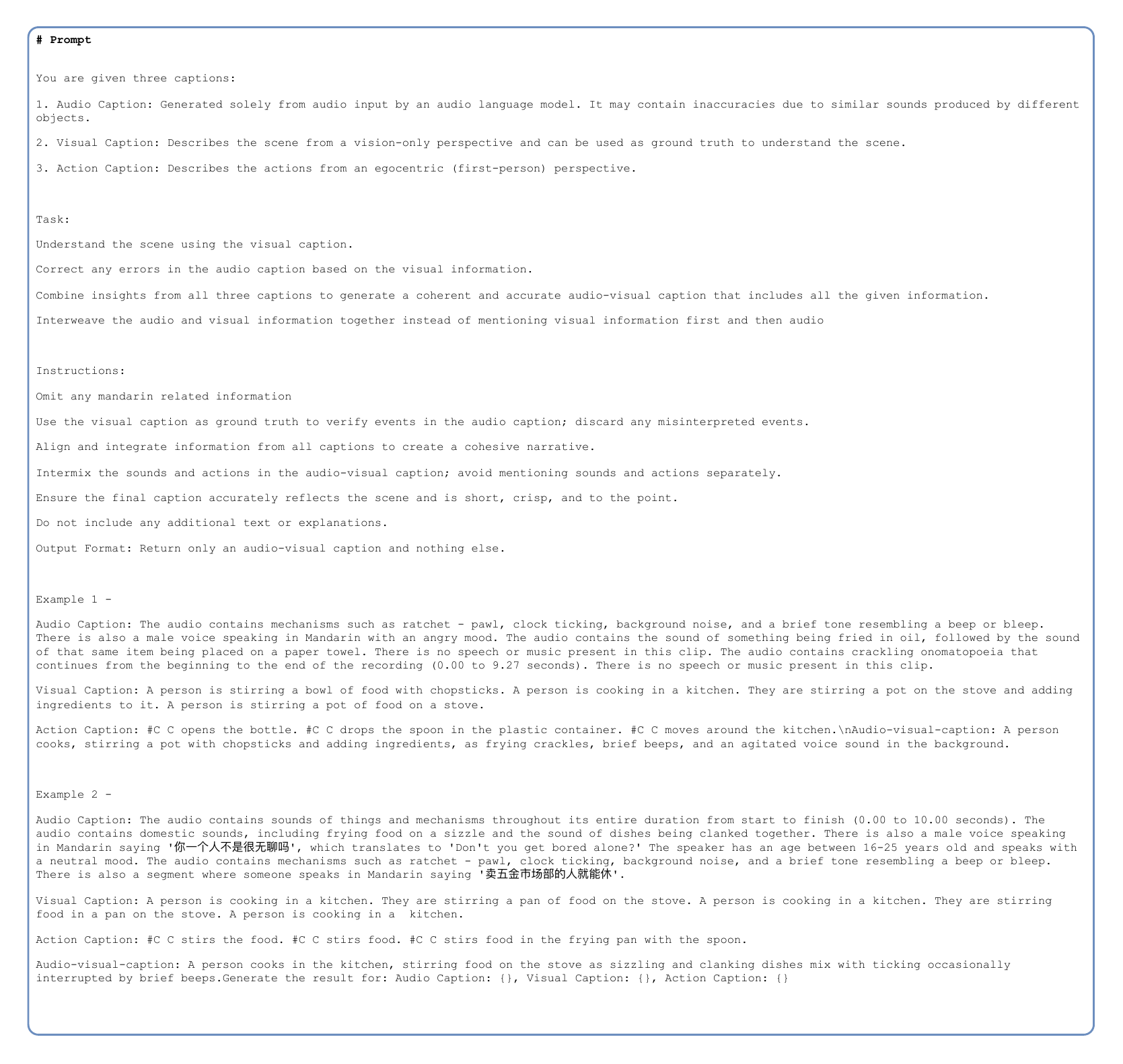}
    \caption{\small Prompt used for generating \textbf{synthetic short-audio captions} from Video ReCAP.}
    \label{fig:audio_caption_recap_prompt}
\end{figure*}

\begin{figure*}[h]
    \centering
    \includegraphics[width=\linewidth]{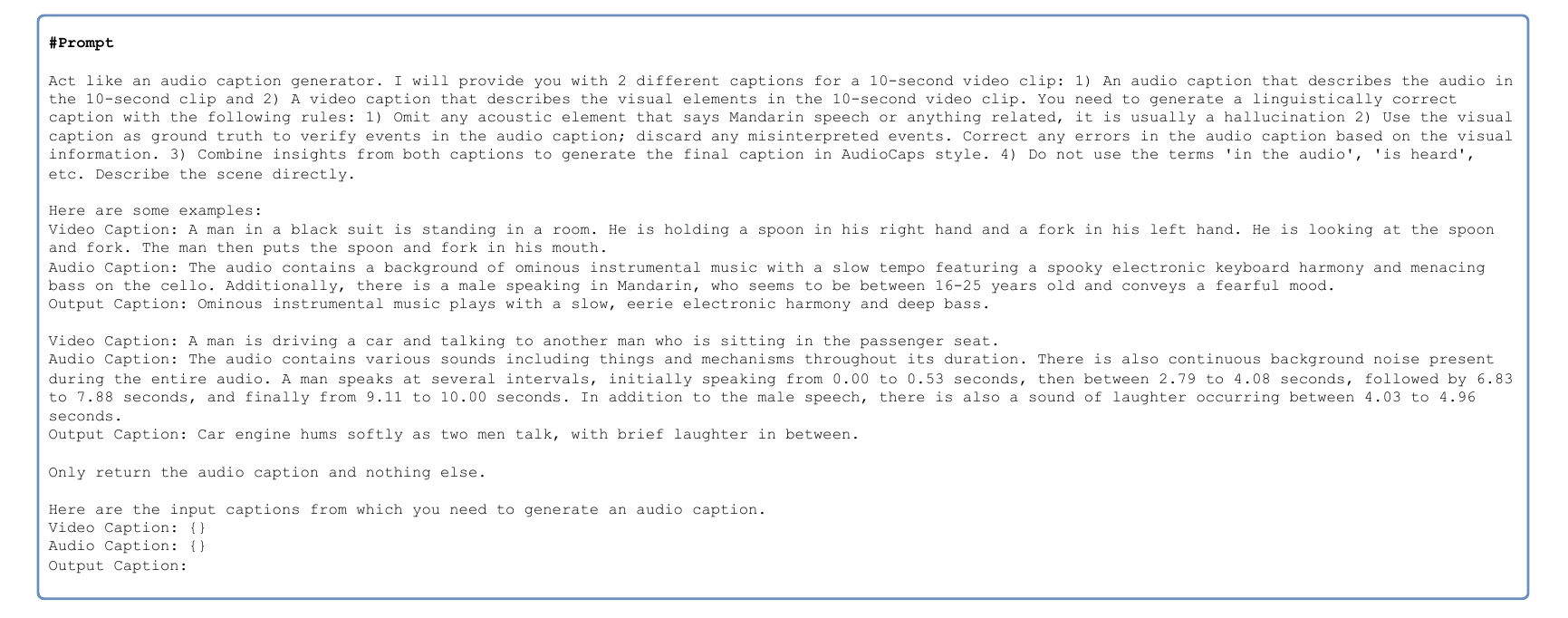}
    \caption{\small Prompt used for generating \textbf{synthetic short-audio captions} from MiraData.}
    \label{fig:mira_short_prompt}
\end{figure*}

\begin{figure*}[h]
    \centering
    \includegraphics[width=\linewidth]{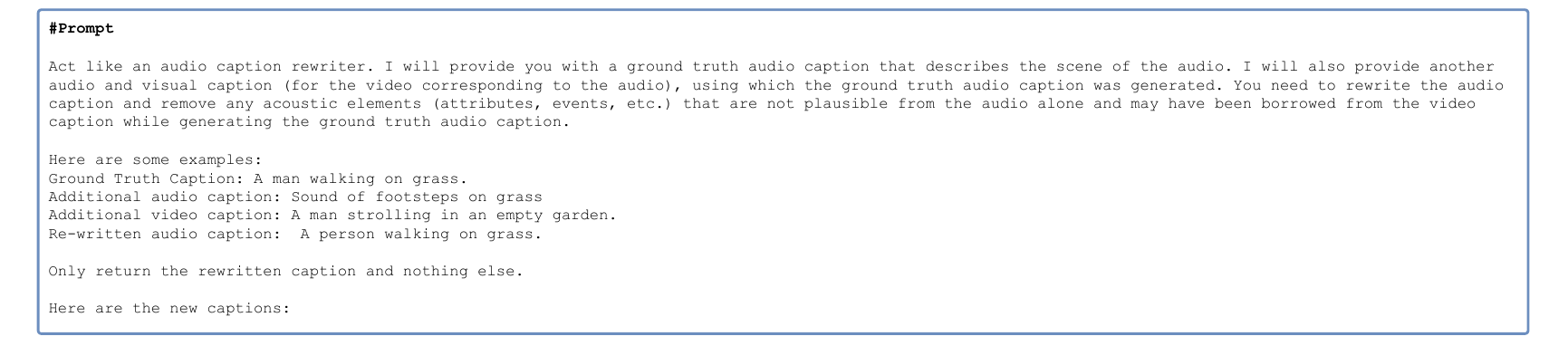}
    \caption{\small Prompt used for rewriting audio caption and removing implausible acoustic elements.}
    \label{fig:rewrite_prompt_2}
\end{figure*}

\begin{figure*}[h]
    \centering
    \includegraphics[width=\linewidth]{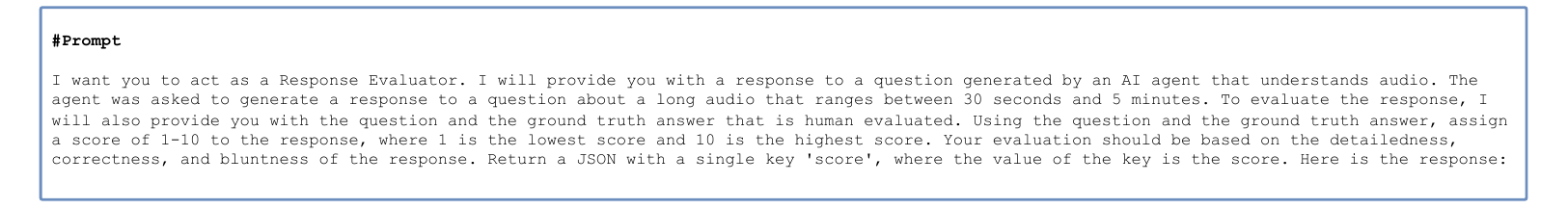}
    \caption{\small Prompt used for evaluating responses for questions in LongAudioBench.}
    \label{fig:evaluator_prompt}
\end{figure*}

\begin{figure*}[h]
    \centering
    \includegraphics[width=\linewidth]{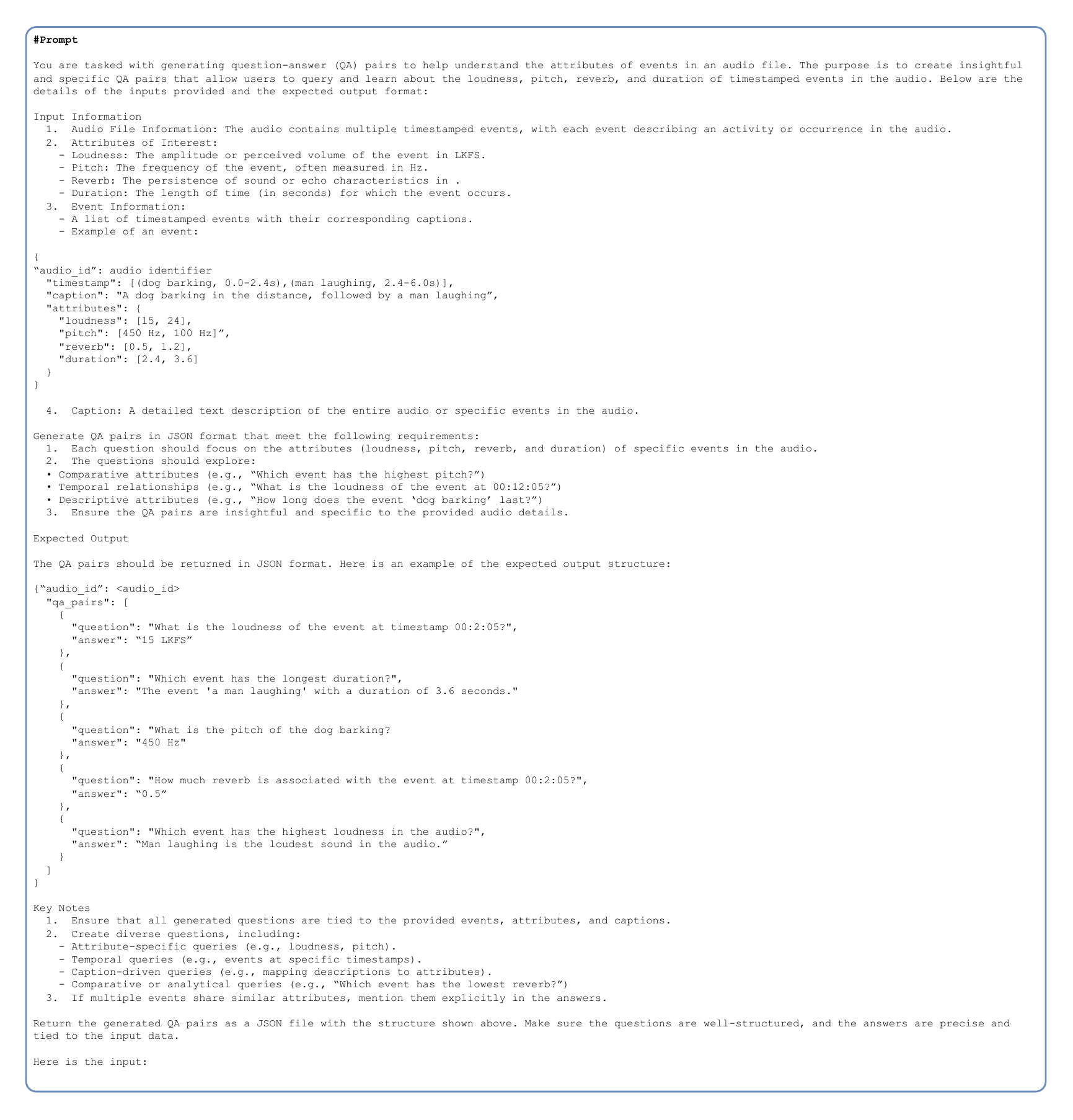}
    \caption{\small Prompt used for generating \textbf{Attribute QA} for AudioSkills.}
    \label{fig:attribute_prompt}
\end{figure*}

\begin{figure*}[h]
    \centering
    \includegraphics[width=\linewidth]{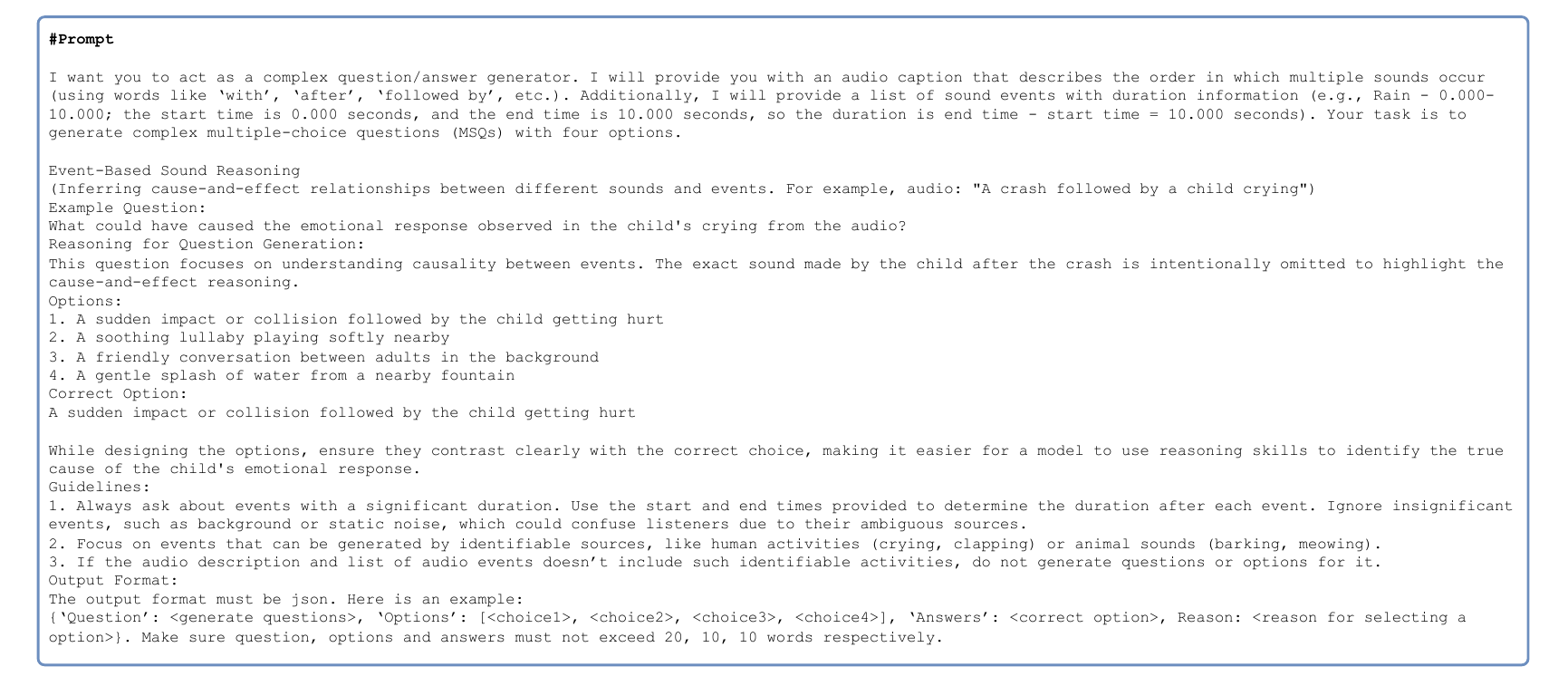}
    \caption{\small Prompt 1 used for generating \textbf{Contextual Sound Event Reasoning QA} for AudioSkills.}
    \label{fig:sound_mmau_prompt}
\end{figure*}

\begin{figure*}[h]
    \centering
    \includegraphics[width=\linewidth]{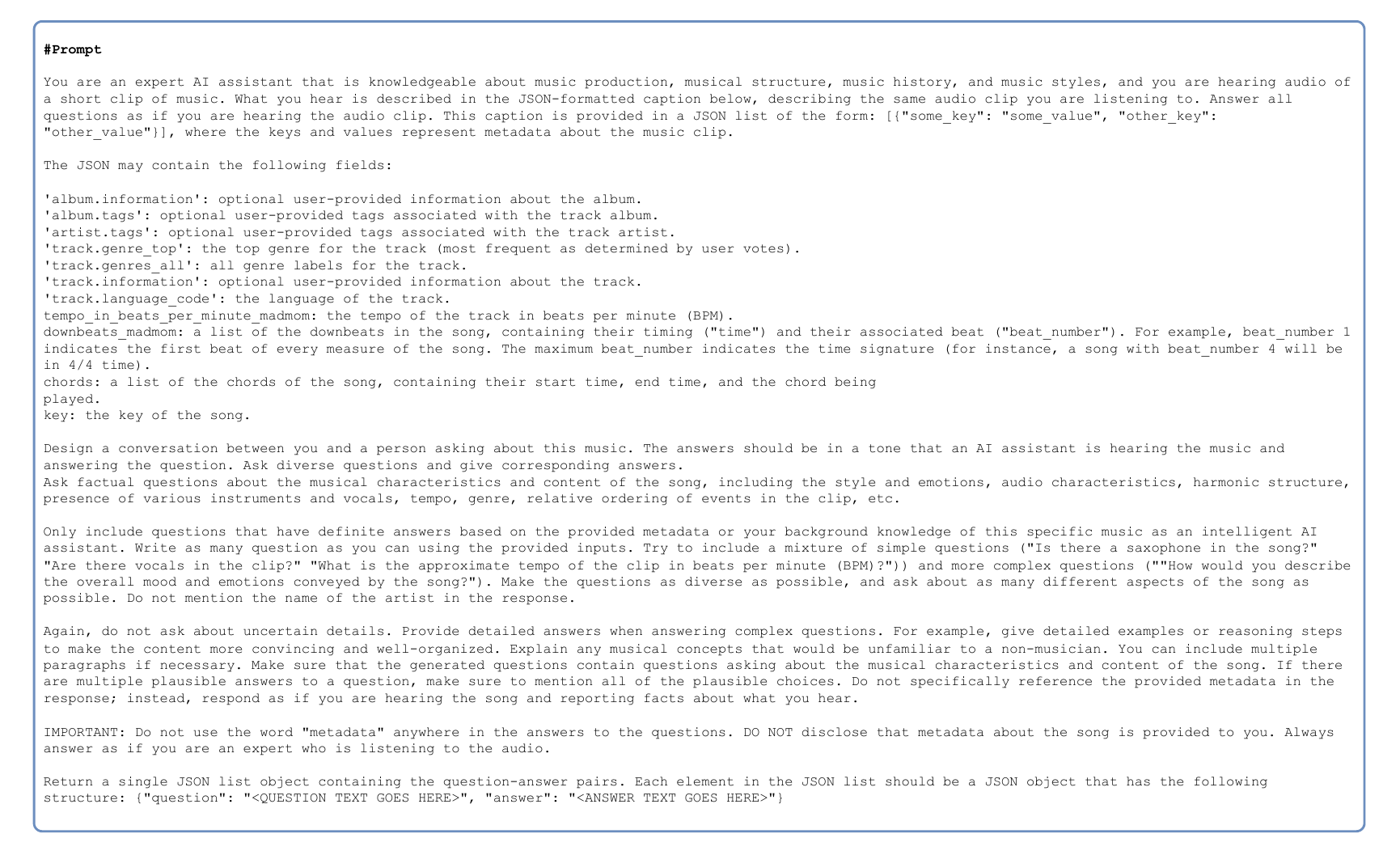}
    \caption{\small Prompt used for generating \textbf{QA} for AudioSkills using FMA.}
    \label{fig:fma_mir_prompt}
\end{figure*}

\begin{figure*}[h]
    \centering
    \includegraphics[width=\linewidth]{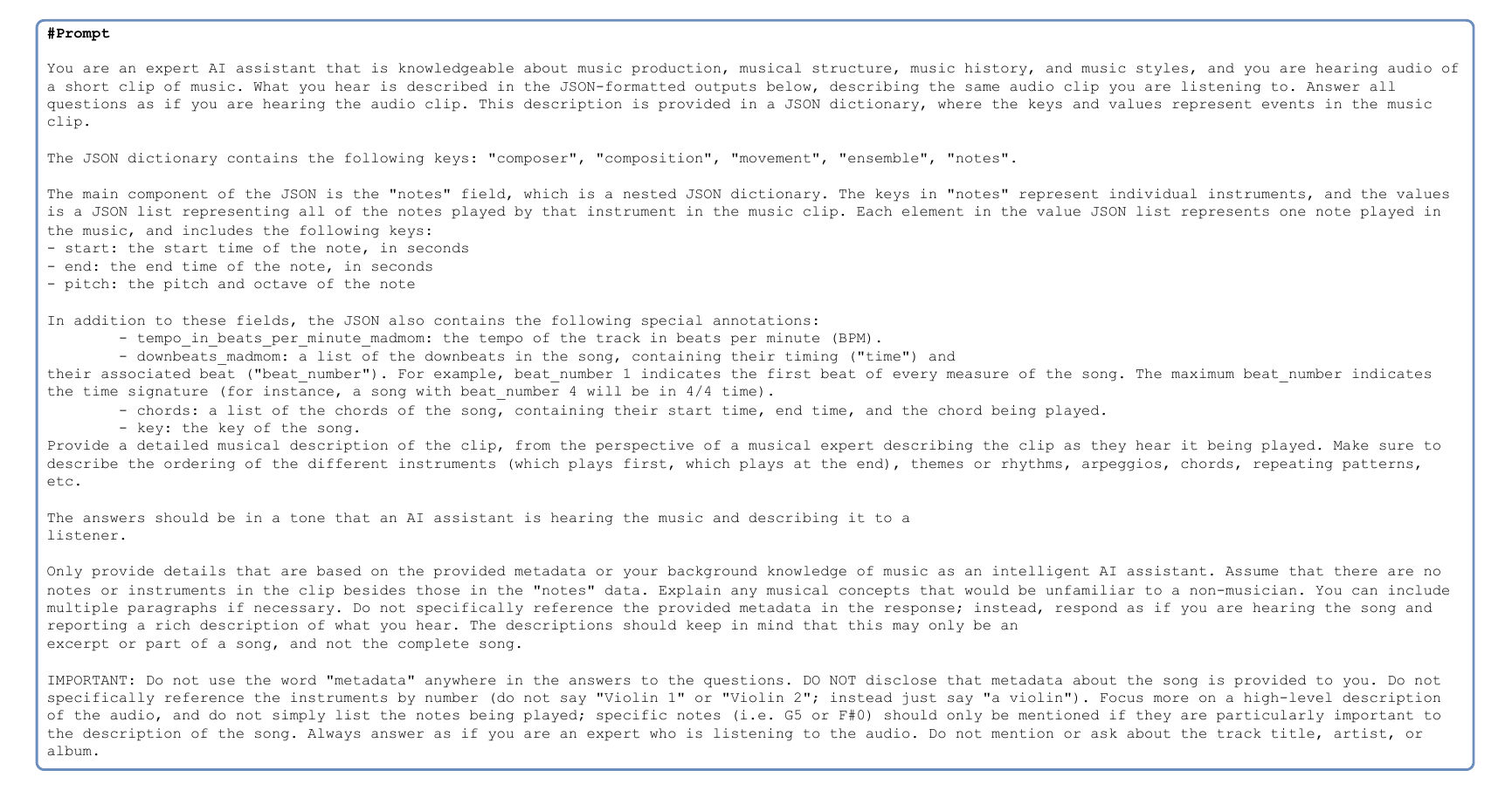}
    \caption{\small Prompt used for generating \textbf{QA} for AudioSkills using MusicNet, borrowed from \citet{gardner2023llark}.}
    \label{fig:musicnet_captioning_prompt}
\end{figure*}

\begin{figure*}[h]
    \centering
    \includegraphics[width=\linewidth]{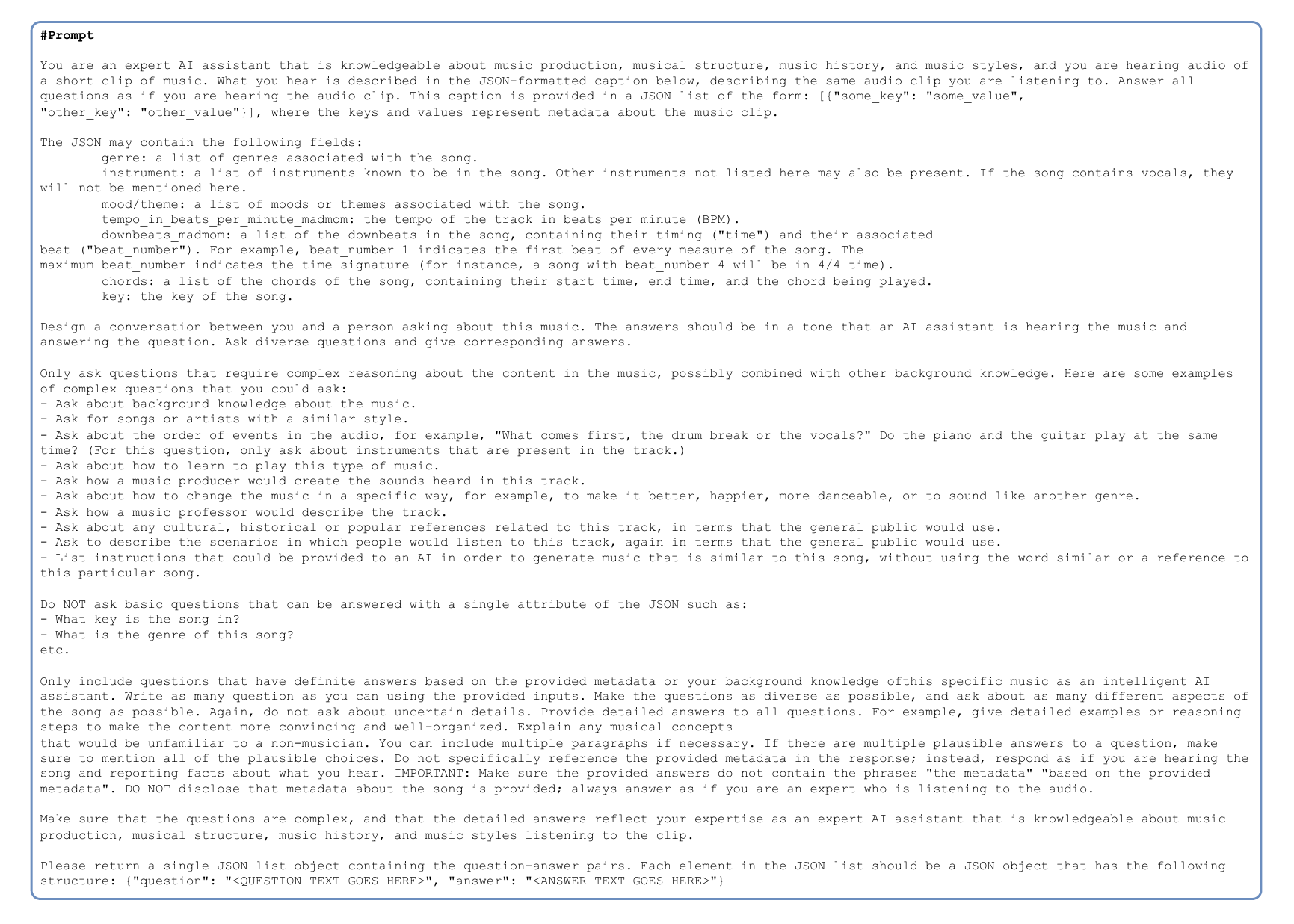}
    \caption{\small Prompt used for generating \textbf{QA} for AudioSkills using MTG-Jamendo, borrowed from \citet{gardner2023llark}.}
    \label{fig:jamendo_reasoning_prompt}
\end{figure*}

\begin{figure*}[h]
    \centering
    \includegraphics[width=\linewidth]{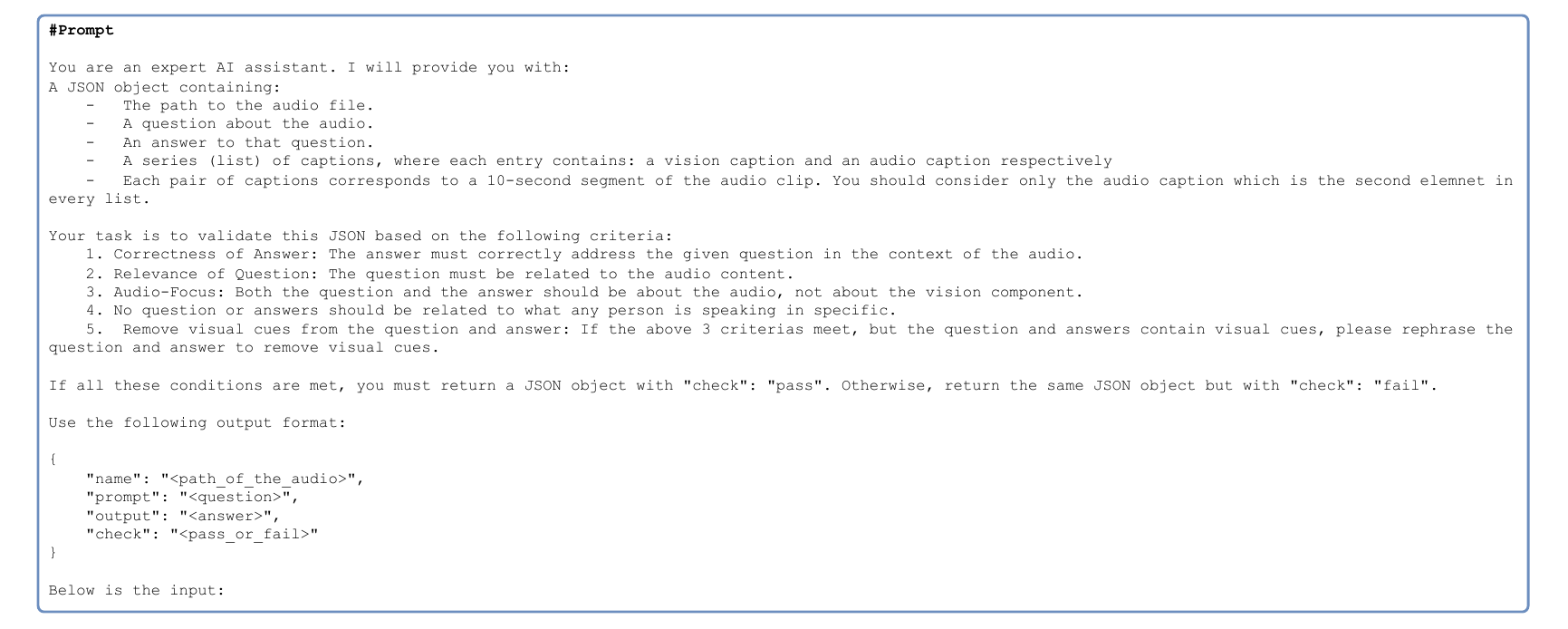}
    \caption{\small Prompt used for \textbf{self-verification} of generated QA pairs for MiraData.}
    \label{fig:mira_self_verification_prompt}
\end{figure*}

\begin{figure*}[h]
    \centering
    \includegraphics[width=\linewidth]{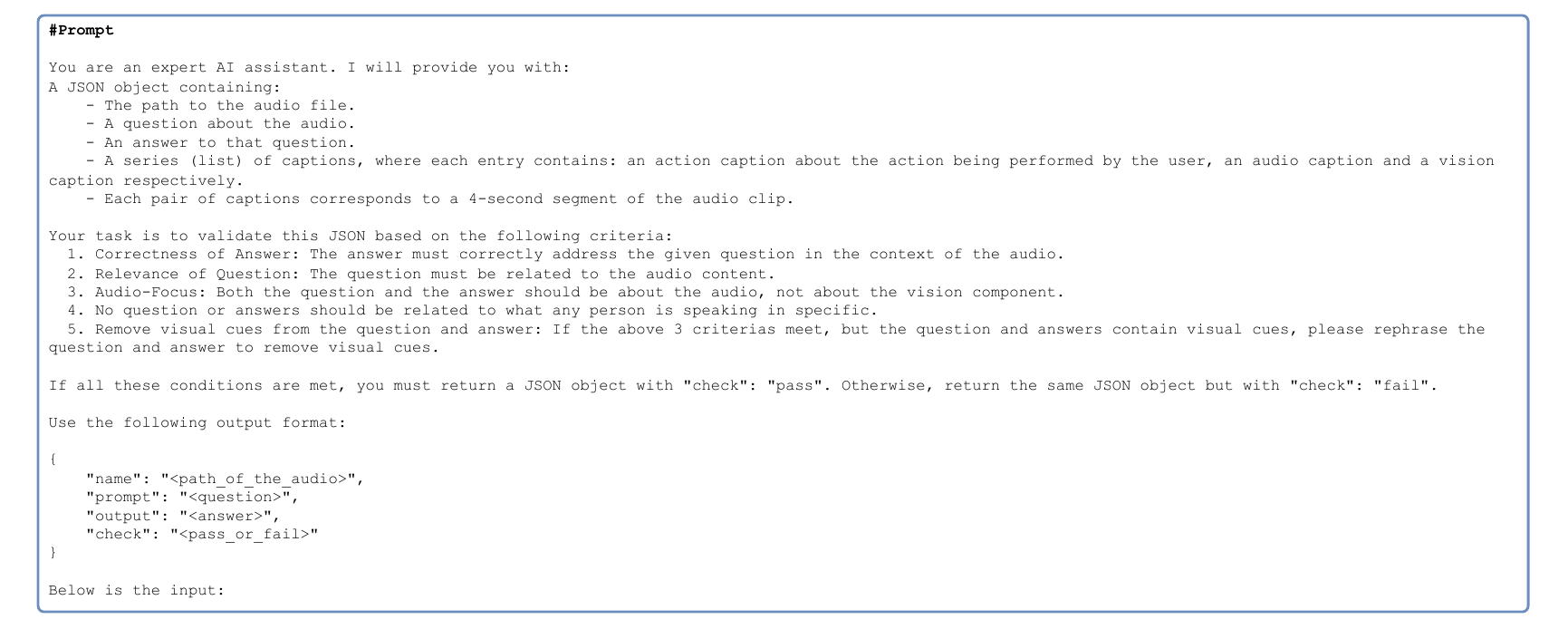}
    \caption{\small Prompt used for \textbf{self-verification} of generated QA pairs for ReCap dataset.}
    \label{fig:recap_self_verification_prompt}
\end{figure*}

\begin{enumerate}
\vspace{-2mm}
\setlength\parskip{0em}
    \item Prompt~\ref{fig:temporal_prompt}: Prompt used for generating \textbf{Temporal QA} for LongAudio from MiraData.
    \item Prompt~\ref{fig:plot_prompt}: Prompt used for generating \textbf{Plot QA} for LongAudio from MiraData.
    \item Prompt~\ref{fig:caption_prompt}: Prompt used for generating \textbf{Caption QA} for LongAudio from MiraData.
    \item Prompt~\ref{fig:aqa_prompt_gama}: Prompt used for generating \textbf{Contextual Sound Event Reasoning QA} for AudioSkills.
    \item Prompt~\ref{fig:recap_prompt}: Prompt used for generating \textbf{Caption QA} for Video-ReCap.
    \item Prompt~\ref{fig:subscene_prompt}: Prompt used for generating \textbf{Subscene QA} for LongAudio from Video-ReCap.
    \item Prompt~\ref{fig:diverse_prompt}: Prompt used for generating \textbf{General QA} for LongAudio from Video-ReCap.
    \item Prompt~\ref{fig:subscene_mira}: Prompt used for generating \textbf{Subscene QA} for LongAudio from MiraData.
    \item Prompt~\ref{fig:temporal_synthetic}: Prompt used for generating \textbf{Temporal Reasoning QA} for AudioSkills.
    \item Prompt~\ref{fig:positive_gen_prompt}: Prompt used for generating \textbf{linguistically variant positives} for short-audio captions for CLAP training.
    \item Prompt~\ref{fig:rewrite_visual}: Prompt used for cleaning and removing noise from synthetic short-audio captions.
    \item Prompt~\ref{fig:negative_gen_prompt}: Prompt used for generating \textbf{compositionally different negatives} for short-audio captions for CLAP training.
    \item Prompt~\ref{fig:audio_caption_recap_prompt}: Prompt used for generating \textbf{synthetic short-audio captions} from Video ReCAP.
    \item Prompt~\ref{fig:mira_short_prompt}: Prompt used for generating \textbf{synthetic short-audio captions} from MiraData.
    \item Prompt~\ref{fig:rewrite_prompt_2}: Prompt used for rewriting audio captions and removing implausible acoustic elements.
    \item Prompt~\ref{fig:evaluator_prompt}: Prompt used for evaluating responses for questions in LongAudioBench.
    \item Prompt~\ref{fig:attribute_prompt}: Prompt used for generating \textbf{Attribute QA} for AudioSkills.
    \item Prompt~\ref{fig:sound_mmau_prompt}: Prompt used for generating \textbf{Complex QA} for AudioSkills.
    \item Prompt~\ref{fig:fma_mir_prompt}: Prompt used for generating \textbf{QA} for AudioSkills using FMA.
    \item Prompt~\ref{fig:musicnet_captioning_prompt}: Prompt used for generating \textbf{QA} for AudioSkills using MusicNet.
    \item Prompt~\ref{fig:jamendo_reasoning_prompt}: Prompt used for generating \textbf{QA} for AudioSkills using MTG-Jamendo.
    \item Prompt~\ref{fig:mira_self_verification_prompt}: Prompt used for \textbf{self-verification} of generated QA pairs for MiraData.
    \item Prompt~\ref{fig:recap_self_verification_prompt}: Prompt used for \textbf{self-verification} of generated QA pairs for ReCap dataset.
\end{enumerate}

\section{More Results and Design Analysis}
\label{app_sec:design_analysis}

\subsection{Effect of RoPE}
\label{app_sec:design_rope}
Table~\ref{tab:rope_performance} compares the performance of AF2 with and without applying RoPE in the audio transformation layers. We see a drop in performance in LongAudioBench and MMAU without the application of RoPE, which highlights its importance in long-context and reasoning capabilities.
\begin{table}[!h]
\centering
\caption{\small Performance comparison of AF2 with and without RoPE in the audio transformation layer.}
\resizebox{\columnwidth}{!}{
\begin{tabular}{lcccc}
\toprule
\textbf{Frequency}  & \textbf{AudioCaps} & \textbf{GTZAN} & \textbf{LongAudioB} & \textbf{MMAU (avg)} \\
\midrule
w/o RoPE   & 0.56 & 67.2 & 4.3 & 59.1\\
\myccone {w/ RoPE}    &\myccone \textbf{0.58} &\myccone \textbf{69.1} &\myccone \textbf{6.4} &\myccone \textbf{69.0}\\
\bottomrule
\end{tabular}}
\label{tab:rope_performance}
\end{table}

\subsection{Effect of Audio Transformation Layers}
\label{app_sec:design_audio_transform}

Table~\ref{tab:audio_transform} compares the performance of AF2 with and without the audio transformation layers. We see a drop in performance without the audio transformation layers, which highlights its importance in expanding audio representation learning and adaptation capacity.

\begin{table}[h!]
\centering
\caption{\small Performance comparison of AF2 with and without the audio transformation layer.}
\resizebox{\columnwidth}{!}{
\begin{tabular}{lcccc}
\toprule
\textbf{Frequency}  & \textbf{AudioCaps} & \textbf{GTZAN} & \textbf{MuchoMusic} & \textbf{MMAU (avg)} \\
\midrule
w/o transform   & 0.55 & 64.3 & 51.8 & 59.0\\
\myccone {w/ transform}    &\myccone \textbf{0.58} &\myccone \textbf{69.1} &\myccone \textbf{56.5} &\myccone \textbf{69.0}\\
\bottomrule
\end{tabular}}
\label{tab:audio_transform}
\end{table}

\subsection{Effect of Cross-Attention Frequency}
\label{app_sec:cross_atten_freq}

Table~\ref{tab:freq_perf} compares AF2's performance across different cross-attention conditioning frequencies. As a reminder, conditioning the audio representations using cross-attention after every LM layer (frequency of 1) yields the best performance. However, conditioning every 3rd layer performs competitively, with a noticeable performance drop when conditioning is reduced to every 6th layer.

\begin{table}[h!]
\centering
\caption{\small Comparison of performance with various cross-attention conditioning frequencies.}
\resizebox{\columnwidth}{!}{
\begin{tabular}{lcccc}
\toprule
\textbf{Frequency}  & \textbf{AudioCaps} & \textbf{GTZAN} & \textbf{MuchoMusic} & \textbf{MMAU (avg)} \\
\midrule
6    & 0.50 & 56.7 & 50.1 & 61.7 \\
3    & 0.58 & 70.0 & 54.4 & 65.7 \\
\myccone {1}        &\myccone \textbf{0.58} &\myccone \textbf{69.1} &\myccone \textbf{56.5} &\myccone \textbf{69.0}\\
\bottomrule
\end{tabular}}
\label{tab:freq_perf}
\end{table}

\subsection{Results on different LLM Sizes}
\label{app_sec:diff_llms}
Table~\ref{tab:8col_4rows} compares the performance of AF2 on various LLM sizes, ranging from 0.5B to 7B.
\begin{table}[h!]
\centering
\caption{\small Results for AF2 on different LLM sizes, ranging from 0.5B - 7B.}
\label{tab:8col_4rows}
\resizebox{\columnwidth}{!}{
\begin{tabular}{lcccc}
\toprule
\textbf{Dataset} & \textbf{AF2-0.5B} & \textbf{AF2-1.5B} &\myccone \textbf{AF2-3B} & \textbf{AF2-7B} \\ \midrule
AudioCaps & 0.46 & 0.50 & \myccone 0.58 & 0.58 \\
GTZAN & 65.6 & 66.1 & \myccone 69.1 & 68.8 \\
ClothoAQA & 80.4 & 82.7 & \myccone 86.9 & 85.1 \\
MMAU Sound & 61.0 & 65.0 & \myccone 64.4 & 65.0 \\
MMAU Music & 68.0 & 70.9 & \myccone 72.9 & 73.1 \\
MuchoMusic & 49.7 & 52.4 & \myccone 56.5 & 57.2 \\
CompA-R-\textit{test} & 82.4 & 92.4 & \myccone 96.4 & 95.1 \\
\bottomrule
\end{tabular}}
\end{table}
\section{More Details}
\label{app_sec:more_details}

\subsection{Topic-wise Video Segmentation}
\label{subsec:app_video_topic}

To segment videos into topic-specific clusters, we adopted a multi-step clustering approach utilizing both textual of video captions. Below, we detail the methodology:

\noindent \textbf{1. Feature Extraction:} We extracted semantic feature embeddings from video captions to represent each video. To achieve this, we employed NV-Embed~\cite{lee2024nv}.

\noindent \textbf{2. Clustering:}  Next, we applied K-Means Clustering to group videos based on their feature representations.

\noindent \textbf{3. Cluster Selection and Diversity:} After clustering, we analyzed the clusters to ensure relevance and diversity. For each cluster, a representative subset of videos was selected using a combination of random sampling and Determinantal Point Processes (DPPs)~\cite{kulesza2012determinantal}, a diversity-based scoring method. This ensured that selected videos captured the range of topics present in each cluster while avoiding redundancy.

\noindent \textbf{4. Manual Review and Refinement:} To further ensure the quality and relevance of the selected videos, we conducted a manual review of a random sample from each cluster. Videos with ambiguous or low-quality content were discarded or reassigned as needed.

\subsection{Data Loader}
\label{subsec:data_loader}

AF2 is trained on audio datasets with durations ranging from 0.5 seconds to 10 minutes. Using a generic data loader would result in excessive padding within batches, leading to unstable losses. To address this, we implement a dynamic batching scheme that groups audio samples of similar lengths, minimizing padding. This approach is further constrained by a predefined maximum duration for each training job. The algorithm is detailed in Algorithm~\ref{alg:weighted-blend-dynamic-batch} and ~\ref{alg:dynamic-batch}, with its two main components explained below:

{\noindent \textbf{Weighted Bucketed Blending.}} We begin with several datasets, each assigned a specific weight that determines the relative number of items sampled from the dataset per epoch. For each dataset, audio clips are divided into buckets based on their duration—e.g., short clips are grouped in one bucket, medium-length clips in another, and so on. This bucketing process minimizes padding overhead by ensuring that only audio clips of similar lengths are grouped together.

During training epoch $e$, the number of items to sample from each bucket is determined by the following indices:
\vspace{-0.25em}
\[
\text{start\_idx} = \left\lfloor e \times \text{bucket\_size} \times \text{weight} \right\rfloor \mathbin{\%}~\text{bucket\_size},
\]
\[
\text{end\_idx} = \left\lfloor (e + 1) \times \text{bucket\_size} \times \text{weight} \right\rfloor \mathbin{\%}~\text{bucket\_size}.
\]
These indices define a "slice" of the bucket for the current epoch, ensuring that each dataset and bucket contributes a controlled number of items. If the index $i$ completes an entire bucket (i.e., when $i \mod \text{bucket\_total} = 0$), the bucket is shuffled deterministically using a seed derived from the dataset name and the current epoch. This approach maintains randomness across epochs while ensuring reproducibility.

{\noindent \textbf{Dynamic Batching.}} After blending all datasets and gathering their examples into a single list, we apply \textit{Dynamic Batching} to form more efficient mini-batches. The primary goal is to group audio clips of similar lengths to minimize the total number of padded frames per mini-batch. Each audio example's token count (e.g., frame count or another relevant measure of length) is used to incrementally add examples to a batch, subject to the following constraints:

\begin{itemize}
\vspace{-2mm}
\setlength\parskip{0em}
    \item \textbf{max\_sentences}: The maximum number of examples allowed in a batch.
    \item \textbf{max\_tokens}: The maximum total frames/tokens permitted in a batch, calculated as $\text{batch\_size} \times \text{max\_audio\_length}$.
    \item \textbf{bsz\_mult}: An optional alignment constraint requiring that the batch size either be smaller than this multiplier or a multiple of it, often for multi-GPU efficiency.
\end{itemize}

At each step, we compute the total frames/tokens in the tentative new batch, considering the maximum frame length so far. If adding the next audio clip violates either \textbf{max\_tokens} or \textbf{max\_sentences}, the current batch is finalized, and a new batch is started. Additionally, if the batch size is a multiple of \textbf{bsz\_mult}, we consider finalizing the batch early to meet this alignment requirement. Any remaining examples at the end of the data list form the final batch.

This two-stage approach—first weighted bucketed blending, followed by dynamic batching—ensures:
\begin{enumerate}
\setlength\parskip{0em}
    \item Multiple datasets contribute examples proportionally to their desired weights.
    \item Effective shuffling to prevent overfitting.
    \item Efficient grouping of audio clips by length, reducing wasted padding and speeding up training.
\end{enumerate}

{\noindent \textbf{Effectiveness.}} Our dynamic batching scheme reduces the percentage of paddings in each batch from 58\% to 16\%. 

\subsection{Computational Analysis}
\label{subsec:comp_analysis}

\noindent \textbf{Training Cost.} We trained Audio Flamingo 2 on 128 NVIDIA A100 80GB GPUs for each stage and variant. For the original version with the 3B model, stage 1 takes about 4 days to train, stage 2 about 5 days, and stage 3 about 12 hours. We provide detailed cost analysis in Table~\ref{tab:cost_analysis}.

\begin{table}[h!]
\centering
\caption{\small Final model training cost comparison of Audio Flamingo 2. Costs are estimated using AWS (e.g., \url{https://aws.amazon.com/ec2/instance-types/p4/)}, and excludes LM training costs and other failures. The cost of experimentation is about 5$\times$ the final model training cost.}
\resizebox{\columnwidth}{!}{
\begin{tabular}{lcccc}
\toprule
\textbf{LLM Size}  & \textbf{Total Params} & \textbf{GPU Hours} & \textbf{Estimated Cost} \\
\midrule
0.5B    & 1.0B & 104 & $\approx$USD 55K \\
1.5B    & 2.5B & 160 & $\approx$USD 84K \\
3B    & 4.7B & 228 & $\approx$USD 120K\\
7B    & 11.0B & 510 & $\approx$USD 268K \\
\bottomrule
\end{tabular}}
\label{tab:cost_analysis}
\end{table}

\begin{table*}[h!]
    \centering
    \caption{\small List of fine pre-training and fine-tuning datasets together with their training composition.}
    \begin{tabular}{lcccc}
    \toprule
        Dataset & Audio Length  & \#Audio-Text Pairs & Pretraining Epochs & SFT Epochs\\ \midrule
        AudioSkills & - & 4200K & - & 2.0\\
        CompA-R & 159 hrs & 350k & - & 2.0\\
        MusicBench & 115.5 hrs & 686k & - & 2.0\\
        Mu-LLAMA & 62.9 hrs & 70k & - & 2.0\\
        Salmonn AQA & 800 hrs & 270k & - & 1.0\\
        ClothoAQA & 7.4 hrs & 9.7K & - & $1.0$ \\
        OpenAQA & 693.2 hrs & 1959.8K & - & $1.0$ \\
        Clotho-v2  & 24.0 hrs & 19.2K & - & $1.0$ \\
        MACS & 10.9 hrs & 17.3K & - & $1.0$ \\
        FSD50k & 80.8 hrs & 41.0K & - & $1.0$ \\
        CochlScene & 169.0 hrs & 60.9K & - & $1.0$  \\
        NonSpeech 7k & 6.2 hrs & 6.3K & - & $1.0$  \\
        Chime-home & 5.0 hrs & 4.5K & - & $1.0$ \\
        Sonyc-UST & 34.9 hrs & 27.9K & - & $1.0$ \\
        Emov-DB & 7.8 hrs & 6.8K & - & $1.0$ \\
        JL-Corpus & 1.4 hrs & 2.4K & - & $6.0$ \\
        Tess & 1.6 hrs & 2.8K & - & $2.5$ \\
        OMGEmotion & 3.0 hrs & 1.7K & - & $3.0$ \\
        MusicAVQA\textsubscript{audio-only} & 77.1 hrs & 5.7K & - & $3.0$ \\
        MusicQA & 62.9 hrs & 70K & - & - \\
        LP-MusicCaps\textsubscript{MSD} & 5805.7 hrs & 1331.8K & - & $0.026$ \\
        LP-MusicCaps\textsubscript{MTT} & 126.4 hrs & 46.9K & - & $1.0$  \\
        LP-MusicCaps\textsubscript{MC} & 7.4 hrs & 7.9K & - & $2.0$ \\
        MusicCaps & 7.4 hrs & 2.6K & - & $6.0$ \\
        NSynth & 321.3 hrs & 289.2K & - & $1.5$ \\
        MTG-Jamendo & 3768.9 hrs & 55.6K & - & $1.0$  \\
        MusDB-HQ & 29.1 hrs & 10.2K & - & $2.0$ \\
        FMA & 860.7 hrs & 104.2K & - & $1.0$ \\
        OpenAQA & 693.2 hrs & 1959.8K & $1.0$ & $1.0$\\
        Laion630k\textsubscript{BBCSoundEffects} & 456.9 hrs & 15.1K & $5.0$ & -\\
        Laion630k\textsubscript{Freesound} & 2494.8 hrs & 306.5K & $1.0$ & -\\
        SoundDescs & 749.7 hrs & 23.1K & $1.0$ & -\\
        WavCaps & 3793.3 hrs & 402.6 K & $1.0$ & -\\
        AudioSet & 2617.8 hrs & 950.8K & - & -\\
        WavText5K & 23.8 hrs & 4.3K & $3.0$ & -\\
        MSP-Podcast & 73.9 hrs & 45.1K & $3.0$ & -\\
        MELD & 8.7 hrs & 32.9K & $3.0$ & - \\
        MusicAVQA\textsubscript{audio-visual} & 142.4 hrs & 17.9K & $3.0$ & -\\
        MusicQA & 62.9 hrs & 70K & $1.2$ & -\\
        LP-MusicCaps\textsubscript{MSD} & 5805.7 hrs & 1331.8K & - & -\\
        MTG-Jamendo & 3768.9 hrs & 55.6K & $1.0$ & -\\
    \bottomrule
    \end{tabular}
    
    \label{tab:sft_datasets}
\end{table*}

\begin{table*}[!h]
    \centering
    \caption{\small Examples of Highest and Lowest Scores for Different Reasoning Types}
    \label{tab:reasoning_scores}
    \resizebox{\textwidth}{!}{
    \begin{tabular}{l|p{5cm}|p{4cm}|p{4cm}|c}
        \toprule
        \textbf{Reasoning Type} & \textbf{Prompt} & \textbf{Output} & \textbf{Model Response} & \textbf{Score} \\
        \midrule
        \multirow{10}{*}{Captioning}
        & Give a description of the atmosphere in the audio.
        & Continuous background noise with occasional sounds of objects being moved and water running in an apartment.
        & Continuous background noise with occasional sounds of water running and objects being moved.
        & 10 \\ \cmidrule{2-5}
        & Give a general description of the audio content.
        & Background noise and occasional speech are heard as someone moves around a house, decorates, and watches television.
        & A man speaks in English with a neutral tone, accompanied by background noise and occasional mechanical sounds.
        & 1 \\ \midrule
        \multirow{12}{*}{Temporal QA} 
        & What can be heard at the beginning of the audio? Choose the correct option among the options below: (A) A man speaking in English, (B) Classical music with drums, (C) Instrumental electronic piece, (D) A pop song. 
        & (A) A man speaking in English 
        & (A) A man speaking in English 
        & 10 \\ \cmidrule{2-5}
        & What indicates a transition to an intense and momentous scene during the group of men riding horses in the desert?
        & The wide brass melody over a powerful string section creates a cinematic and epic atmosphere, suggesting an important moment.
        & The sound of a train horn honking followed by a train passing by.
        & 1 \\
        \midrule
        \multirow{7}{*}{Subscene QA}
        & What do you hear between the piano playing and the guitar strumming? 
        & I hear a male voice speaking in English with a sad mood.
        & A male voice speaking in English with a sad mood.
        & 10 \\ \cmidrule{2-5}
        & What is happening between the sound of game music and the idling engine?
        & The scene consists of a person playing with the remote control with background whirring and clicking sounds.
        & A man is speaking in English.
        & 1 \\ \midrule
        \multirow{8}{*}{Needle QA}
        & What sudden auditory change occurs that might indicate a shift in the person's physical state while navigating the snow?
        & A significant increase in breathing and panting sounds, suggesting physical exertion or difficulty.
        & The sound of breathing becomes prominent, suggesting the person may be exerting effort or experiencing discomfort.
        & 9 \\ \cmidrule{2-5}
        & What does the woman mention about the view from the bedroom?
        & She mentions seeing the red and yellow cliffs from her bedroom in the morning.
        & She mentions that the view from the bedroom is amazing.
        & 2 \\ \midrule
        \multirow{7}{*}{Plot QA}
        & What is the mood conveyed by the male voice singing in the audio?
        & Sad.
        & The male voice singing in the audio conveys a deep and resonant mood.
        & 10 \\ \cmidrule{2-5}
        & What innovative feature does the man highlight about the Pulse smart watch?
        & It can DJ without using a phone.
        & The man highlights the smartwatch's ability to monitor heart rate and sleep patterns.
        & 1 \\ \midrule
        \multirow{4}{*}{General QA}
        & What natural sound is consistently present throughout the entire audio?
        & Birds chirping lightly.
        & Birds chirping lightly.
        & 10 \\ \cmidrule{2-5}
        & What sound is heard consistently throughout the audio clip?
        & Continuous background noise.
        & The sound of a rooster crowing.
        & 1 \\
        \bottomrule
    \end{tabular}
    }
\end{table*}

\begin{algorithm*}[h]
\caption{Weighted Bucketed Blending \& Dynamic Batching}
\label{alg:weighted-blend-dynamic-batch}
\begin{algorithmic}
\Require 
  \(\text{all\_data}\): A dictionary of datasets, each with \(\text{contents}\) (bucketed examples) and \(\text{weight} > 0\)
\Require 
  \(\text{epoch}\): Current epoch index
\Require 
  \(\text{IsBrokenFile}(\cdot)\): Checks if an audio file is unreadable
\Require 
  \(\text{ShuffleDictFixedRand}(\cdot, \text{seed})\): Shuffles the bucketed data deterministically
\Require 
  \(\text{batch\_by\_size\_fn}(\cdot)\): Function implementing dynamic batching by size (Algorithm~\ref{alg:dynamic-batch})

\Function{BlendData}{all\_data, epoch}
  \State \(\text{self\_data} \gets \{\text{data}: \{\},\, \text{total\_num}: 0\}\)
  \For{\( \text{dataset\_name} \in \text{all\_data}\)}
    \State \(\text{contents} \gets \text{all\_data}[\text{dataset\_name}].\text{contents}\)
    \State \(\text{weight} \gets \text{all\_data}[\text{dataset\_name}].\text{weight}\)
    \For{\( (\text{bucket\_idx}, \text{bucket\_data}) \in \text{contents}\)}
      \State \( \text{bucket\_total} \gets |\text{bucket\_data}|\)
      \If{ \( \text{bucket\_total} = 0 \)}
        \State \textbf{continue} \Comment{Skip empty buckets}
      \EndIf
      \State \( \text{start\_idx} \gets 
         \big\lfloor \text{epoch} \times \text{bucket\_total} \times \text{weight} \big\rfloor 
         \bmod \text{bucket\_total}\)
      \State \( \text{end\_idx} \gets 
         \big\lfloor (\text{epoch} + 1) \times \text{bucket\_total} \times \text{weight} \big\rfloor 
         \bmod \text{bucket\_total}\)
      \For{\( \text{idx} \gets \text{start\_idx} \text{ to } \text{end\_idx}-1\)}
        \If{ \((\text{idx} > 0)\) \(\wedge\) \((\text{idx} \bmod \text{bucket\_total} = 0)\)}
          \State \(\text{seed} \gets \text{SumOfCharCodes}(\text{dataset\_name} \,\Vert\, 
                \text{"epoch"} \,\Vert\, \lfloor \text{idx}/\text{bucket\_total} \rfloor )\)
          \State \(\text{bucket\_data} \gets \text{ShuffleDictFixedRand}(\text{bucket\_data}, \text{seed})\)
        \EndIf
        \State \( \text{key} \gets \text{idx} \bmod \text{bucket\_total}\)
        \State \( \text{item} \gets \text{bucket\_data}[\text{key}]\)
        \If{ \(\Call{IsBrokenFile}{\text{item}[\text{"name"}]}\)}
          \State \textbf{continue} \Comment{Skip broken files}
        \EndIf
        \State \(\text{self\_data}[\text{"data"}][\text{self\_data}[\text{"total\_num"}]] \gets \text{item}\)
        \State \(\text{self\_data}[\text{"total\_num"}] \gets \text{self\_data}[\text{"total\_num"}] + 1\)
      \EndFor
    \EndFor
  \EndFor
  \State \Return \(\text{self\_data}\)
\EndFunction

\Function{DynamicBatching}{\text{indices}, \text{num\_tokens\_fn}, \text{max\_tokens}, 
                           \text{max\_sentences}, \text{bsz\_mult}}
  \Comment{High-level wrapper for dynamic batching}
  \State \(\text{batches} \gets \text{batch\_by\_size\_fn}(\text{indices}, \text{num\_tokens\_fn}, 
         \text{max\_tokens}, \text{max\_sentences}, \text{bsz\_mult})\)
  \State \Return \(\text{batches}\)
\EndFunction
\end{algorithmic}
\end{algorithm*}

\begin{algorithm*}[h]
\caption{Dynamic Batching by Size (Pseudocode for \(\text{batch\_by\_size\_fn}\))}
\label{alg:dynamic-batch}
\begin{algorithmic}
\Require 
  \(\text{indices}\): Array of example indices
\Require 
  \(\text{numTokensVec}[i]\): Number of tokens (or frames) for \(\text{indices}[i]\)
\Require 
  \(\text{max\_tokens}, \text{max\_sentences}\): Batch constraints
\Require 
  \(\text{bsz\_mult}\): Required multiple of batch size (e.g., for multi-GPU alignment)

\Function{batch\_by\_size\_vec}{\text{indices}, \text{numTokensVec}, 
                                \text{max\_tokens}, \text{max\_sentences}, \text{bsz\_mult}}
  \State \(\text{batches} \gets [\,]\) \Comment{List of subarrays of \(\text{indices}\)}
  \State \(\text{batchStart} \gets 0\)
  \State \(\text{tailMaxTokens} \gets 0\)
  \State \(\text{batchMaxTokens} \gets 0\)

  \For{\( \text{pos} \gets 0 \text{ to } |\text{indices}| - 1\)}
    \State \(\text{tailMaxTokens} \gets \max(\text{tailMaxTokens}, \text{numTokensVec}[\text{pos}])\)
    \State \(\text{newBatchEnd} \gets \text{pos} + 1\)
    \State \(\text{newBatchMaxTokens} \gets \max(\text{batchMaxTokens}, \text{tailMaxTokens})\)
    \State \(\text{newBatchSentences} \gets \text{newBatchEnd} - \text{batchStart}\)
    \State \(\text{newBatchNumTokens} \gets \text{newBatchSentences} 
                 \times \text{newBatchMaxTokens}\)

    \State \(\text{overflow} \gets 
       (\text{newBatchSentences} > \text{max\_sentences} > 0) \;\lor\;
       (\text{newBatchNumTokens} > \text{max\_tokens} > 0)\)
    \State \(\text{sizeOk} \gets (\text{newBatchSentences} < \text{bsz\_mult}) \;\lor\;
       (\text{newBatchSentences} \bmod \text{bsz\_mult} = 0)\)

    \If{\(\text{overflow}\)}
      \Comment{Finalize the current batch and start a new one}
      \State \(\text{batches}.\Call{append}{\text{indices}[\text{batchStart}:\text{pos}]}\)
      \State \(\text{batchStart} \gets \text{pos}\)
      \State \(\text{batchMaxTokens} \gets \text{numTokensVec}[\text{pos}]\)
      \State \(\text{tailMaxTokens} \gets \text{numTokensVec}[\text{pos}]\)
    \ElsIf{\(\text{sizeOk}\)}
      \Comment{Optionally finalize if batch size is a multiple of \(\text{bsz\_mult}\)}
      \State \(\text{batches}.\Call{append}{\text{indices}[\text{batchStart}:\text{newBatchEnd}]}\)
      \State \(\text{batchStart} \gets \text{newBatchEnd}\)
      \State \(\text{batchMaxTokens} \gets 0\)
      \State \(\text{tailMaxTokens} \gets 0\)
    \Else
      \Comment{Continue accumulating examples in the current batch}
      \State \(\text{batchMaxTokens} \gets \text{newBatchMaxTokens}\)
    \EndIf

  \EndFor

  \Comment{Add any leftover examples in the final batch}
  \If{\(\text{batchStart} < |\text{indices}|\)}
    \State \(\text{batches}.\Call{append}{\text{indices}[\text{batchStart}:]}\)
  \EndIf

  \State \Return \(\text{batches}\)
\EndFunction
\end{algorithmic}
\end{algorithm*}

\begin{table*}[!t]
    \small
    \centering
    \caption{All datasets used to train our model. We mark the datasets used in Stage 1\textsuperscript{1}, Stage 2\textsuperscript{2}, Stage 3\textsuperscript{3} or multiple stages of training\textsuperscript{1,2,3}. Datasets marked with $*$ were added to AF2 over the ones already present in Audio Flamingo.}
    \resizebox{\linewidth}{!}{
    \begin{tabular}{cccc}
        \toprule
        Audio Type & Task & Datasets & \#Audio-Text Pairs \\ 
        \toprule
        \multirow{10}{*}{\shortstack{General\\Sound}} & \multirow{3}{*}{CAP} & WavCaps\textsuperscript{1} \citep{mei2024wavcaps}, Macs\textsuperscript{2} \citep{martin2021diversity}, & \multirow{3}{*}{$\sim$829 K} \\
        & & SoundDescs\textsuperscript{1}~\citep{oncescu2021audio}, Clotho-v2\textsuperscript{2} \citep{drossos2020clotho}, & \\
        & & WavText5K\textsuperscript{1}~\citep{deshmukh2022audio}, Laion-630k\textsuperscript{1}~\citep{10095969} & \\ 
        \cmidrule{2-4} 
        
        &  & Clotho-AQA\textsuperscript{2}~\citep{lipping2022clotho}, Open-AQA\textsuperscript{2}~\citep{10389742}, & \multirow{3}{*}{$\sim$1970 K} \\
        & AQA & Salmonn AQA\textsuperscript{2*}~\citep{tang2023salmonn}, AudioEntailment\textsuperscript{2*}~\citep{audioentail} \\
         & & CompA-R\textsuperscript{2*}~\citep{ghosh-etal-2024-gama}, AudioSkills\textsuperscript{1,2*} (\textit{ours}), LongAudio\textsuperscript{3*} (\textit{ours}) &\\
        \cmidrule{2-4} 
        & \multirow{3}{*}{CLS} & AudioSet \textsuperscript{2}~\citep{gemmeke2017audio}, FSD50k\textsuperscript{2} \citep{fonseca2021fsd50k}, & \multirow{3}{*}{$\sim$1091 K} \\
        & & CochlScene\textsuperscript{2} \citep{jeong2022cochlscene}, NonSpeech7K \textsuperscript{2}\citep{rashid2023nonspeech7k}, & \\
        & & Chime-Home\textsuperscript{2} \citep{foster2015chime}, Sonyc-UST\textsuperscript{2} \citep{cartwright2019sonyc} & \\ 
        \midrule
        
        \multirow{6}{*}{Music} & CAP & LP-MusicCaps\textsuperscript{2} \citep{doh2023lp}, MusicCaps\textsuperscript{2} \citep{agostinelli2023musiclm} & $\sim$1389 K \\ \cmidrule{2-4} 
        & & MusicQA\textsuperscript{2} \citep{liu2024music}, MusicAVQA\textsuperscript{2}~\citep{li2022learning} \\ & AQA & MusicBench\textsuperscript{2*}~\cite{melechovsky2023mustango}, Mu-LLAMA\textsuperscript{2*}~\cite{liu2024music} & $\sim$94 K  \\ 
        & & AudioSkills\textsuperscript{1,2*} (\textit{ours}), LongAudio\textsuperscript{3*} (\textit{ours}) \\ \cmidrule{2-4} 
        & \multirow{2}{*}{CLS} & NSynth\textsuperscript{2} \citep{engel2017neural}, MTG-Jamendo\textsuperscript{2}  \citep{bogdanov2019mtg}, & \multirow{2}{*}{$\sim$459 K} \\
        & & FMA\textsuperscript{2}   \citep{defferrard2016fma}, MusDB-HQ\textsuperscript{2} \citep{rafii2019musdb18}, & \\ \midrule

        \multirow{3}{*}{Speech} & \multirow{3}{*}{CLS} & MSP-Podcast\textsuperscript{1}  \citep{lotfian2017building}, Emov-DB\textsuperscript{2} \citep{adigwe2018emotional} & \multirow{3}{*}{$\sim$92 K} \\
        & & JL-Corpus\textsuperscript{2} \citep{james2018open}, Tess\textsuperscript{2} \citep{pichora2020toronto}, & \\
        & & MELD\textsuperscript{2} \citep{poria2018meld}, OMGEmotion\textsuperscript{2} \citep{barros2018omg} & \\ \midrule
    \end{tabular}}
    \label{tab:dataset_by_type}
\end{table*}

\begin{table*}[!t]
    \centering
    \caption{All datasets used for evaluation in AF2, with additional usage/contextual notes for foundational datasets.}
    \label{tab:inference_data}
    \begin{tabular}{cccp{9.5cm}}
        \toprule
        \textbf{Dataset Type} & \textbf{Audio Type} & \textbf{Task} & \textbf{Datasets} \\
        \toprule
       \multirow{16}{*}{Foundational} 
       &  \multirow{12}{*}{Sound} 
          & \multirow{1}{*}{CAP} 
             & Clotho-v2~\cite{drossos2020clotho}, AudioCaps~\cite{kim2019audiocaps}  
       \\ \cmidrule(lr){3-4}
       &  & \multirow{8}{*}{CLS} 
             & UrbanSound8K (USD8K)~\cite{salamon2014dataset} (\emph{environmental sound classification}), \\
       &  &  
             & ESC50~\cite{piczak2015dataset} (\emph{environmental sound classification}), \\
       &  &  
             & CochlScene~\cite{jeong2022cochlscene} (\emph{acoustic scene classification}), FSD50k~\cite{fonseca2021fsd50k} (\emph{sound event classification}) \\
       &  &  
             & CREMA-D~\cite{cao2014crema} (\emph{emotion classification}), Ravdess~\cite{livingstone2018ryerson} (\emph{emotion classification}), \\
       &  &  
             & NonSpeech7k~\cite{rashid2023nonspeech7k} (\emph{non-speech audio classification}) 
       \\ \cmidrule(lr){3-4}
       &  & \multirow{1}{*}{AQA} 
             & Clotho-AQA~\cite{lipping2022clotho} (\emph{audio question answering}) 
       \\ \cmidrule(lr){2-4}
       & \multirow{5}{*}{Music} 
          & \multirow{3}{*}{CLS} 
             & NSynth (NS)\textsubscript{source}~\cite{nsynth2017} (\emph{instrument classification}), \\
       &  &  
             & NSynth (NS)\textsubscript{instrument}~\cite{nsynth2017} (\emph{instrument classification}), \\
       &  &  
             & GTZAN~\cite{sturm2013gtzan} (\emph{genre classification}), Medley-solos-DB~\cite{lostanlen_2019_1344103} (\emph{instrument classification}) 
       \\ \cmidrule(lr){3-4}
       &  & \multirow{1}{*}{AQA} 
             & MusicAVQA\textsubscript{audio}~\cite{li2022learning} (\emph{music audio question answering}) 
       \\ 
       \midrule
       \multirow{7}{*}{Reasoning} 
       & \multirow{4}{*}{Sound} 
          & \multirow{4}{*}{AQA} 
             & OpenAQA~\cite{gong2024listen}, MMAU Sound~\cite{sakshi2024mmau}, \\
       &  &  
             & AudioEntailment (AE) Clotho~\cite{audioentail}, AudioEntailment (AE) AudioCaps~\cite{audioentail}, \\
       &  &  
             & CompA-R-\textit{test}~\cite{ghosh-etal-2024-gama} 
       \\ \cmidrule(lr){2-4}
       & \multirow{3}{*}{Music} 
          & \multirow{3}{*}{AQA} 
             & MMAU Music~\cite{sakshi2024mmau}, MuchoMusic~\cite{weck2024muchomusic}, \\
       &  &  
             & MusicInstruct (MI) (Long)~\cite{deng2023musilingo}, MusciQA~\cite{liu2024music} \\
        \bottomrule
    \end{tabular}
\end{table*}

\begin{table*}[h]
    \centering
    \caption{More examples from AudioSkills.}
    \begin{tabular}{p{3cm}p{11.5cm}}
        \toprule
        \textbf{Category} & \textbf{Example} \\
        \midrule
        {Temporal Relationship Identification} & \textbf{Order:} In what sequence do the sounds appear in the audio? (A) Car engine, honk, music (B) Music, honk, car engine \\
        & \textbf{Attribute:} When does the breathing sound change over time? (A) Disappear (B) Get louder (C) Get soft \\
        & \textbf{Grounding:} When does the sound of a bird chirp occur in the audio? (A) Beginning (B) Middle (C) End \\
        & \textbf{Referring:} What sound appears first in the audio? (A) Music (B) Car honk (C) Dog bark \\ \cmidrule{2-2}
        & \textbf{Order:} In what sequence do the footsteps appear in the audio? (A) Slow footsteps, running, door slam (B) Running, door slam, slow footsteps (C) Door slam, slow footsteps, running \\
        & \textbf{Attribute:} How does the volume of the rain sound change over time? (A) Increases (B) Decreases (C) Stays the same \\
        & \textbf{Grounding:} When does the sound of a baby crying occur in the audio? (A) Beginning (B) Middle (C) End \\
        & \textbf{Referring:} What sound appears immediately after the thunder? (A) Wind blowing (B) Rainfall (C) Car alarm \\
        \midrule
        {Contextual Speech Event Reasoning} &  How might the playful interaction between the boy and the goat be affected by the tone of the speech and the background sounds? \\ \cmidrule{2-2}
        & What can be deduced about the environment and the relationship between the boy and the goat from the audio events? \\
        \midrule
        {Contextual Sound Event Reasoning} & What might the presence of the music and its continuous play suggest about the overall atmosphere of the scene? \\
        & Based on the timing and sequence of impact sounds and male speech, identify the possible work environment depicted in the audio. \\
        \midrule
        {Counting} & \textbf{Level 1:} How many times was the hammer sound heard? \\ \cmidrule{2-2}
        & \textbf{Level 2:} How many times did the first sound occur in the entire audio? \\
        \midrule
        {Information Extraction} & What key is the main melody of the audio played in? (A) C major (B) A major (C) A\# major (D) G\# major \\ \cmidrule{2-2}
        & How does the melody in the audio contribute to the hypnotic effect of the music? (A) By changing frequently (B) By maintaining a consistent and repetitive loop (C) By using multiple key changes \\
        \midrule
        {General Reasoning} & Considering the chord sequence Bm7, D, Bm7, D in the audio, what is the role of the melody in relation to these chords? (A) It follows the progression (B) It decorates the progression \\ \cmidrule{2-2}
        & How does the melody in the audio contribute to the hypnotic effect of the music? (A) By changing frequently (B) By maintaining a consistent loop \\
        \midrule
        {Attribute Identification} & Which event has the highest pitch in the audio? \\ \cmidrule{2-2}
        & Which is the least loud event in the audio? \\
        \bottomrule
    \end{tabular}
    \label{tab:audioskills_exp}
\end{table*}

\begin{figure*}[h]
    \centering
    \includegraphics[width=\linewidth]{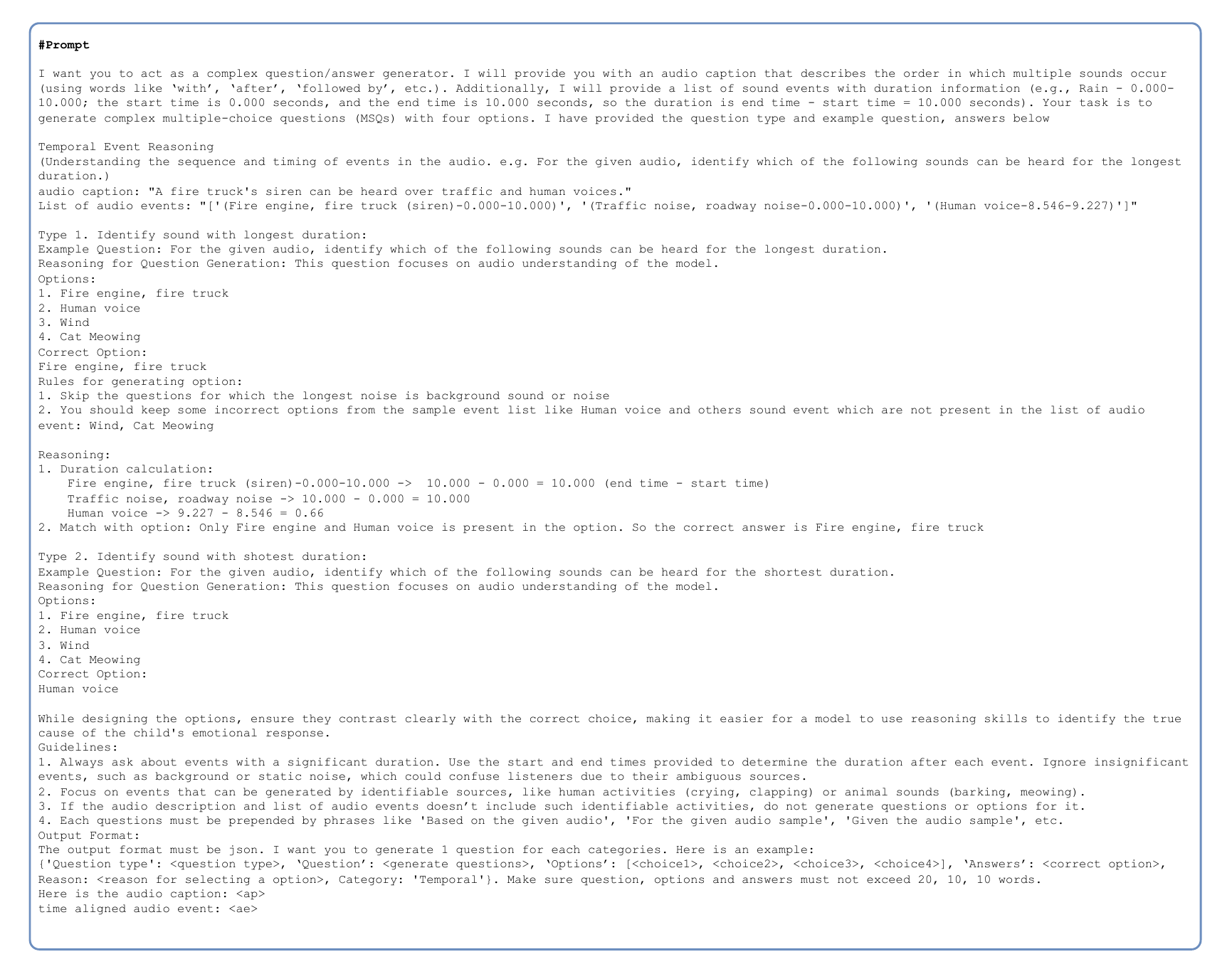}
    \caption{\small Prompt 2 used for generating \textbf{Temporal QA}.}
    \label{fig:mmau_temporal_prompt}
\end{figure*}

\begin{figure*}[h]
    \centering
    \includegraphics[width=\linewidth]{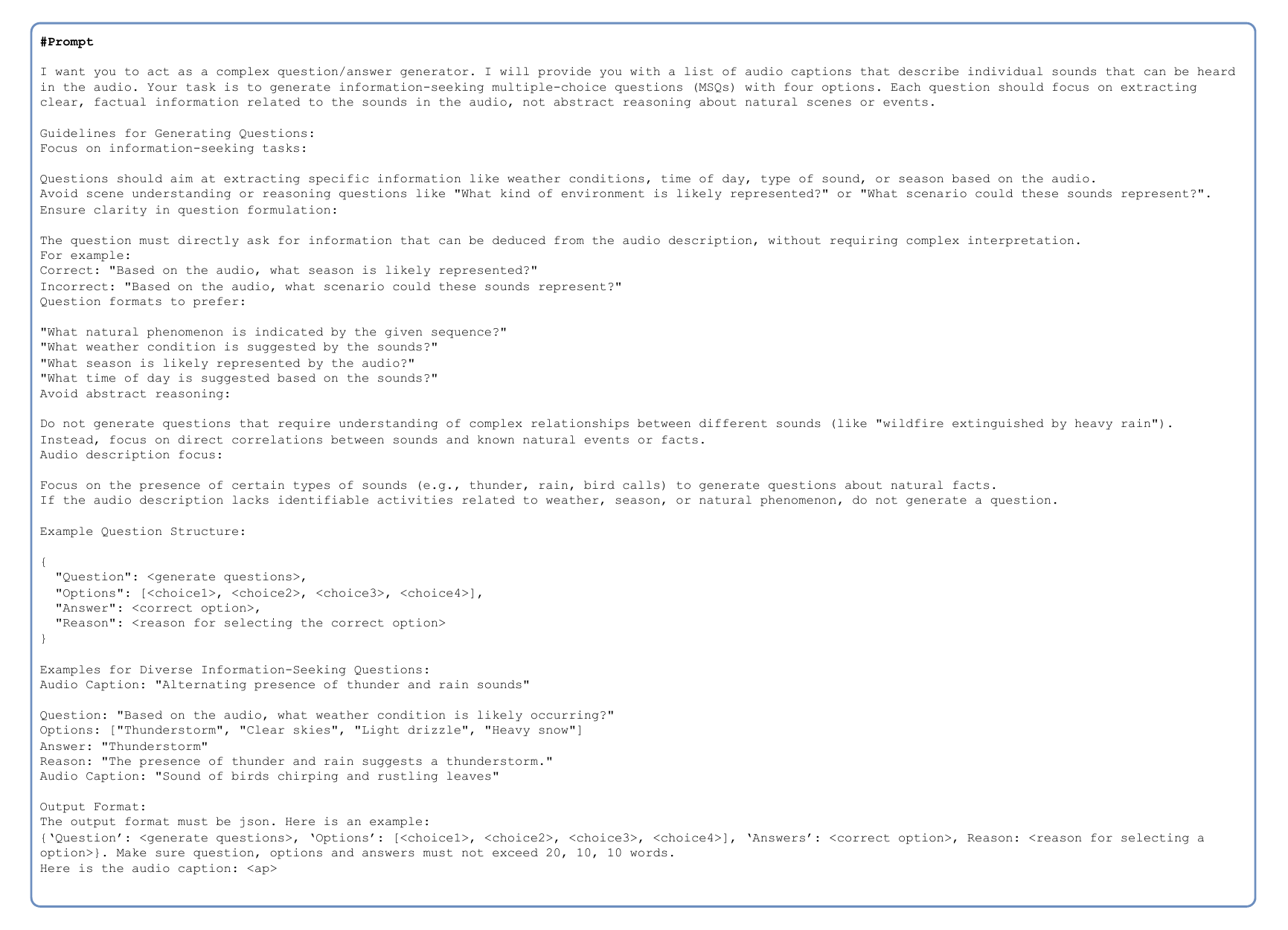}
    \caption{\small Prompt 2 used for generating \textbf{Contextual Sound Event Reasoning QA} for AudioSkills.}
    \label{fig:eco_acoustic_prompt}
\end{figure*}

\begin{figure*}[h]
    \centering
    \includegraphics[width=\linewidth]{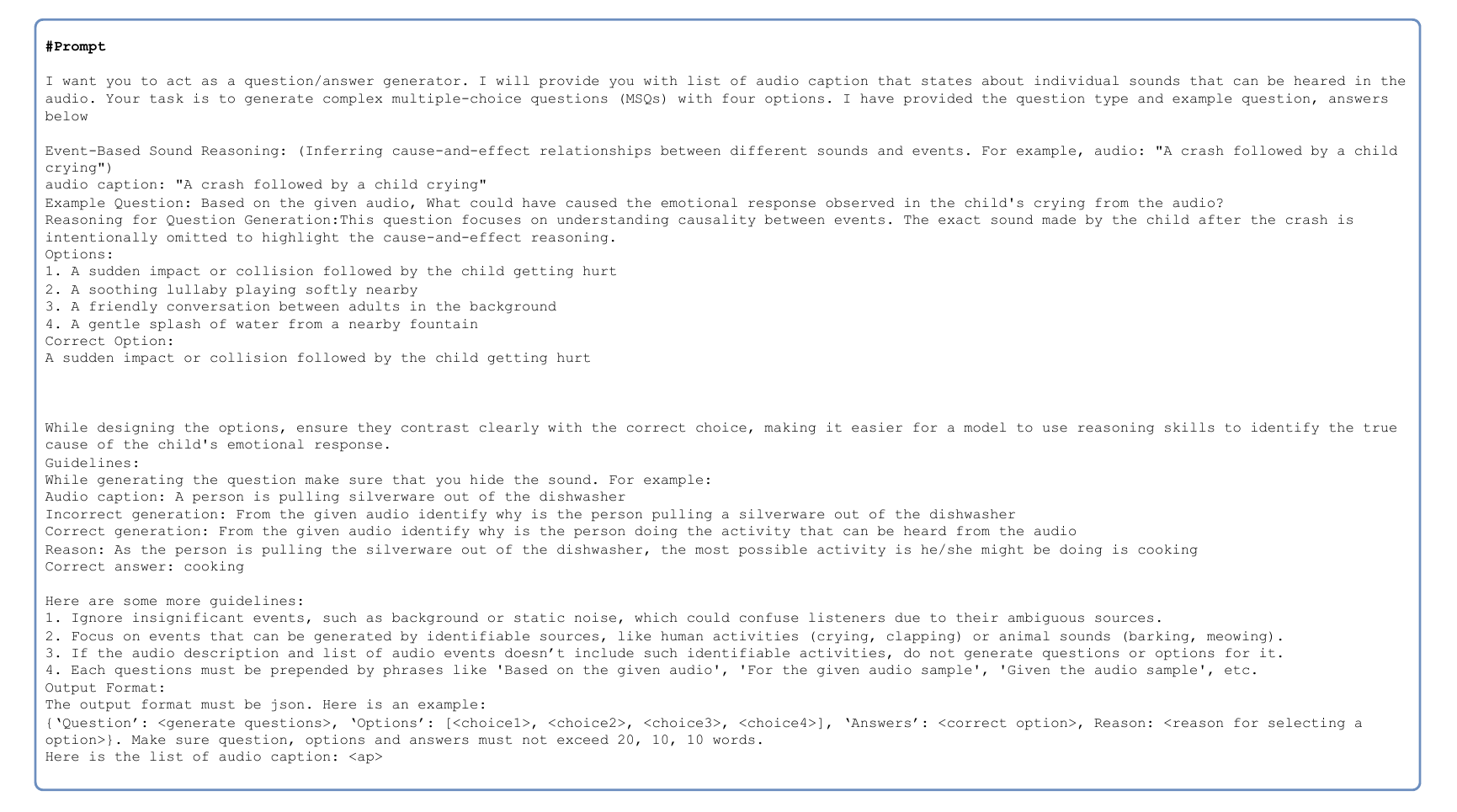}
    \caption{\small Prompt 3 used for generating \textbf{Contextual Sound Event Reasoning QA} for AudioSkills.}
    \label{fig:mmau_complex_3_prompt}
\end{figure*}

\begin{figure*}[h]
    \centering
    \includegraphics[width=\linewidth]{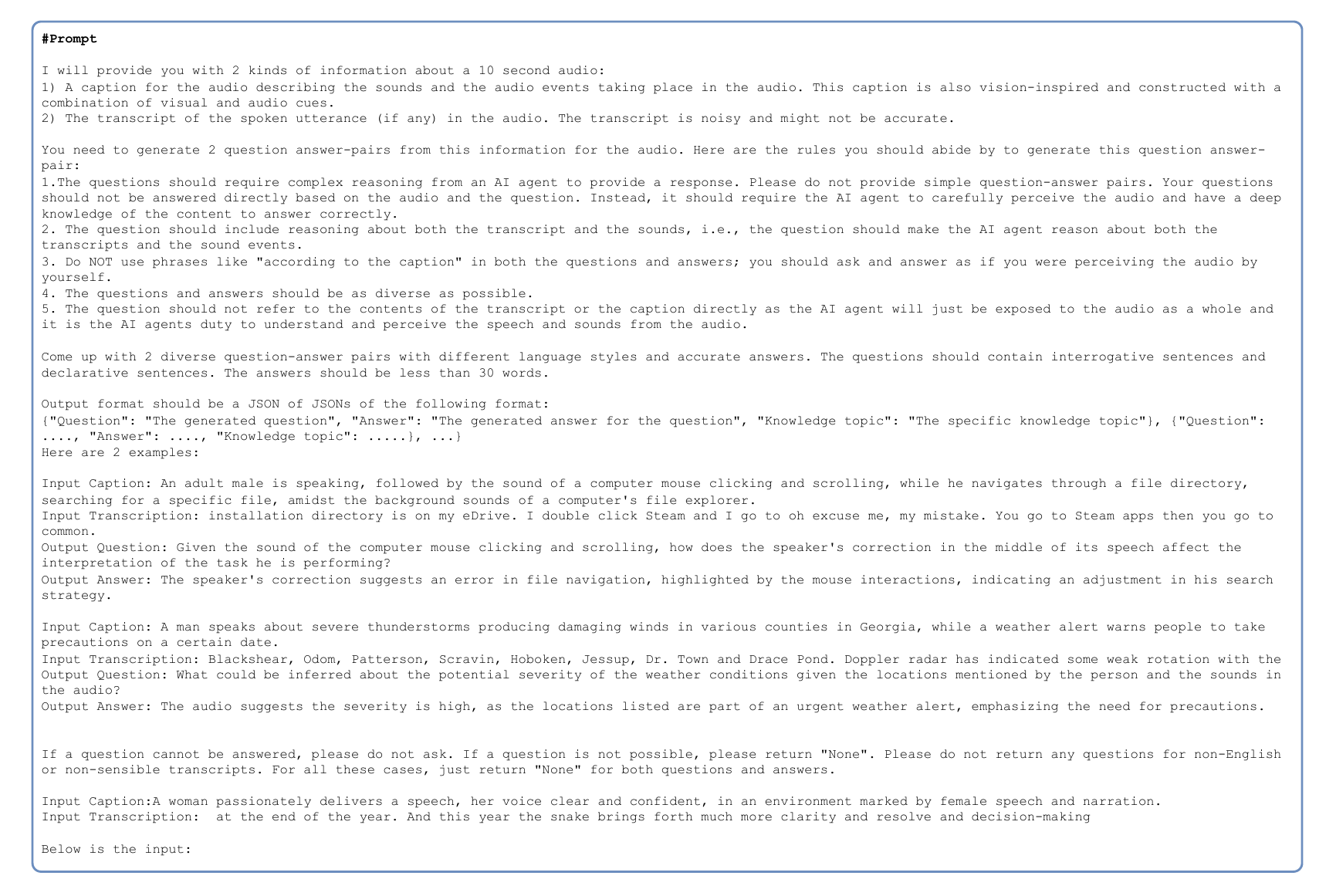}
    \caption{\small Prompt 4 used for generating \textbf{Contextual Speech Event Reasoning QA} for AudioSkills.}
    \label{fig:mmau_speech_prompt}
\end{figure*}

\begin{figure*}[h]
    \centering
    \includegraphics[width=\linewidth]{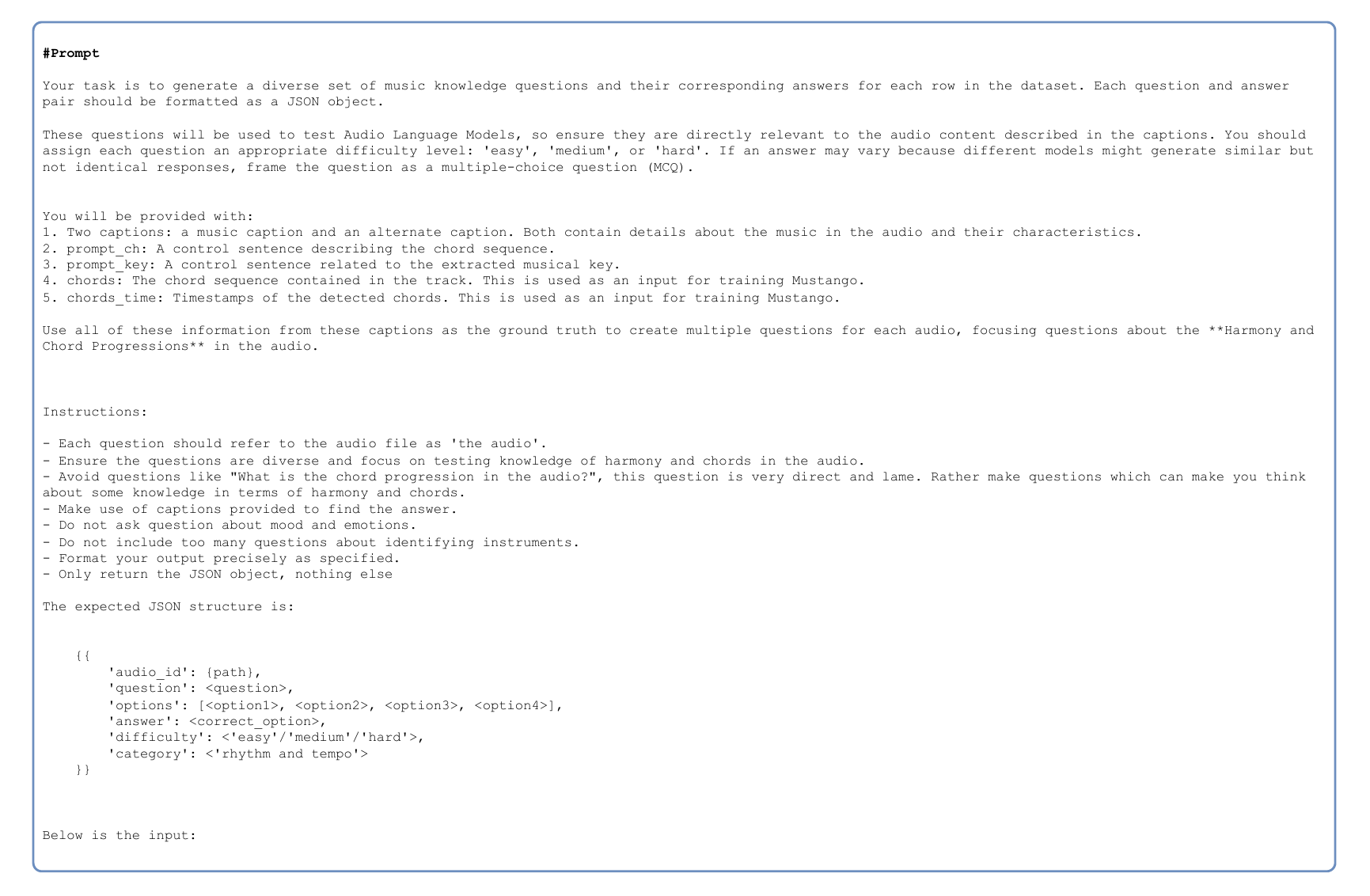}
    \caption{\small Prompt 1 used for generating \textbf{Information Extraction QA} for AudioSkills.}
    \label{fig:harmony_mmau_prompt}
\end{figure*}

\begin{figure*}[h]
    \centering
    \includegraphics[width=\linewidth]{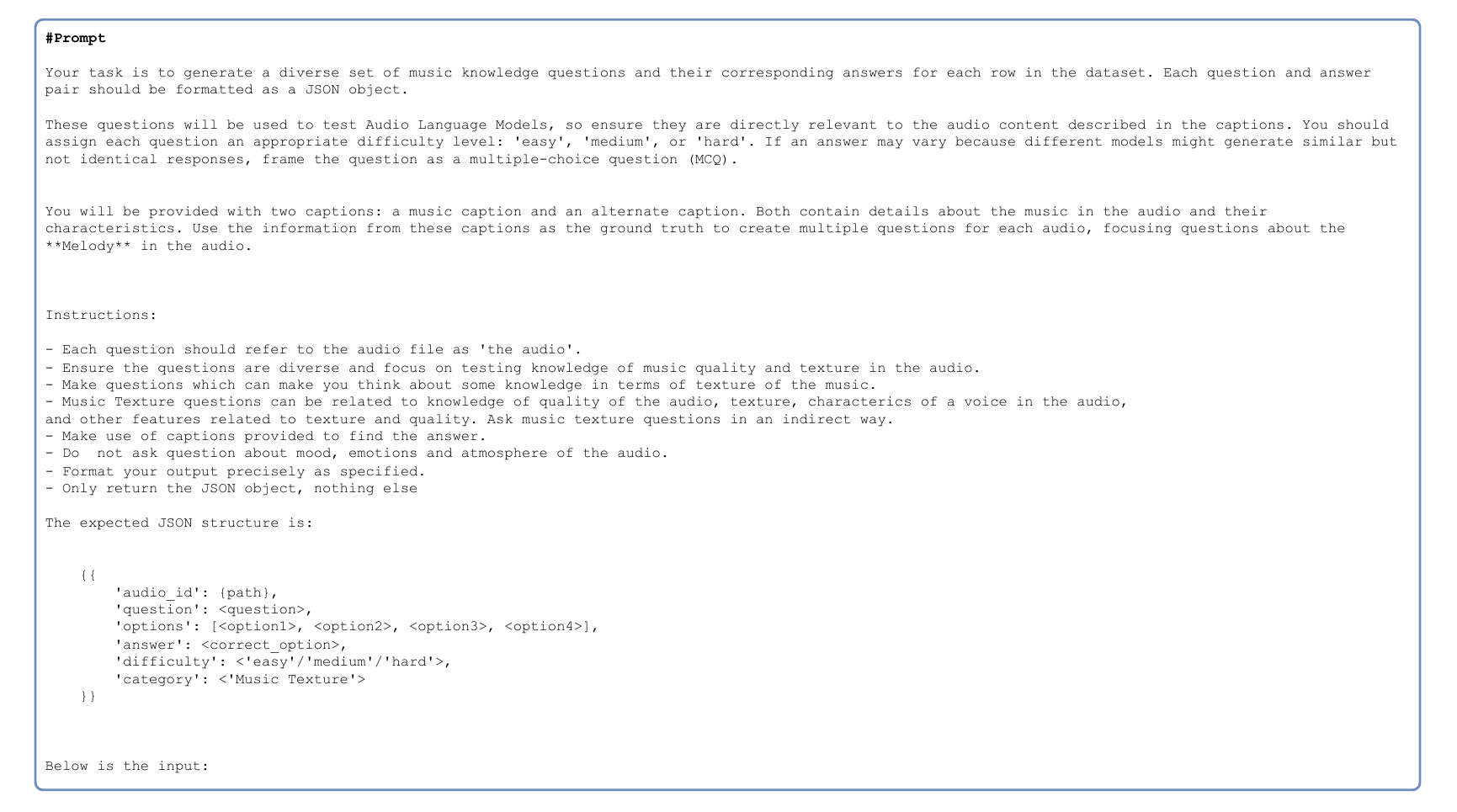}
    \caption{\small Prompt 3 used for generating \textbf{Information Extraction QA} for AudioSkills.}
    \label{fig:texture_mmau_prompt}
\end{figure*}

\begin{figure*}[h]
    \centering
    \includegraphics[width=\linewidth]{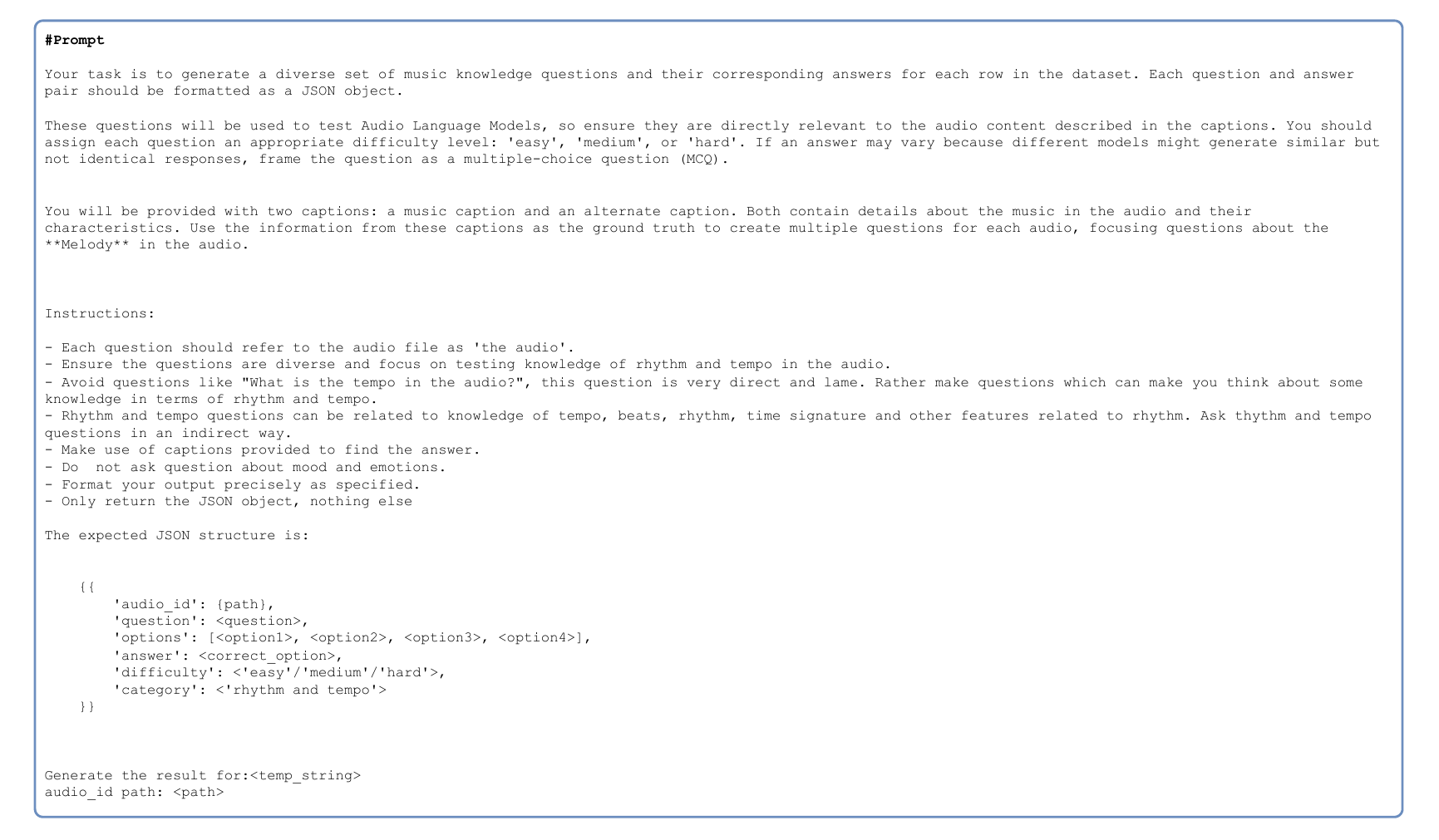}
    \caption{\small Prompt 4 used for generating \textbf{Information Extraction QA} for AudioSkills.}
    \label{fig:rhythm_prompt}
\end{figure*}

\begin{figure*}[h]
    \centering
    \includegraphics[width=\linewidth]{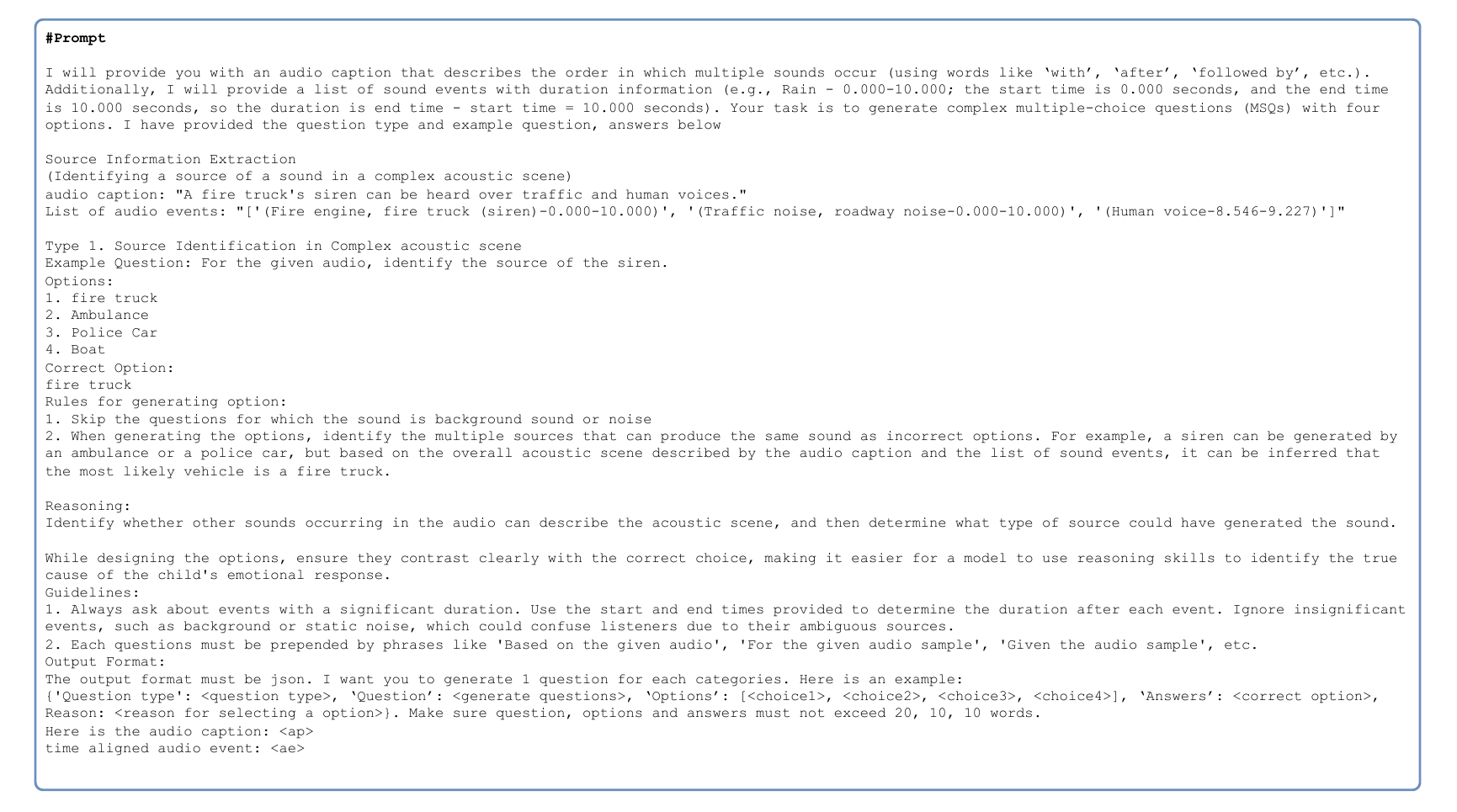}
    \caption{\small Prompt 5 used for generating \textbf{Information Extraction QA} for AudioSkills.}
    \label{fig:sound_source_prompt}
\end{figure*}

\begin{table*}[ht]
\centering
\caption{Prompts designed for specific audio processing tasks, used to transform foundational audio understanding datasets into QA pairs.}
\begin{tabular}{|l|p{12cm}|}
\hline
\textbf{Task Name} & \textbf{Prompts} \\ \hline
Instrument Classification & What instrument is playing in the audio?, Identify the instrument in this audio clip., Classify the instrument heard in the music., What is the primary instrument in this track?, Provide the instrument tag for this audio., Which instrument is most prominent in the clip?, What instrument is featured in this music?, What is the instrument label for this audio., Identify the type of instrument being played., Classify the instrument based on the audio content., What is the main instrument in this track?, What musical instrument is audible in the audio?, Provide the instrument classification for this audio., What type of instrument is represented in this clip?, What instrument is most recognizable in this music?, Classify the music by its primary instrument., Which instrument defines the sound of this track?, Identify the instrument most prominently heard in the audio., What instrument category does this audio belong to?, Determine the instrument used in the music clip. \\ \hline
Audio Captioning & Caption the input audio., Describe the sounds in the audio., Provide a caption for the audio., What is happening in this audio clip?, Summarize the audio content., Describe the events in the audio., What sounds can you hear in this audio?, Give a detailed description of the audio scene., Caption the sounds in the audio., What is the main action or event in the audio?, How would you describe the sounds in this audio?, Describe the background sounds in the audio., What can be heard in the audio?, Give a brief description of the sounds in this audio., Describe the setting based on the audio., Provide a caption describing the audio scene., What events are occurring in this audio?, Give a general description of the audio content., What's going on in the audio clip?, Describe any notable sounds in the audio., Provide a summary of the audio sounds., What is taking place in the audio?, Give a description of the atmosphere in the audio., Describe the main sounds in the audio., What does the audio depict?, Describe the audible events in this audio., Summarize what you hear in the audio., Provide a descriptive caption for the audio clip., What is the soundscape in this audio?, How would you describe the scene from the audio?, Describe the key elements heard in the audio. \\ \hline
Music Captioning & Describe the music in the audio., Provide a caption for the music., Summarize the characteristics of the music., Summarize the music content in a sentence., Caption the input music. \\ \hline
Speech Emotion Classification & Identify the emotion in the utterance., What is the emotion of the utterance?, Describe the emotional tone in this audio., What emotion is expressed in the audio?, What is the primary emotion in this recording?, Identify the feeling conveyed in the utterance., How would you describe the emotion in the audio?, What emotion stands out in this audio clip?, Describe the mood of the speaker., What sentiment is present in the audio?, What is the dominant emotion in this audio?, Classify the emotion expressed in the clip., What is the emotional state of the speaker?, What feeling does the speaker convey?, Identify the mood in this audio., What is the overall emotion of this recording?, How would you classify the emotion in the clip?, What kind of emotion is detectable in the audio?, Describe the emotional expression of the speaker., What is the perceived emotion in the recording?, Determine the emotion conveyed by the speaker., What is the underlying emotion in this utterance?, Classify the speaker's emotional tone., What emotional state is reflected in the audio?, Describe the speaker's feeling in this clip., What is the prevailing emotion in this recording?, Identify the primary mood of the utterance., What emotional quality does the audio suggest?, What is the affective tone of the speaker?, What emotion can be inferred from the speaker's tone? \\ \hline
\end{tabular}
\label{tab:task_prompts_1}
\end{table*}

\begin{table*}[ht]
\centering
\caption{\small Prompts designed for specific audio processing tasks, used to transform foundational audio understanding datasets into QA pairs.}
\begin{tabular}{|l|p{12cm}|}
\hline
\textbf{Task Name} & \textbf{Prompts} \\ \hline
Audio Event Classification & What are the unique sounds in the audio?, Provide a comma separated list of all sounds you hear in the input audio., What sounds can you hear in the audio?, List all the sounds present in the audio., Identify the distinct sounds in the audio clip., What specific sounds are detectable in the audio?, Give a list of sounds you can identify in this audio., What types of sounds are included in the audio?, List the audible elements in the audio., What are the main sounds you notice in this audio?, Provide a list of all recognizable sounds in the audio., What sounds are prominent in the audio clip?, List all distinct sounds in the audio., What are the different sounds occurring in this audio?, Identify and list each sound in the audio., Provide a detailed list of sounds heard in the audio., What sounds stand out in the audio?, List each unique sound in this audio clip., Which sounds are identifiable in the audio?, What sounds are repeated in the audio?, What environmental sounds can you hear in this audio?, List all background sounds in the audio., Identify the primary sounds in the audio., What are the foreground sounds in the audio?, List any musical or rhythmic sounds present., Provide a list of natural sounds in the audio., What artificial or mechanical sounds can be heard?, What ambient sounds are in the background?, What noticeable sounds can be identified in this clip?, List all sound sources in the audio. \\ \hline
Sentiment Classification & Identify the sentiment of the utterance., What is the sentiment of the utterance?, What is the primary sentiment of the utterance in this recording? \\ \hline

Music Understanding and Classification & Generate music tags including genre, instrument, and mood, Identify the genre, instrument, and theme of this music, What are the tags for genre, instrument, and mood for this track, Describe the genre, instruments, and theme of the audio, What is the genre and mood of the music, Provide tags for the genre, mood, and instruments in the track, What music tags best describe this audio, including genre and theme, Generate tags for the genre and atmosphere of the music, What genre and instruments are featured in this track, What are the primary genre and theme of this audio, Describe the mood, genre, and instrumentation in the music, Identify the genre and musical style of the audio, What genre tags and mood fit this music, What genre and theme are represented in this track, Provide tags for genre, style, and mood of the audio, Generate descriptive tags for genre and atmosphere in the music, Identify the genre, theme, and instrumental tags for this track, What tags describe the genre and emotional tone of this audio, Provide genre and theme tags for the music, What genre, mood, and instruments define this audio clip, What genre and style does the music represent, Describe the tags for genre and instruments in the audio, What genre and theme best describe this track, Provide tags that include genre and emotional tone for the music, What genre and mood characterize the music in this audio, Generate tags for genre, style, and mood in the track, What is the genre and overall theme of the audio, Identify genre and instrumentation tags for this music, What genre, theme, and mood does the audio convey, Provide tags that describe the genre and atmosphere of the track.\\ \hline

Genre Classification & What is the genre of the music, Identify the genre of this audio clip, Classify the genre of the music in the audio, What musical genre does this clip represent, Provide the genre tag for this music, Which genre best describes the audio, What is the genre label for this music clip, Classify the genre based on the audio content, What category of music does this audio belong to, Determine the genre of the audio, Identify the genre of this track, What is the musical genre for the given audio, What kind of music genre is this, Provide the genre classification for this audio, What type of genre is represented in this audio clip, What genre does this music fall under, Classify the music by its genre, What style or genre does the audio represent, What is the most suitable genre for this track, Provide a genre tag for the music in this clip.\\ \hline
\end{tabular}
\label{tab:task_prompts_2}
\end{table*}

\end{document}